\documentclass[pra,twocolumn,superscriptaddress,aps,showpacs]{revtex4-1}
\usepackage{graphicx}
\usepackage{amsmath}
\usepackage{amsfonts}
\usepackage{amssymb}
\usepackage{wasysym}
\usepackage{times}
\usepackage{mathtools}
\usepackage{lineno}
\usepackage{hyperref}
\usepackage{longtable}
\usepackage{subfigure}
\usepackage[usenames,dvipsnames]{xcolor}
\usepackage{bm}
\everymath{\displaystyle}
\hypersetup{
  colorlinks=true,
  citecolor=blue,
  linkcolor=blue,
  urlcolor=blue}

\begin{document}
\title{Quantum phases of dipolar bosons in multilayer optical lattice}

\author{Soumik Bandyopadhyay}
\affiliation{Physical Research Laboratory,
             Ahmedabad - 380009, Gujarat,
             India}
\affiliation{Indian Institute of Technology Gandhinagar,
             Palaj, Gandhinagar - 382355, Gujarat,
             India}
\affiliation{INO-CNR BEC Center and Department of Physics, 
             University of Trento, Via Sommarive 14, I-38123 Trento, Italy}
\author{Hrushikesh Sable}
\affiliation{Physical Research Laboratory,
             Ahmedabad - 380009, Gujarat,
             India}
\affiliation{Indian Institute of Technology Gandhinagar,
             Palaj, Gandhinagar - 382355, Gujarat,
             India}
\author{Deepak Gaur}
\affiliation{Physical Research Laboratory,
             Ahmedabad - 380009, Gujarat,
             India}
\affiliation{Indian Institute of Technology Gandhinagar,
             Palaj, Gandhinagar - 382355, Gujarat,
             India}
\author{Rukmani Bai}
\affiliation{Physical Research Laboratory,
             Ahmedabad - 380009, Gujarat,
             India}
\affiliation{Institute for Theoretical Physics III and
             Center for Integrated Quantum Science and Technology,\\
             University of Stuttgart, 70550 Stuttgart,
             Germany}
\author{Subroto Mukerjee}
\affiliation{Department of Physics,  
             Indian Institute of Science,  
             Bangalore - 560012, 
             India}             
\author{D. Angom}
\affiliation{Physical Research Laboratory,
             Ahmedabad - 380009, Gujarat,
             India}
\affiliation{Department of Physics, Manipur University,
             Canchipur - 795003, Manipur, India}	     
\date{\today}

\begin{abstract}
We consider a minimal model to investigate the quantum phases of hardcore,
polarized dipolar atoms confined in multilayer optical lattices. The model is
a variant of the extended Bose-Hubbard model, which incorporates intralayer
repulsion and interlayer attraction between the atoms in nearest-neighbour
sites. We study the phases of this model emerging from the
competition between the attractive interlayer interaction and the interlayer
hopping. Our results from the analytical and cluster-Gutzwiller mean-field 
theories reveal that multimer formation occurs in the regime of weak intra and 
interlayer hopping due to the attractive interaction. In addition, intralayer 
isotropic repulsive interaction results in the checkerboard ordering of the 
multimers. This leads to an incompressible checkerboard multimer phase at 
half-filling. At higher interlayer hopping, the multimers are destabilized to 
form resonating valence-bond like states. 
Furthermore, we discuss the effects of thermal fluctuations on the
quantum phases of the system.
\end{abstract}

\maketitle


\section{Introduction}\label{Introduction}
The experimental demonstration of the superfluid to Mott-insulator phase transition in 
the 1D, 2D and 3D optical lattices with bosonic 
atoms~\cite{greiner_2002_1, greiner_2002_2, stoferle_2004, folling_2006, 
spielman_2007, baier_2016} marks a paradigm shift in the exploration of
quantum phase transitions. High tunability of the system parameters and almost 
near complete isolation of ultracold atoms in the optical lattice potentials 
have thus opened a new avenue to explore quantum many-body 
physics~\cite{jaksch_2005, bloch_2008_1, bloch_2008_2}. At present, optical lattices are considered as macroscopic quantum simulators of condensed matter 
systems~\cite{lewenstein_2007, gross_2017}. They have been employed to 
understand properties 
of equilibrium quantum phases~\cite{jaksch_1998, goral_2002, danshita_2009, yi_2007, zhang_2015, bandyopadhyay_2019, pal_2019, suthar_2020, suthar_2020_2}, 
characteristics of collective excitations~\cite{suthar_2015,suthar_2016, 
suthar_2017, krutitsky_2010, krutitsky_2011, saito_2012} , 
non-equilibrium dynamics of quantum phase 
transitions~\cite{chen_2011, braun_2015, shimizu_misf_2018, shimizu_dwss_18, 
shimizu_dwsf_18, zhou_2020, sable_21},
quantum thermalisation~\cite{kaufman_2016, bohrdt_2017}, the many-body 
localisation transition~\cite{sierant_2017, sierant_2018}, 
driven and dissipative dynamics~\cite{tomita_2017, roy_2019}, etc. Optical 
lattices, when loaded with Bose-Einstein condensed atoms, can simulate the 
Bose-Hubbard model (BHM) ~\cite{fisher_1989, jaksch_1998}. This model is the 
bosonic counterpart of the Hubbard model~\cite{hubbard_1963}. The latter has 
been considered as the prototypical model to understand the properties of 
interacting electrons in the tight binding regime. The BHM considers only 
nearest-neighbour hopping and onsite contact interaction. These basic 
considerations are sufficient to describe the properties of the superfluid 
and Mott insulator phases of neutral atoms. But, BHM can not describe phases 
arising from the long-range inter-atomic interactions. 

 A minimal extension of the BHM includes nearest-neighbour (NN) interactions. 
The model is, then, referred to as the extended BHM. Several theoretical 
studies have analyzed the quantum phases of this model~\cite{scarola_2005, 
kovrizhin_2005, sengupta_2005, mazzarella_2006, iskin_2011, dutta_2015, 
suthar_2020}. This interaction can induce periodic density modulation, 
which is  a type of diagonal order. The system can also host a supersolid, 
which is characterized by the coexistence of diagonal and off-diagonal 
long-range order, and has been a topic of extensive 
research~\cite{boninsegni_2012}. The extended BHM has been experimentally 
realized in a 3D optical lattice loaded with magnetic dipolar 
atoms~\cite{baier_2016}. But, to be precise, the  inter-atomic dipole-dipole 
interaction is long-range and anisotropic. Consequently, several theoretical 
studies have investigated the effects of these features on the quantum 
phases~\cite{goral_2002, danshita_2009, yi_2007, menotti_2007, zhang_2015, 
bandyopadhyay_2019, wu_2020}, leading to a further generalisation of BHM. In 
this context, the dimensionality of the lattice plays an important role. For 
example, for dipoles polarized perpendicular to the lattice plane, the 
intralayer NN interaction is repulsive and isotropic. But, the interlayer NN 
interaction is attractive. Such an anisotropy can stabilize additional quantum 
phases in a 3D system. A simplified or minimal 3D lattice system is stacking 
two layers of a 2D lattice. While with the increase in number of layers, 3D 
properties of the system become more prominent.  

Recent observations of unconventional superconductivity~\cite{cao_2018_1}
and correlated insulating phase~\cite{cao_2018_2} in twisted bilayer graphene
have provided an impetus for studies on bilayer 
systems~\cite{lian_2019, wu_2018, gonzalez_2019_1, gonzalez_2019_2,
pizarro_2019, mahapatra_2020}. There are proposals to simulate the physics of 
twisted bilayers using optical 
lattice set-ups~\cite{gonzalez_2019_2, luo_2021}. Further more, the bilayer 
optical lattice loaded with polarized dipolar atoms, can host a multitude of 
quantum phases, which are absent in the monolayer lattice. In particular, due 
to the attractive interlayer interaction, there can be a pairing between the 
atoms in the different layers. The system can, then, host phases like the 
pair superfluid~\cite{trefzger_2009, naini_2013, macia_2014}, 
pair supersolid~\cite{trefzger_2009, naini_2013}, etc. But, the interlayer 
hopping can destabilize the pairs, and lead to the formation of phases like the valence bond solid and supersolid~\cite{ng_2015}. 

 In this work, we investigate the quantum phases of polarised dipolar atoms confined to a multi-layer optical lattice. In particular, we focus 
on bi- and tri-layer lattices. We study the ground state quantum phases of 
the system using the cluster Gutzwiller mean-field 
theory~\cite{Buonsante_04,Yamamoto_09,Pisarski_11,McIntosh_12,luhmann_13}. Our 
study reveals the existence of exotic quantum phases arising from the 
competitions of the intra- and interlayer interactions and hoppings. This 
study encapsulates one of the important aspects of the multi-layer systems by varying the interlayer hopping from the weak to the strong domain. We supplement the 
numerical phase boundaries between the incompressible and compressible phases 
through an analytical approach based on mean-field perturbation theory. 
Unlike earlier studies based on the site-decoupling scheme
\cite{oosten_2001, iskin_2009, bandyopadhyay_2019, bai_2020}, here 
we include the interlayer hopping in the unperturbed Hamiltonian and apply the method with respect to multiple sites. This approach can be thought of 
as a cluster generalisation of the site-decoupling method. Furthermore, we extend our study to the finite temperature domain in order to understand the stability of quantum phases against thermal fluctuations.

We have organized the remainder of this paper as follows.  In 
Sec.~\ref{sec_model}, we discuss the extended BHM apt for a description of 
polarized dipolar atoms in multi-layers of 2D square optical lattice. In 
Secs.~\ref{cgmf_theory}-\ref{quant_phases}, we give a brief account of the 
cluster Gutzwiller mean-field theory, adapted numerical procedure to solve the 
model, and the quantum phases of the system. In Sec.~\ref{mf_phase_boundaries}, 
we discuss the mean-field perturbation analysis to obtain analytical phase 
boundaries between incompressible and compressible phases. 
In Sec.~\ref{sec_results}, we present and elaborate the phase diagrams of the 
bi and tri-layer lattice systems, and then discuss the effects of finite 
temperature on the quantum phases. In Sec.~\ref{conclusions}, we summarize our 
key results and conclude.
\begin{figure}[ht]
\includegraphics[width = 9.0cm]{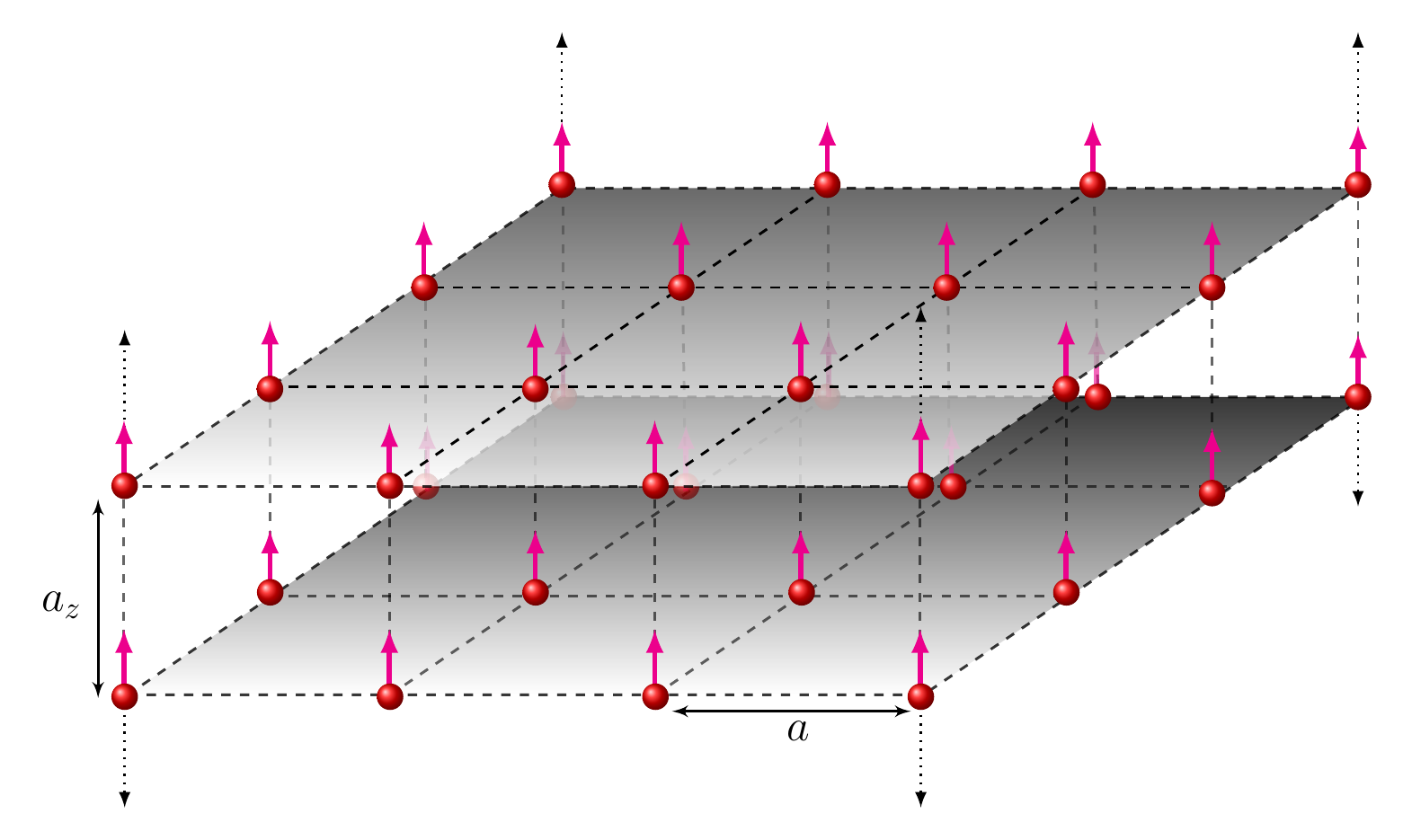}
        \caption{Schematic representation of the dipolar bosonic 
            atoms in a bilayer optical lattice. The lattice spacing within
            a layer is $a$, and inter-layer spacing is $a_z$. The red spheres 
            represent atoms, and the arrows indicate the orientation of the 
            dipole. The dotted arrows indicate possible extension to
            multilayered optical lattice.}
\label{schm_multlay}
\end{figure}


\section{Theoretical Model}\label{sec_model}

 We consider polarized dipolar atoms confined to bi- and tri-layers of 
a two-dimensional (2D) square optical 
lattice~\cite{goral_2002,yi_2007,danshita_2009,zhang_2015,baier_2016,
bandyopadhyay_2019} with in-plane lattice constant $a$. The $x$ and $y$ 
coordinates of the lattice sites with index $i\equiv (p, q)$ in layer-$1$, and 
$2$ (and $3$) coincide, but, these lattice sites are separated by a distance 
$a_{z}$ along the $z$-direction as shown in Fig~(\ref{schm_multlay}). So, in 
general, the unit cell of the multilayer lattice system is a cuboid and cube 
for the special case of $a_{z} = a$. We fix the dipole moments of the atoms 
to be along the $z$-axis. In our study, we limit the range of the dipole-dipole 
interaction to nearest-neighbour (NN) 
sites~\cite{baier_2016,bandyopadhyay_2019,suthar_2020}. This simplified model
is a good representation to provide qualitative features of dipolar atoms in 
cubic lattices. Then, the intralayer NN dipole-dipole interaction is repulsive 
and isotropic. The interlayer NN interaction is, however, attractive. For 
compact notations, let us denote the strengths of the intralayer and 
interlayer NN interactions by $V\propto d^{2}/a^{3}$ and
$V^{\prime}\propto 2d^{2}/a_{z}^{3}$, respectively, where $d$ is the magnitude
of the induced dipole-moment. In addition, we consider the atoms to be hardcore,
that is, not more than one atom can occupy each site of the lattice. So, the
local Fock space has dimension $N_{b} = 2$ with basis states
$|0\rangle$ and $|1\rangle$ corresponding to zero and single occupation,
respectively. Then, the grand-canonical Hamiltonian describing the bilayer
system is
\begin{eqnarray}
	\hat{H}_{\rm bi}&=&-J\sum_{\langle ij\rangle, r}
               \Big(\hat{b}_{i, r}^{\dagger}\hat{b}_{j, r} 
               + {\rm H.c.}\Big) - J'\sum_i\left(\hat{b}_{i, 1}^{\dagger}
               \hat{b}_{i, 2} + {\rm H.c.}\right ) 
	   \nonumber\\
	&+& V\sum_{\langle ij \rangle, r} \hat{n}_{i, r}\hat{n}_{j, r}
           + V' \sum_i\hat{n}_{i, 1}\hat{n}_{i, 2}
           - \mu\sum_{i, r}\hat{n}_{i, r},
 \label{ml_hamil}           
\end{eqnarray}
where $r \in \{1, 2\}$ is the layer index; $\hat{b}_{i, r}^{\dagger}$, 
$\hat{b}_{i, r}$ and $\hat{n}_{i, r}$ are the local bosonic creation,
annihilation and occupation number operators. Here, $J$ is the strength of 
intralayer hopping considered identical for all the layers, $J^{\prime}$ is the interlayer hopping strength, and $\mu$ the chemical potential which fixes 
the total number of atoms in the system. It is important to note that the 
interlayer hopping strength can be varied by changing the depth of the optical 
lattice along $z$-direction. In experiments this is done by tuning the 
intensity of counterpropagating laser beams along the
$z$-direction~\cite{baier_2016}. To extend the model to a larger number of 
layers, define the Hamiltonian for one lattice plane as
\begin{equation}
	\hat{H}_r=\sum_{\langle ij\rangle}
               \left [ -J\Big(\hat{b}_{i, r}^{\dagger}\hat{b}_{j, r} 
               + {\rm H.c.}\Big) + V\hat{n}_{i, r}\hat{n}_{j, r} \right ]        
               - \mu\sum_{i}\hat{n}_{i, r},
 \label{layer_hamil}           
\end{equation}
then, the bilayer Hamiltonian in Eq.(\ref{ml_hamil}) assumes the form
\begin{equation}
  \hat{H}_{\rm bi}= \sum_{r=1}^2 \hat{H}_r - J'\sum_i\left(
                    \hat{b}_{i, 1}^{\dagger} \hat{b}_{i, 2} + {\rm H.c.}\right ) 
                   + V' \sum_i\hat{n}_{i, 1}\hat{n}_{i, 2}.
\end{equation}
The Hamiltonian can be generalized to the case of $M$ layers as
\begin{eqnarray}
  \hat{H}_M&=&\sum_{r=1}^M \hat{H}_r - J'\sum_i\sum_{r=1}^{M-1}\left(
                  \hat{b}_{i, r}^{\dagger} \hat{b}_{i, r+1} + {\rm H.c.}\right ) 
                  \nonumber \\
           & & + V' \sum_i\sum_{r=1}^{M-1}\hat{n}_{i, r}\hat{n}_{i, r+1}.
 \label{mult_hamil}
\end{eqnarray}
In the present study we restrict ourselves to bi and trilayer systems.

The bilayer system with attractive interlayer NN interaction and suppressed 
interlayer hopping can exhibit the PSF phase~\cite{trefzger_2009, naini_2013,
macia_2014}. In this case, the pair formation is between the atoms in two
different layers, and the pairs hop around the lattice. In addition,
the intralayer repulsive and isotropic NN interaction can induce checkerboard 
order in the system. Then, a bilayer optical lattice system of polarized 
dipolar bosons can simultaneously exhibit superfluidity of pairs and 
crystalline order, which corresponds to pair supersolid (PSS) 
phase~\cite{trefzger_2009, naini_2013}. With the increase in interlayer 
hopping, the enhanced interlayer kinetic energy destabilizes the pairs. 
But, the resonating particle pair states can be stabilized when the interlayer 
hopping is large~\cite{ng_2015}. So, the hardcore dipolar atoms in bilayer 
system can exhibit a triplet state of the form
\begin{equation}
 |t_{0}\rangle_{p,q} =
 \frac{1}{\sqrt{2}}\left(|10\rangle_{p,q}+|01\rangle_{p,q}\right),
\label{t0_state}
\end{equation}
where $|m_{1}m_{2}\rangle_{p,q}$ denotes a dimer state with $m_{1}$ and $m_{2}$
atoms at lattice site $(p, q)$ of first and second layers, respectively. This
triplet state is the bosonic counterpart of the familiar valence bond
antisymmetric singlet state of electrons.  Like in the case of PSS, the 
attractive and repulsive inter and intralayer interactions can induce
checkerboard order to the valence bond like states as well. Then, the ground 
state of the system is referred to as the valence bond checkerboard 
solid (VCBS). In addition, the ground state of the system can 
exhibit superfluidity and valence bond checkerboard order simultaneously, 
which is referred to as valence bond supersolid (VSS) phase. It is worth noting
that the pair expectation calculated with respect to the triplet state in 
Eq.~(\ref{t0_state}) is,
\begin{equation}
  _{p,q}\langle t_{0}|\hat{n}_{p,q,1}\hat{n}_{p,q,2}|t_{0}\rangle_{p,q} = 0.
\end{equation}

 An important symmetry of the Hamiltonian in Eq.~(\ref{ml_hamil}) is manifested 
through the particle-hole transformation of the creation and 
annihilation operators. This transformation leads to replacing the particle 
creation operator $\hat{b}^{\dagger}_{i,r}$ by hole annihilation operator 
$\hat{a}_{i,r}$, and particle annihilation operator $\hat{b}_{i,r}$ 
by hole creation operator $\hat{a}_{i,r}^{\dagger}$ in the Hamiltonian. 
Then, by employing canonical anticommutation relations between these operators
(details are in Appendix~\ref{append_a}), we obtain the particle-hole symmetry 
point along the $\mu$ axis of the phase diagram as
$\mu = [4V + V^{\prime}]/2$ for a bilayer system.


\section{Theoretical methods}\label{sec_theory}
\subsection{Cluster Gutzwiller mean-field theory}\label{cgmf_theory} 
 
 We solve the  model using the cluster mean-field theory with Gutzwiller 
ansatz~~\cite{fisher_1989,rokhsar_1991,sheshadri_1993,bai_2018,pal_2019,
bandyopadhyay_2019,suthar_2020, malakar_2020}. For this, we subdivide the 
$K\times L\times M$ lattice system ($K$ and $L$ are the number of lattice 
sites along $x$ and $y$ directions) into $W$ small clusters of 
$k\times l\times m$ lattice sites, that is, 
$W = (K\times L\times M)/(k\times l\times m)$. Then, like in the
site-decoupled mean-field theory, we can define a local Hamiltonian of the
clusters~\cite{luhmann_2013,bai_2018,pal_2019,suthar_2020}, and the total 
Hamiltonian is the sum of the cluster Hamiltonians.

In the present work, we limit ourselves to the case of bilayer ($M = 2$) and
trilayer ($M = 3$) systems. The  Hamiltonian of a cluster in the bilayer 
system is
\begin{eqnarray}
	\hat{H}_C &=& \sum_r\Bigg\{\sum_{p,q \in C}'\Big[
                -J\Big(\hat{b}_{p+1,q,r}^{\dagger} \hat{b}_{p,q,r} 
                + \hat{b}_{p,q+1,r}^{\dagger} \hat{b}_{p,q,r}
	                  \nonumber\\
	     &&+ {\rm H.c.}\Big) + V\Big(\hat{n}_{p+1,q,r}\hat{n}_{p,q,r}        
                + \hat{n}_{p,q+1,r}\hat{n}_{p,q,r}\Big)\Big]
	        \nonumber\\ 
	     &&+\sum_{p,q\in \delta C} \Big[-J\Big(\phi_{p+1,q,r}^* 
	            \hat{b}_{p,q,r} + \phi_{p,q+1,r}^* 
	            \hat{b}_{p,q,r}+ {\rm H.c.}\Big)
                \nonumber\\
         &&+ V\Big(\langle\hat{n}_{p+1,q,r}\rangle\hat{n}_{p,q,r} 
                + \langle\hat{n}_{p,q+1,r} \rangle\hat{n}_{p,q,r}\Big)
                \Big]
                \nonumber\\
	     &&-\mu\sum_{p, q \in C}\hat{n}_{p,q,r}\Bigg\} 
	            -\sum_{p, q \in C}\Big[ -J^{\prime}
                \Big(\hat{b}^{\dagger}_{p,q,1} \hat{b}_{p,q,2} 
                + {\rm H.c.}\Big) 
                \nonumber \\                
         &&+V^{\prime}\hat{n}_{p,q,1}\hat{n}_{p,q,2} \Big],
 \label{ml_clus_hamil} 
\end{eqnarray}
where the prime in the summation denotes sum over lattice sites $(p, q)\in C$,
such that, $(p+1, q)$ and $(p, q+1)\in C$, and $(p,q)\in\delta C$ denote the
lattice sites at the boundary of the cluster $C$. The mean-field
$\phi^{*}_{p+1,q,r} = \langle\hat{b}^{\dagger}_{p+1,q,r}\rangle$
and average occupancy $\langle\hat{n}_{p+1,q,r}\rangle$ with
$(p + 1, q)\notin C$ are computed at the boundary of the neighbouring cluster
along $x$-direction, and are required to describe the intercluster hopping and
NN interaction, respectively. Similarly, $\phi^{*}_{p,q+1,r} = 
\langle\hat{b}^{\dagger}_{p,q+1,r}\rangle$ and
$\langle\hat{n}_{p,q+1,r}\rangle$ with $(p, q + 1)\notin C$ are required
to describe the intercluster hopping and NN interaction along $y$-direction
between adjacent clusters. Thus, within a cluster the hopping and NN
interaction terms are exact, and the intercluster hopping and NN
interactions are accounted through the mean-fields and average
occupancies, respectively. It is important to note that the interlayer
hopping and NN interaction terms are exact in the cluster
Hamiltonian. Now, like in the site-decoupled mean-field theory, the cluster
Hamiltonian matrix can be calculated using the Fock basis states of the 
cluster $\{|n_{0}n_{1}...n_{m^{\prime}}\rangle\}$, where 
$m^{\prime} = (k\times l\times m)-1$. So, these basis states are
the direct products of the local Fock states of the $(k\times l\times m)$ 
lattice sites within the cluster. By diagonalizing the Hamiltonian matrix, the 
ground state of the cluster can be obtained in the form
\begin{equation}
 |\psi^{C}_{\alpha}\rangle = \sum_{n_{0}n_{1}...n_{m^{\prime}}}
 c^{(\alpha)}_{n_{0}n_{1}...n_{m^{\prime}}}|n_{0}n_{1}...n_{m^{\prime}}\rangle,
 \label{ml_gs_clus}
\end{equation}
where $\alpha$ is the cluster index, and
$c^{(\alpha)}_{n_{0}n_{1}...n_{m^{\prime}}}$ are the complex coefficients
of the ground state $|\psi^{C}_{\alpha}\rangle$. Then, employing the Gutzwiller
ansatz, the ground state of the entire lattice is
\begin{equation}
 |\Psi_{\rm GW}\rangle = \prod_{\alpha=1}^{W}|\psi^{C}_{\alpha}\rangle 
 = \prod_{\alpha=1}^{W}\sum_{n_{0}n_{1}...n_{m^{\prime}}} 
 c^{(\alpha)}_{n_{0}n_{1}...n_{m^{\prime}}}|n_{0}n_{1}...n_{m^{\prime}}\rangle.
 \label{ml_gs_sys}
\end{equation}
The local superfluid order parameter and average occupancy at the $(p, q)$
lattice site of the $r$th layer are
\begin{eqnarray}
 \phi_{p,q,r} 
  &=& \langle\Psi_{\rm GW}|\hat{b}_{p,q,r}|\Psi_{\rm GW}\rangle, \nonumber \\
  n_{p,q, r} 
  &=& \langle\Psi_{\rm GW}|\hat{n}_{p,q,r}|\Psi_{\rm GW}\rangle. 
 \label{ml_sf_op} 
\end{eqnarray}
A relevant parameter of a quantum phase is the average occupancy per lattice
site,
\begin{equation}
   \rho = \frac{1}{(K\times L\times M)}
	\sum_{p=1, q=1,r=1}^{K, L,M}n_{p,q, r}.
 \label{ml_av_num}
\end{equation}
In this work, we study quantum phases with the hard-core approximation. So, we
have $\rho\leq 1$. An important quantity to probe the valence bond order in the
bilayer system is the average pair density
\begin{equation}
      \tilde{\rho} =  \frac{1}{(K\times L)}
      \sum_{p, q=1}^{K, L}
      \langle\Psi_{\rm GW}|\hat{n}_{p,q,1}\hat{n}_{p,q,2}
      |\Psi_{\rm GW}\rangle .
 \label{bil_av_pair_den}
\end{equation}
For valence bond checkerboard solid phases $\tilde{\rho}$ is zero while $\rho$ 
is finite. The zero superfluid order parameter indicates the incompressibility 
of these solid phases and the checkerboard order in each layer is identified by 
the structure factor
\begin{equation}
   S_{r}(\pi,\pi)= \frac{1}{K\times L}
   \sum_{\substack{{p',q'}\\p,q}}e^{{\rm i}\pi
   \{(p-p')+(q-q')\}}
   \langle\hat{n}_{p,q,r}\hat{n}_{p',q',r}
   \rangle.
  \label{st_fac}        
\end{equation}
In the phases with checkerboard order $S_{r}(\pi,\pi)$ is finite, while
it is zero in the phases with uniform density distribution.


\subsection{Numerical methods}\label{sec_numeric}

 The starting point of our cluster mean field, hereafter, referred as cluster
Gutzwiller mean field (CGMF) theory, is to choose an appropriate initial guess 
state $|\Psi_{\rm GW}\rangle$. From this we compute the initial 
$\phi_{p,q,r}$ and $n_{p,q,r}$. In general, we choose 
the same initial guess state $|\Psi^{C}_{\alpha}\rangle$ for all the $W$ 
clusters and consider 
$ c^{(\alpha)}_{n_{0}n_{1}...n_{m^{\prime}}}=  1/\sqrt{2^{m^{\prime}+1}}$. 
Then, using corresponding $\phi_{p,q,r}$ and $n_{p,q,r}$, we 
calculate the Hamiltonian matrix of a cluster given in 
Eq.~(\ref{ml_clus_hamil}), which we diagonalize~\cite{bai_2018,pal_2019,
suthar_2020}. We update the state $|\Psi_{\rm GW}\rangle$ using the new 
ground state of the cluster obtained from the diagonalization. Afterwards, we compute $\phi_{p,q,r}$ and $n_{p,q,r}$ using this updated 
$|\Psi_{\rm GW}\rangle$, and advance to the next cluster to repeat the same 
steps. We sweep the entire lattice system by continuing the procedure, and one 
such sweep constitutes an iteration. We continue the iterations until desired 
convergence in $\phi_{p,q,r}$ and $n_{p,q,r}$ is obtained. In our 
computations, we consider clusters ranging in size from 
$1\times 1\times 2$ to $4\times 4\times 2$ to tile lattice systems
ranging in size from $8\times 8\times 2$ to $16\times 16\times 2$. To model an 
uniform infinite lattice system, we employ periodic boundary conditions in 
$\phi_{p,q, r}$ and $n_{p,q,r}$ along $x$ and $y$-directions. 
We also corroborate the stability of the obtained ground states with respect to 
different initial guess states having inhomogeneous distribution in 
$n_{p,q,r}$ and $\phi_{p,q,r}$. The initial guess states considered
have checkerboard and random density patterns.


\subsection{Quantum phases}\label{quant_phases}
  
The system admits particle and hole vacuum states. And, these
states correspond to $\rho = 0$  and $\rho =1$, respectively. Using the dimer 
notation these states can be represented as 
\begin{equation}
|\Psi\rangle_{\rm VAC}^{\rho = 0} 
= \prod_{(p,q)}|00\rangle_{p,q},
\label{pvac}
\end{equation}
and 
\begin{equation}
|\Psi\rangle_{\rm VAC}^{\rho = 1} 
= \prod_{(p,q)}|11\rangle_{p,q}.
\label{hvac}
\end{equation}
Thus, in these states the system either has no atoms or has uniform 
distribution of one atom per lattice site throughout the system, respectively. 
The interlayer attractive NN interaction energetically favors the dimer state 
$|11\rangle_{p,q}$. In addition, the intralayer repulsive NN interaction can 
induce checkerboard density order. This interaction induced spatially periodic 
intralayer density modulation may be considered as two interpenetrating 
sublattices, $A$ and $B$. It is important to note that $n_{p,q,1}$ and 
$n_{p,q,2}$ have identical distributions or  the checkerboard structure 
in both the layers are aligned. This is due to the attractive interlayer NN 
interaction, and is in contrast to the case when $V^{\prime}$ is repulsive. As 
mentioned earlier, the Hamiltonian of this system with $J'=0$ is equivalent to 
the two species lattice Hamiltonian in 2D. In which case, $V^{\prime}$ is 
identified as the onsite interspecies interaction strength. For this 
two-species system with repulsive onsite interspecies interaction the 
checkerboard structure can have phases with spatially separated density or 
phase-separated~\cite{bai_2020}.  

\begin{figure}[ht]
   \includegraphics[height = 3.0cm]{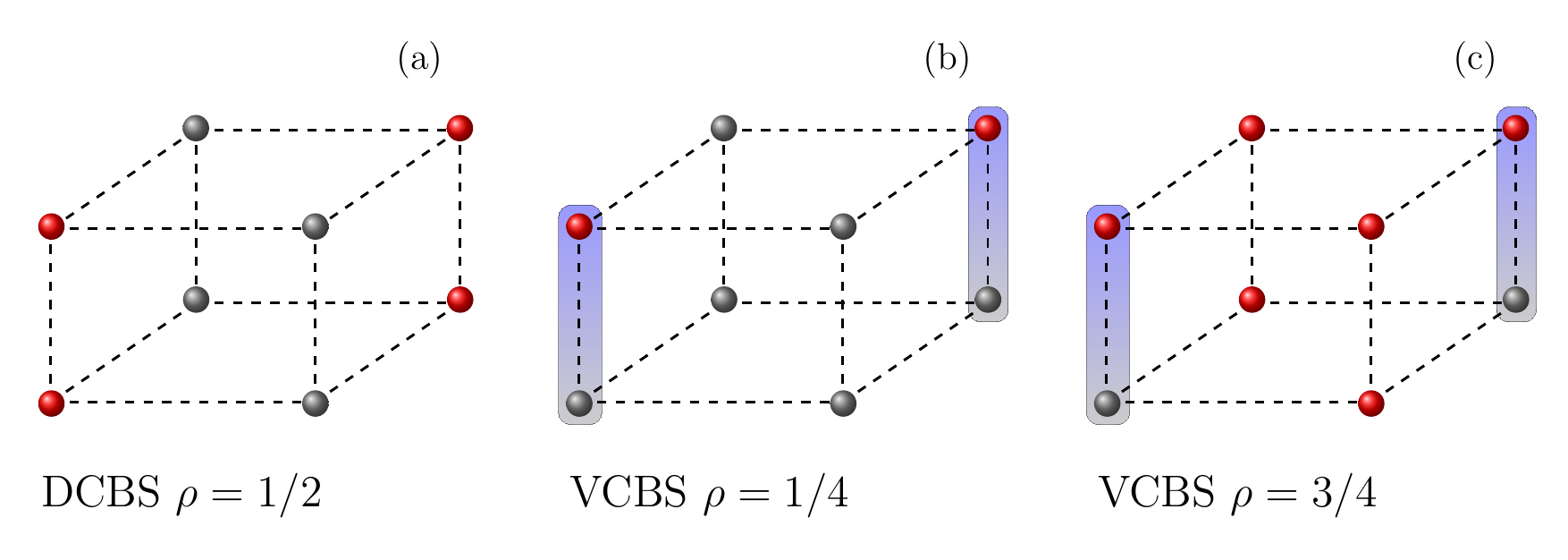}
   \caption{Schematic representation of the incompressible phases. 
	     The red (gray) spheres represent particles or atoms (holes).
	     Panel (a), (b) and (c) correspond to the DCBS, VCBS $\rho = 1/4$ 
	     and VCBS $\rho = 3/4$ phases, respectively. The blue shading across
	     the interlayer bonds in (b) and (c) denote the entangled 
	     state, as given by Eq.~\ref{t0_state}.}
\label{phase_schm}
\end{figure}

 Due to the intralayer repulsive NN interaction, the bilayer system can exhibit
a state of the dimers with solid or diagonal order, which is referred to as 
dimer checkerboard solid (DCBS). Then, considering the sublattice description
of the checkerboard order, the DCBS state can be expressed as 
\begin{equation}
|\Psi\rangle_{\rm DCBS}^{\rho =1/2} 
= \prod_{(p,q)\in A}|11\rangle_{p,q}
  \prod_{(p^{\prime},q^{\prime})\in B}|00\rangle_{p^{\prime},q^{\prime}}.
\label{dcbs_1by2}
\end{equation}
On the other hand, the interlayer hopping can stabilize the triplet state as in 
Eq.~(\ref{t0_state}). As discussed before, this state with the checkerboard 
ordered density is the VCBS state. Depending on the average 
occupancy the VCBS states have the forms 
\begin{equation}
 |\Psi\rangle_{\rm VCBS}^{\rho =1/4} 
  = \prod_{(p,q)\in A}|t_{0}\rangle_{p,q}
  \prod_{(p^{\prime},q^{\prime})\in B}|00\rangle_{p^{\prime},q^{\prime}},
 \label{vcbs_1by4}
\end{equation}
and 
\begin{equation}
 |\Psi\rangle_{\rm VCBS}^{\rho =3/4}
  = \prod_{(p,q)\in A}|t_{0}\rangle_{p,q}
  \prod_{(p^{\prime},q^{\prime})\in B}|11\rangle_{p^{\prime},q^{\prime}}.
\label{vcbs_3by4}
\end{equation}
It is important to note that the states described in 
Eqs.~(\ref{pvac})--~(\ref{vcbs_3by4}) correspond to the incompressible phases 
of the bilayer system and the SF order parameter $\phi_{p,q,r}$ is zero in 
these phases. In addition, the checkerboard order in the DCBS and VCBS states 
are quantified by the nonvanishing $S_{r}(\pi, \pi)$. The average pair 
density $\tilde{\rho}$ given in Eq.~(\ref{bil_av_pair_den}) serves as an order 
parameter to distinguish the VCBS states from the DCBS state. Note that 
$\tilde{\rho}$ is zero in the VCBS states, but it is nonzero in the DCBS state. 
The system also exhibits supersolid and superfluid phases in which 
$\phi_{p,q,r}$ is finite. In the SF phase the system has uniform 
$n_{p,q,r}$ and $\phi_{p,q,r}$. In contrast, the supersolid phase has 
checkerboard order in $n_{p,q,r}$ and $\phi_{p,q,r}$.
We show the schematic representation of the quantum phases discussed here, in 
Fig.~\ref{phase_schm}. In the figure, the dimer in the DCBS phase is 
recognizable, and the blue shade over the bonds indicate the resonating 
structure of the VCBS states.

Like in the triplet state in the bilayer system, the dipolar atoms in a
trilayer optical lattice can exhibit states of the form
\begin{equation}
 |w_{0}\rangle_{p,q} =
 \frac{1}{\sqrt{3}}\left(|100\rangle_{p,q}+|010\rangle_{p,q}
        + |001\rangle_{p,q}\right),
\label{w0_state}
\end{equation}
and
\begin{equation}
 |w_{1}\rangle_{p,q} =
 \frac{1}{\sqrt{3}}\left(|011\rangle_{p,q}+|101\rangle_{p,q}
        + |110\rangle_{p,q}\right).
\label{w1_state}
\end{equation}
These states have a resonating structure in one of the two sublattices, and
are analogous to the VCBS states, given in Eq.(\ref{vcbs_1by4}) and
Eq.(\ref{vcbs_3by4}), for bilayer optical lattice. They are stabilized by the
interlayer hopping. They resemble the \textit{W-state} of the three qubits
that is studied in the context of the quantum information
theory ~\cite{dur_2000}. The density of these states can vary as
$\rho = 1/6$, $2/6$, $4/6$ and $5/6$ based on the filling in sublattice A and
B. For the $|w_{0} \rangle$ state in sublattice A, the density of these 
states can be $\rho=1/6$ and $4/6$ for the zero and unit filling in sublattice 
B, respectively. In a similar way, for the $|w_{1} \rangle$ state in sublattice
A, the density can be $\rho=2/6$ and $5/6$.
On the other hand, the state corresponding to the density
$\rho = 3/6 = 1/2$ does not have a resonating structure unlike others,
and is referred as the trimer checkerboard solid (TCBS) state. The form of
this state, in terms of the sublattice description, is
\begin{equation}
|\Psi\rangle_{\rm TCBS}^{\rho =1/2} 
= \prod_{(p,q)\in A}|111\rangle_{p,q}
  \prod_{(p^{\prime},q^{\prime})\in B}|000\rangle_{p^{\prime},q^{\prime}}.
\label{tcbs_1by2}
\end{equation}
This is a generalization of the DCBS $\rho = 1/2$ state, given in
Eq.(\ref{dcbs_1by2}), for a bilayer optical lattice. It has a checkerboard
pattern between the sublattice A and B, aligned between the three layers.
The same occupancy between the three layers at a given lattice site is the
result of the strong attractive interlayer interaction.


\subsection{Mean-field phase boundaries}\label{mf_phase_boundaries}

  We calculate phase boundaries between the incompressible and 
compressible phases of the bilayer system using the mean-field theory. We
do this by adapting the site-decoupling scheme, and perform perturbative 
analysis of the mean-field Hamiltonian. In this method we decompose the 
creation, annihilation and occupation number operators of a lattice site in 
terms of the local mean-fields and fluctuation operators as 
$\hat{b}_{p,q,r} = \phi_{p,q,r} +\delta\hat{b}_{p,q,r}$, 
$\hat{b}^{\dagger}_{p,q,r} = \phi^{*}_{p,q,r} 
+\delta\hat{b}^{\dagger}_{p,q,r}$, and 
$\hat{n}_{p,q,r} = n_{p,q,r}+\delta\hat{n}_{p,q,r}$  Then, the 
bilinear terms of the Hamiltonian in Eq.~(\ref{ml_hamil}) are sum of linear 
terms of the local operators. In the perturbation analysis, we consider the 
interaction terms as the unperturbed Hamiltonian. These are diagonal with 
respect to the single site Fock basis states. The off diagonal hopping terms 
are considered as perturbations. The incompressible to compressible 
phase boundaries are, then, marked by the vanishing superfluid order parameter, 
that is $\phi_{p,q,r}\rightarrow 0^{+}$. The $\phi_{p,q,r}$ being 
small and associated with the off diagonal terms, it can be considered as the 
perturbation parameter.

 To perform the first order perturbation analysis, we write the 
site-decoupled unperturbed Hamiltonian as
\begin{eqnarray} 
 \hat{H}^{(0)}&=&\sum_{p,q,r}\hat{h}^{(0)}_{p,q,r}
                   \nonumber \\
	     &=&\sum_{p,q,r}\hat{n}_{p,q,r}\left(V'n_{p,q,3-r}
	        +V\overline{n}_{p,q,r} -\mu\right), 
 \label{unperturb_hamil}	     
\end{eqnarray}
where $\overline{n}_{p,q,r} = (n_{p+1,q,r} + n_{p-1,q,r} 
+ n_{p,q+1,r} + n_{p,q-1,r})$, with 
$n_{p,q,r} = \langle\hat{n}_{p,q,r}\rangle$ representing the ground state 
occupancy at the site $(p,q,r)$. Now, consider the sublattice 
description of incompressible states with aligned checkerboard order, the 
unperturbed energy of the two different sublattices are
\begin{equation} 
 E^{A}_{n^{A}_{1}n^{A}_{2}}= V'n^{A}_{1}n^{A}_{2}  
	+ \sum_{r=1}^{2}\left(4V n^A_r n^B_r - \mu n^A_r\right),
 \label{subA_energy}	
\end{equation}
for $(p,q)\in A$, and
\begin{equation} 
 E^{B}_{n^{B}_{1}n^{B}_{2}}= V'n^B_1 n^B_2  
        + \sum_{r=1}^{2}\left( 4Vn^B_r n^A_r - \mu n^B_r\right),
 \label{subB_energy}	
\end{equation}
for $(p,q)\in B$. Here, $n^A_r$ and $n^B_r$ are the occupancies in the 
$r$th layer. It is important to note that $E^{A}_{00} = E^{B}_{00} = 0$. As 
mentioned earlier, the hopping terms in the Hamiltonian are treated as the 
perturbations. Then, the site-decoupled perturbation Hamiltonians are 
\begin{eqnarray}
  \hat{T}^{A} &=& - 4J\sum_{r=1}^{2}\phi_r^B \left (\hat{b}^A_r
	            +{\hat{b}^{A^\dagger}_r}\right ) - J'\sum_{r=1}^{2}
		    \phi_{3-r}^{A}
	            \left(\hat{b}^A_r+\hat{b}^{A^{\dagger}}_r\right), 
	            \nonumber \\
 \hat{T}^{B} &=& -4J\sum_{r=1}^{2}\phi_{r}^{A}\left(\hat{b}^{B}_{r}
	   +\hat{b}^{B^{\dagger}}_{r}\right) 
           -J^{\prime}\sum_{r=1}^{2}\phi_{3-r}^{B}
	    \left(\hat{b}^{B}_{r}+\hat{b}^{B^{\dagger}}_{r}\right), 
        \nonumber 	
\end{eqnarray}
for the sublattice-$A$ and $B$, respectively. Now, to the first order in the 
superfluid order parameters, the perturbed ground state is
\begin{equation}
 |\chi^{\alpha}\rangle = |n^{\alpha}_{1}n^{\alpha}_{2}\rangle 
	+ \sum_{\substack{(m^{\alpha}_{1},m^{\alpha}_{2})\\ 
	        \neq (n^{\alpha}_{1},n^{\alpha}_{2})}} 
	  \frac{\langle m^{\alpha}_{1}m^{\alpha}_{2}|\hat{T}^{\alpha}
	  |n^{\alpha}_{1}n^{\alpha}_{2}\rangle}
	  {(E^{\alpha}_{n^{\alpha}_{1}n^{\alpha}_{2}}
	   -E^{\alpha}_{m^{\alpha}_{1}m^{\alpha}_{2}})} 
	  |m^{\alpha}_{1}m^{\alpha}_{2}\rangle,
 \label{perturb_gs}	
\end{equation}
where $\alpha\in\{A,B\}$. Then, the superfluid order parameters are
$\phi^{\alpha}_{\kappa} =\langle\chi^{\alpha}|
\hat{b}^{\alpha}_{\kappa}|\chi^{\alpha}\rangle$. In our study, we have 
considered that the intralayer parameters are same for both the layers--the 
layers are identical. Therefore, 
$\phi^{A}_1 = \phi^{A}_{2} = \varphi^{A}$ and 
$\phi^{B}_1 = \phi^{B}_{2} = \varphi^{B}$.

 We first obtain the phase boundary separating the DCBS phase from a 
compressible phase, which can either be superfluid or supersolid. The DCBS 
state in Eq.~(\ref{dcbs_1by2}) is an eigenstate of $\hat{H}^{(0)}$ in 
Eq.~(\ref{unperturb_hamil}), with occupancies $(n^{A}_1,n^{A}_{2}) = (1, 1)$ 
and $(n^{B}_1,n^{B}_{2}) = (0, 0)$. Then, the superfluid order parameters 
calculated with respect to the perturbed ground state in Eq.~(\ref{perturb_gs})
are 
$(p,q)\in A$ \begin{equation}
  \varphi^{A} = -\frac{4J\varphi^{B}}{(E^{A}_{11}- E^{A}_{01})+J^{\prime}}, 
 \label{subA_sfop}
 \end{equation}
and	
 \begin{equation}
  \varphi^{B} = -\frac{4J\varphi^{A}}{(E^{B}_{00}- E^{B}_{10})+J^{\prime}}.
 \label{subB_sfop}
 \end{equation}
We solve Eqs.~(\ref{subA_sfop}) and ~(\ref{subB_sfop}) simultaneously, and 
then, we take the limit $\{\varphi^{A}, \varphi^{B}\}\rightarrow0^{+}$ to 
get
\begin{equation}
 16J^{2} =\left[(E^{A}_{11}- E^{A}_{01})+J^{\prime}\right]
	  \times
	  \left[(E^{B}_{00}- E^{B}_{10})+J^{\prime}\right].
\end{equation}	
Now, substituting the values of $E^{A}_{11}$, $E^{A}_{01}$ from 
Eq.~(\ref{subA_energy}), and $E^{B}_{10}$ from 
Eq.~(\ref{subB_energy}), we obtain the DCBS phase boundary as a solution of 
\begin{equation}
 16J^{2} = (V^{\prime}-\mu+J^{\prime})(\mu-4V+J^{\prime}).	
 \label{pb_dcbs_1by2}
\end{equation}
From this, the DCBS lobe in the plane of $J/V-\mu/V$ can be obtained for 
different values of $J^{\prime}$ and $V^{\prime}$.

Next, we calculate the equations of the phase boundaries separating the 
parameter domains of the VCBS states described in Eqs.~(\ref{vcbs_1by4}) and 
~(\ref{vcbs_3by4}) from the compressible phases of the system. It is important
to notice that these two states are eigenstates of $\hat{H}^{(0)}$ in
Eq.~(\ref{unperturb_hamil}). But, as emphasized earlier, the interlayer hopping 
is essential to stabilize the $|t_{0}\rangle_{p,q}$ triplet state of the VCBS
states. And, this term is not present in the site-decoupled unperturbed 
Hamiltonian $\hat{H}^{(0)}$. So, in order to obtain the triplet state 
$|t_{0}\rangle$ as an eigenstate, we define local unperturbed Hamiltonian 
of the sublattice-$A$ as 
\begin{eqnarray}
 \mathcal{H}^{(0)}_{A} &=& -J'(\hat{b}^{\dagger}_{p,q,1}\hat{b}_{p,q,2} 
                     + {\rm H.c.})
                     + V'\hat{n}_{p,q,1}\hat{n}_{p,q,2}  \nonumber\\
		   &&+ \sum_{r=1}^{2}\left(V\hat{n}_{p,q,r}\overline{n}_{p,q,r}
		     -\mu\hat{n}_{p,q,r}\right),	
 \label{unperturb_vcbs_hamil}	
\end{eqnarray}
for $(p,q)\in A$. Then, we can express the unperturbed ground state energy for 
the sublattice-A as
\begin{equation}
  E^{A}_{t_{0}}= \langle t_{0}|\mathcal{H}^{(0)}_{A}|t_{0}\rangle  
      = -J^{\prime} + \sum_{r=1}^{2}\left(4Vn^A_r n^B_r - \mu n^A_r\right).
 \label{subA_t0_energy}	
\end{equation}
This has a contribution from interlayer hopping process, and in contrast to 
the Eq.~(\ref{subA_energy}), it does not have interlayer interaction energy. 
This is because the pair expectation, as mentioned earlier, is zero with 
respect to $|t_{0}\rangle$. It is to be noted that in 
Eq.~(\ref{subA_t0_energy}), $n^{A}_1= n^{A}_{2} = 0.5$, which are the 
occupancies calculated with respect to the $|t_{0}\rangle$ state. On the other 
hand, unlike in the previous case, the perturbation Hamiltonian now contains
only the intralayer hopping terms, that is, 
\begin{equation}
 \hat{T^{\prime}}^{A} 
	= -4J\sum_{r=1}^{2}\phi_r^B (\hat{b}^A_r +\hat{b}^{A^{\dagger}}_r).
 \label{pert_hamil_vcbs}
\end{equation}
Then, the perturbed ground state is
\begin{equation}
|\chi^{A}_{t_0}\rangle = |t_{0}\rangle + \sum_{m^{A}_{1},m^{A}_{2}} 
	  \frac{\langle m^{A}_{1}m^{A}_{2}|\hat{T^{\prime}}^{A}
	  |t_{0}\rangle}{(E^{A}_{t_{0}}-E^{A}_{m^{A}_{1}m^{A}_{2}})} 
	  |m^{A}_{1}m^{A}_{2}\rangle,
 \label{perturb_gs}	
\end{equation}
where $(m^{A}_{1},m^{A}_{2})\in\{(0,0),(1,1)\}$.
Using this state the superfluid order parameters in the sublattice-$A$ can be 
calculated as $\phi^A_r=\langle\chi^{A}_{t_0}|\hat{b}^A_r|
\chi^{A}_{t_0}\rangle$. Similar to the previous case 
$\phi^{A}_{1} = \phi^{A}_{2} = \varphi^{A}$. 
Then, the superfluid order parameter is
\begin{equation}
 \varphi^{A} = -4J\phi^{B}\left[\frac{1}{(E^{A}_{t_{0}}-E^{A}_{00})}
	           + \frac{1}{(E^{A}_{t_{0}}-E^{A}_{11})}\right].	
 \label{subA_sfop_t0}		   
\end{equation}

 For the VCBS state with $\rho = 1/4$, we calculate the superfluid order 
parameter as given in Eq.~(\ref{subA_sfop_t0}) from the perturbative correction 
to $|t_{0}\rangle_{p,q}$ in sublattice-$A$. But, we perform the perturbation 
analysis of the site-decoupled Hamiltonian in sublattice-$B$. So, we obtain 
the superfluid order parameter as given in Eq.~(\ref{subB_sfop}) from the 
perturbative correction to $|00\rangle_{p,q}$ state in sublattice-$B$. We 
solve the Eqs.~(\ref{subA_sfop_t0}) and ~(\ref{subB_sfop}) simultaneously, and 
then take the limit $\{\varphi^{A}, \varphi^{B}\}\rightarrow0^{+}$ to obtain
\begin{eqnarray}
 \frac{1}{16J^{2}} 
	= &&\left[\frac{1}{(E^{A}_{t_{0}}-E^{A}_{00})}
        + \frac{1}{(E^{A}_{t_{0}}-E^{A}_{11})}\right]\nonumber\\
	&&\times
	\left[\frac{1}{(E^{B}_{00}- E^{B}_{10})+J^{\prime}}\right].
\end{eqnarray}
Note that, for the unperturbed ground state the occupancies 
$(n^{A}_1, n^{A}_{2}) = (0.5, 0.5)$ and $(n^{B}_1, n^{B}_{2}) = (0, 0)$. 
Now, by substituting the values of $E^{A}_{t_{0}}$ from 
Eq.~(\ref{subA_t0_energy}), $E^{A}_{11}$ from Eq.~(\ref{subA_t0_energy}), and 
$E^{B}_{10}$ from Eq.~(\ref{subA_t0_energy}), 
we obtain the VCBS ($\rho=1/4$) phase boundary as a solution of
\begin{equation}
 16J^{2}(2J^{\prime}+V^{\prime}) 
	= (\mu-2V+J^{\prime})(J^{\prime}+\mu)(\mu-J^{\prime}-V^{\prime}).
 \label{pb_vcbs_1by4}	
\end{equation}
It is worth mentioning that $E^{A}_{11}$ from Eq.~(\ref{subA_t0_energy}) is 
equal to $\langle11|\mathcal{H}^{(0)}_{A}|11\rangle$. Similarly, for the VCBS 
state with $\rho = 3/4$, we perform the perturbative analysis to state 
$|11\rangle_{p,q}$ in the sublattice-$B$. We obtain the expression of the 
superfluid order parameter $\varphi^{B}$, which is similar to the 
Eq.~(\ref{subB_sfop}) with the superscripts $A$ and $B$ interchanged. We, then 
simultaneously solve this equation and Eq.~(\ref{subA_sfop_t0}), and take the 
limit $\{\varphi^{A}, \varphi^{B}\}\rightarrow0^{+}$ like in the previous case.
Then, we obtain the VCBS ($\rho=3/4$) phase boundary as a solution of
\begin{eqnarray}
 16J^{2}(2J^{\prime}+V^{\prime})
	= &&(2V+V^{\prime}-\mu+J^{\prime})(\mu-4V+J^{\prime})\nonumber\\
	&&\times(\mu-J^{\prime}-4V-V^{\prime}).
 \label{pb_vcbs_3by4}
\end{eqnarray}
From Eqs.~(\ref{pb_vcbs_1by4}) and~(\ref{pb_vcbs_3by4}) the VCBS lobes with 
$\rho=1/4$ and $3/4$ can be obtained in the $J/V-\mu/V$ plane for different 
values of $J^{\prime}$ and $V^{\prime}$.

 The above formalism can be generalized to the trilayer system and the details
of the derivation are given in the Appendix \ref{append_b}. The equation 
defining the phase boundary between the TCBS phase and the compressible phase is
\begin{equation}
 16J^{2} = (2V^{\prime}-\mu + 2J^{\prime})(\mu-4V + 2J^{\prime}).
 \label{pb_tcbs_1by2}
\end{equation}
Based on this, the VCBS $(\rho = 1/6)$ phase to compressible phase boundary 
is given by 
\begin{equation}
  16J^{2}(3V^{\prime} + \mu + 8J^{\prime}) = (3\mu -4V +  6J^{\prime})
	                      (\mu + 2J^{\prime})(\mu- V^{\prime}),
 \label{pb_vcbs_1by6}
\end{equation}
and that of VCBS $(\rho = 2/6)$ is 
\begin{equation}
  16J^{2}(5V^{\prime} -\mu +  8J^{\prime}) = (3\mu - 8V +  6J^{\prime})
			      (\mu -2V^{\prime}- 2J^{\prime})(\mu- V^{\prime}).
 \label{pb_vcbs_2by6}
\end{equation}
Invoking the particle-hole symmetry of the model, we can write the phase
boundaries between the VCBS $\rho = 4/6$ and the compressible phase as 
\begin{eqnarray}
  16J^{2}&&\left(3V^{\prime} -4V + \mu +  8J^{\prime}\right) = 
	\left(6J^{\prime}+4V+6V^{\prime}-3\mu\right) \nonumber \\ 
	&& \times \left(4V - \mu -  2J^{\prime}\right)
	\left(-\mu +  V^{\prime} + 4V \right),
 \label{pb_vcbs_4by6}
\end{eqnarray}
and that of VCBS $\rho = 5/6$ as 
\begin{eqnarray}
  16J^{2}&&\left(5V^{\prime} + 4V - \mu +  8J^{\prime}\right) = 
	\left(6J^{\prime}+8V+6V^{\prime}-3\mu\right) \nonumber \\ 
	&& \times \left(4V - \mu +  2J^{\prime} + 2V^{\prime}\right)
	\left(-\mu +  V^{\prime} + 4V \right).
 \label{pb_vcbs_5by6}
\end{eqnarray}
The phase diagram obtained based on these equations, shown in the results
section, is in good agreement with the one obtained numerically.


\section{Results and discussions}\label{sec_results}

To find the ground state of the bilayer system, we first scale the Hamiltonian 
in Eq.~(\ref{ml_hamil}) by the intralayer NN interaction strength $V$.
This choice yields four independent parameters, $J/V$, $J^{\prime}/V$, 
$V^{\prime}/V$ and $\mu/V$, which can be varied to probe different quantum 
phases of the system. We present the parameter domains of these quantum 
phases in the $J/V-\mu/V$ plane for fixed values of $V^{\prime}/V$ and 
$J^{\prime}/V$. To begin with we consider the case when $V^{\prime}/V = -1$, 
and examine the quantum phases in three broad domains of interlayer hopping. 
These are weak ($J^{\prime}/V = 0$ and $0.5$), moderate ($J^{\prime}/V = 0.8$ 
and $1.0$), and strong ($J^{\prime}/V = 1.2$ and $1.5$) interlayer hopping. 
We present and discuss the corresponding phase diagrams in Sec.~\ref{vpm1}. Then,
we also examine the effects of the quantum correlations on the parameter 
domains of the quantum phases. We then consider the case of 
$|V^{\prime}|\neq V$ in Sec~\ref{vp_ne_v}. As representative cases we 
consider $V^{\prime} = -0.25V$ and $-2.0V$. The effect of the thermal 
fluctuations is discussed in Sec.~\ref{fint_temp}. Considering the influence
of thermal fluctuations on the quantum phases is essential to relate our
results to the possible experimental realizations.


\subsection{$V^{\prime} = -1.0V$} \label{vpm1} 
  This is a suitable choice to probe the interplay between different 
hopping and interaction energies theoretically. In terms of lattice
constants, it corresponds to $a_{z} = \sqrt[3]{2}a$. We extend the theoretical 
insights gained from this case to the $|V^{\prime}| \neq V$ regime.
\begin{figure}[ht]
   \includegraphics[height = 5.4cm]{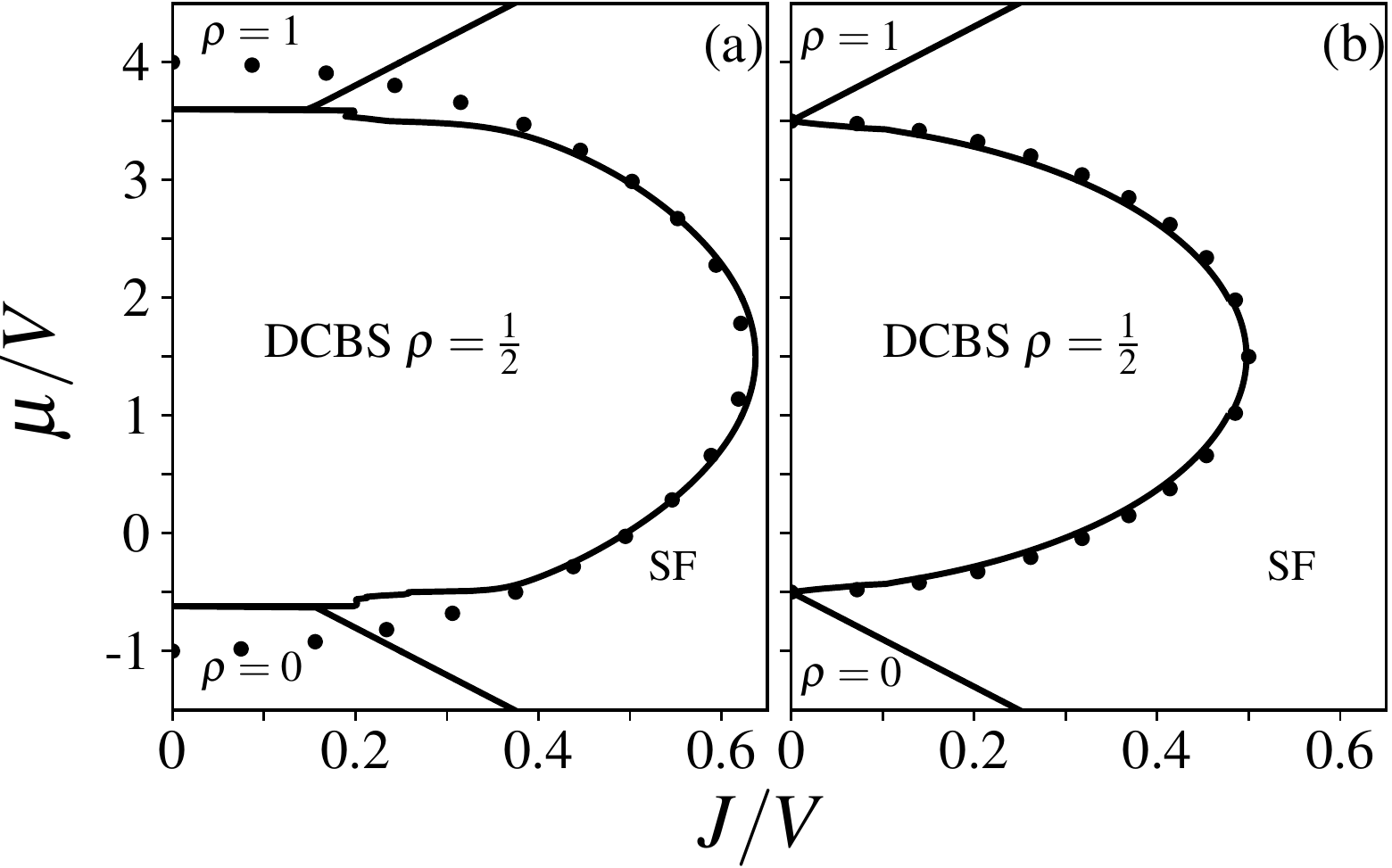}
   \caption{Phase diagrams in the $J/V-\mu/V$ plane for 
	     $J'/V = 0$ and $0.5$, respectively for $V' = -1.0V$. The black 
	     lines represent the phase boundaries between the incompressible 
	     and compressible phases. And, the filled black circles denote the 
	     analytical phase boundaries obtained from the site-decoupled mean 
	     field method, as given in Eq.(\ref{pb_dcbs_1by2}).}
    \label{comb_bil1}
\end{figure}


\subsubsection{Weak interlayer hopping $J^{\prime}/V < 1$}
In this domain, we consider two values of $J^{\prime}/V=0, 0.5$. The 
corresponding phase diagrams are shown in Figs.~\ref{comb_bil1}(a) and
(b), respectively. For $J^{\prime}/V = 0$, the interlayer hopping is absent,
and the two layers are coupled only through the NN attractive interaction. As 
mentioned earlier, this is equivalent to the two-species extend Bose-Hubbard 
model for hardcore bosons with attractive onsite and zero offsite interspecies 
interaction strengths~\cite{bai_2020}. In the Fig.~\ref{comb_bil1}, the solid
lines represent the phase boundaries obtained numerically with 
$2\times2\times2$ clusters in the CGMF method. Here after, for compact 
notation, we shall refer to the $2\times2\times2$ cluster as the $2^3$ cluster.
The filled circles mark the incompressible-compressible phase boundaries 
obtained analytically from the site-decoupled mean-field theory. In particular,
the filled circles in Figs.~\ref{comb_bil1}(a) and (b) represent the phase 
boundary of the DCBS phase obtained by solving the Eq.~(\ref{pb_dcbs_1by2}) with $V^{\prime} = -1.0V$, $J^{\prime} = 0V$ and $0.5V$, respectively. 
It is evident from the Fig.~\ref{comb_bil1}(a) that the numerical 
and analytical results are in good agreement for $J/V \geqslant 0.34$. In this 
parameter regime the difference between the numerical and analytical results
are below $0.1\mu/V$. It is, however, larger for $J/V<0.34$. This is due to 
merging of the incompressible lobes ~\cite{bandyopadhyay_2019, naini_2012} and 
the applicability of site-decoupled mean field to discern only 
incompressible-compressible phase boundaries. Here, the merger of the 
incompressible lobes is a result of the attractive interlayer interaction but 
no interlayer hopping. More importantly, the merging of the parameter domains 
leads to emergence of triple points ~\cite{bandyopadhyay_2019} at 
$\mu/V = 3.57$ and $\mu/V = -0.57 $ for $J/V = 0.138$ in 
Fig.~\ref{comb_bil1}(a). With the increase in the $J^{\prime}/V$ the triple 
points shift towards left due to enhanced kinetic energy of the system. And, 
the points reside on the $\mu/V = 0$ axis for $J^{\prime}/V \apprge 0.3 $. 

 From Fig.~\ref{comb_bil1}(a), it is evident that the ground state is 
either MI with $\rho = 0$ and $\rho = 1$, or a DCBS state with $\rho = 1/2$. 
The DCBS state has checkerboard order of 
the dimers, and can be described as in Eq.~(\ref{dcbs_1by2}). The checkerboard 
ordering is due to the intralayer NN repulsive interaction, which disfavours 
phases with density like the MI phase. It is important to note that 
MI phases sandwich the DCBS phase. This is because, the DCBS state can be 
obtained through dimer creation and annihilation from the MI $\rho = 0$ and 
$\rho = 1$ states, respectively. This feature of the DCBS state is also 
evident from the comparison between the Eqs.~(\ref{pvac})-(\ref{dcbs_1by2}). At 
higher $J/V$, the atoms in the lattice acquire enough kinetic energy, and 
they become itinerant. So, the system exhibit SF phase with uniform density 
distribution. In this phase $\phi_{p,q,r}$ is finite and uniform throughout 
the lattice.

\begin{figure}[ht]
   \includegraphics[height = 6.4cm]{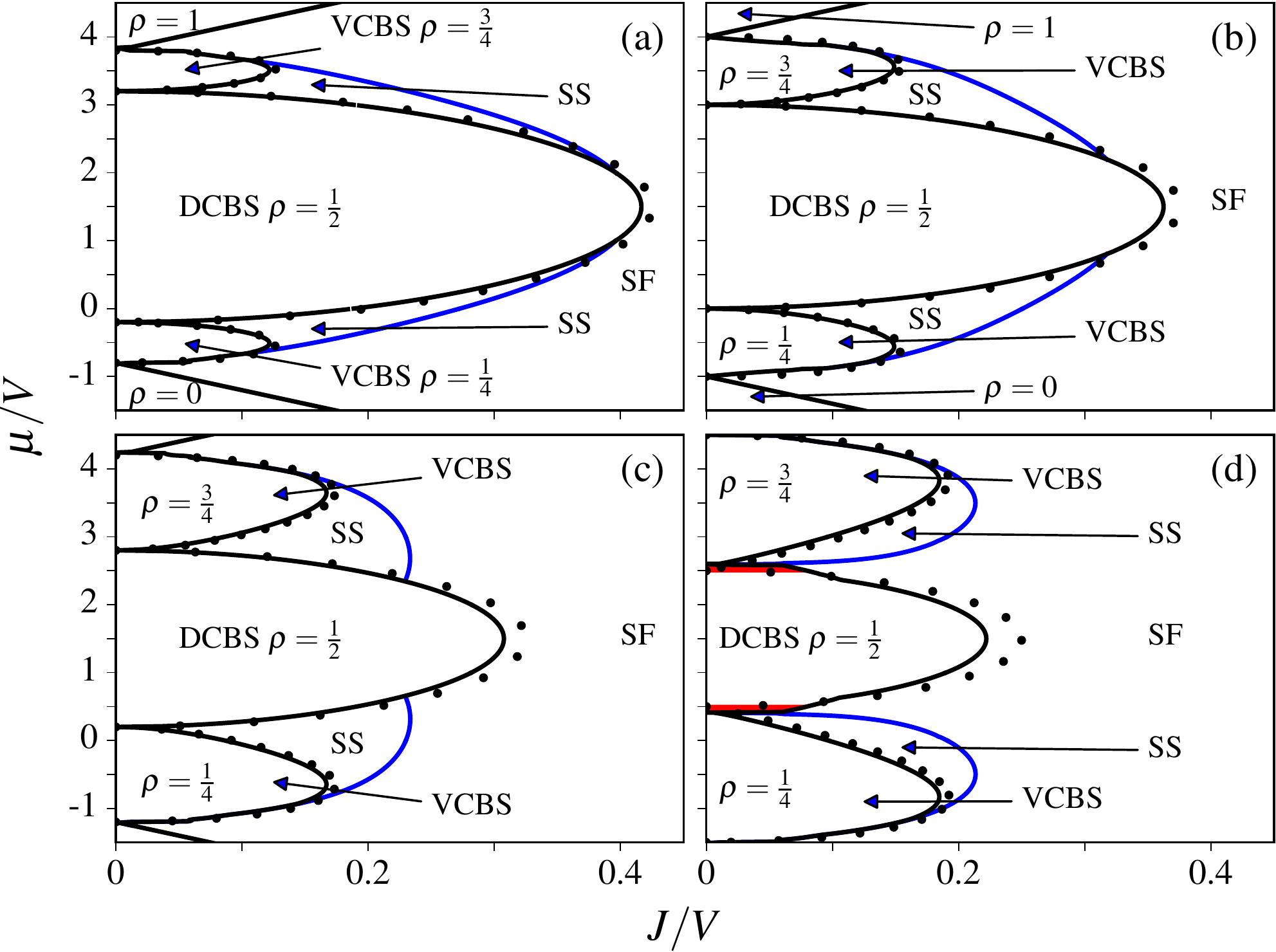}
   \caption{Phase diagrams in the $J/V-\mu/V$ plane for 
	    $J^{\prime}/V = 0.8$, $1.0$, $1.2$ and $1.5$ for 
	    $V^{\prime}/V = -1.0$ are shown in the panels
	    (a), (b), (c) and (d), respectively. The black lines indicate 
	    the phase boundaries between the incompressible and compressible 
	    phases, 
	    and the blue lines represent the SS-to-SF phase boundary. 
	    Filled black circles are the phase boundaries obtained from the 
	    perturbative analysis of the mean-field Hamiltonian, as given in 
	    the Eq.(\ref{pb_dcbs_1by2}), (\ref{pb_vcbs_1by4}) and 
	    (\ref{pb_vcbs_3by4}). The red shaded region in the (d) corresponds 
	    to an incompressible phase with no dimer structure. 
	    }
    \label{comb_bil2}
\end{figure}

 For $J^{\prime}/V = 0.5$, the qualitative features of the phase diagram are
similar to $J^{\prime}/V = 0$. But, the non-zero value of the interlayer 
hopping disfavours the dimer state, and the DCBS $\rho = 1/2$ lobe shrinks. 
This can be observed from  a comparison between Fig.~\ref{comb_bil1} (a) and 
(b). The interlayer hopping is, however, not sufficiently strong to favour the
triplet state as in Eq.~(\ref{t0_state}). The shrinking of incompressible 
lobes like the DCBS at higher hopping strength is a common feature of 
optical lattice system. The system tends to exhibit compressible phases as 
the total hopping energy is enhanced. So, the DCBS lobe continues to shrink 
with further increase in $J^{\prime}/V$, which can also be read off from 
Fig.~\ref{comb_bil2}.  


\subsubsection{Moderate interlayer hopping $J^{\prime}/V \approx 1$}
  We consider $J^{\prime}/V = 0.8$ and $1$ as representative cases for this 
domain, and the phase diagrams are shown in Figs.~\ref{comb_bil2}(a) and 
(b), respectively. As in the previous case, the solid black lines are the
numerical phase boundaries. And, the filled circles are the analytical 
solutions of Eqs.~(\ref{pb_dcbs_1by2}), ~(\ref{pb_vcbs_1by4}), 
and~(\ref{pb_vcbs_3by4}) for the DCBS $\rho = 1/2$, VCBS $\rho = 1/4$, and 
VCBS $3/4$ states, respectively. One striking feature of the phase diagrams is 
the presence of the VCBS phase. In these states, the occupancy can be thought 
as resonating between two NN lattice sites in different layers, which 
corresponds to the triplet state in Eq.~(\ref{t0_state}). This triplet state 
is stabilized by the interlayer hopping, and the VCBS states  appear when 
$J^{\prime}/V > 0.5$. They emerge as small lobes above and below the DCBS 
lobe, and grows with the increase in $J^{\prime}/V$. This is because, the 
$\rho = 1/4$ and $3/4$  VCBS states are the hole and particle excitations of 
the DCBS $\rho = 1/2$ state, respectively. This is also evident from the
comparison of the Eqs.~(\ref{dcbs_1by2})-(\ref{vcbs_3by4}). Thus, besides
the MI and DCBS states, the system hosts the VCBS states at low values of 
$J/V$. In addition, the compressible SS phases also appears in between the
DCBS and VCBS phases. In the SS phase, both the diagonal and off-diagonal 
long-range order are present. That is, it has the superfluid 
characteristics and the periodic modulation in $n_{p,q,r}$ distribution. These
are characterized by the finite values of $\phi_{p,q,r}$ and $S_{r}(\pi,\pi)$, 
respectively. For larger values of $J/V$, the system is in the SF phase with 
uniform $n_{p,q,r}$ and $\phi_{p,q,r}$.

 Comparing Figs.~\ref{comb_bil2}(a) and (b), we observe an enhancement in the
domains of the VCBS and SS phases. This implies that the VCBS phase, even 
though incompressible, is stabilized by the large interlayer hopping. This 
needs to be contrasted with the DCBS phase, for which the domain shrinks
with increased hopping strength. The enhancement of the SS domain indicates 
that the SS phase originates from the intralayer hopping induced 
particle-hole excitations to the VCBS states, but not to the DCBS state. This 
is corroborated later by considering stronger interlayer hopping strength. 
We also observe good agreement between the analytical and numerical phase 
boundaries of the VCBS states. Whereas, for the DCBS phase, the analytical 
results tend to over estimate the phase boundary. This is because, the VCBS 
states are eigenstates of the unperturbed Hamiltonian in 
Eq.~(\ref{unperturb_vcbs_hamil}), which has the interlayer hopping term. So, 
as in the CGMF method, the interlayer hopping term is treated exactly in the 
analytical approach. But, for the DCBS state, the site-decoupled mean field 
analysis is applicable. It treats all the intra and interlayer hopping as 
perturbations, and fails to capture the domain with large hopping.

 
\subsubsection{Strong interlayer hopping $J^{\prime}/V > 1.0$ }

 We consider $J^{\prime}/V = 1.2$ and $1.5$ as representative cases of this
domain. And, the phase diagrams are shown in Fig.~\ref{comb_bil2} (c) and (d), 
respectively. The VCBS states are more prominent with higher $J^{\prime}/V$.
The DCBS $\rho = 1/2$, on the other hand, continues to shrink with the 
increase in $J^{\prime}/V$. And, this is consistent with the earlier 
observation. For $J^{\prime}/V = 1.5$ the extent of the DCBS and VCBS lobes
are comparable. It is important to note that the SS phases are sandwiched 
between the VCBS and DCBS lobes. But, the extent of these domains around the 
DCBS lobe shrink with the increase in $J^{\prime}/V$. That is, the domains 
detach from the DCBS lobe. And, the SS phase surround only the VCBS lobes when 
$J^{\prime}/V = 1.5$. This implies that the SS phase in our study is created 
through the particle-hole excitations to the VCBS states. Hence, this state 
may be referred to as the valence bond SS (VSS) state.

 An emergent feature of the strong inter-layer hopping is the
quantum phase in the shaded parameter domain in the phase diagram. This occurs
for $\mu/V \in [0.4, 0.5]$ and $\mu/V \in [2.5, 2.6]$, in the phase diagram 
as shown in the Fig.~\ref{comb_bil2} (d). The discernible distortion of the 
phase boundary in this region is due to a new incompressible phase which 
replaces the DCBS phase. The intralayer and interlayer density distributions 
exhibit checkerboard order with a two sublattice structure, and average 
density is $\rho = 0.5$. The occupancies, however, are real for both the 
layers. Unlike the DCBS phase, the pair expectation between the two layers of 
this phase is $\approx10^{-4}$, which highlights the non-dimer structure. The
stability of this phase is the strong interlayer hopping, which prefers the 
basis states allowing maximal interlayer hopping. For $\mu/V = 0.5$ to $0.6$, 
this new phase is metastable and competes with the energetically lower DCBS 
phase. Further studies, in the strong interlayer hopping domain are underway, 
to understand the properties of the quantum phases in this domain.

An important feature of all the phase diagrams in Figs.~\ref{comb_bil1} and 
~\ref{comb_bil2} is the vertical symmetry of the quantum phases around the
$\mu/V = 1.5$ axis. This is due to the underlining particle-hole symmetry of 
the system Hamiltonian. From the Eq.~(\ref{par_hole_sym}), it is evident that 
the particle-hole symmetry point is at $\mu/V = 1.5$ for $V^{\prime} = -1.0 V$,


\subsubsection{Quantum fluctuations} 
\label{clus_role}

 The phase diagrams considered so far are computed with the $2^3$ clusters. 
To understand the effects of the quantum fluctuations to the quantum phases, 
we perform computations by varying the cluster sizes from single-site to
$4\times1\times2$. In the CGMF method, quantum correlations are better 
described with increasing cluster size. Thus, the method provides better
better results with larger clusters.

\begin{figure}[ht]
\includegraphics[height = 6.0cm]{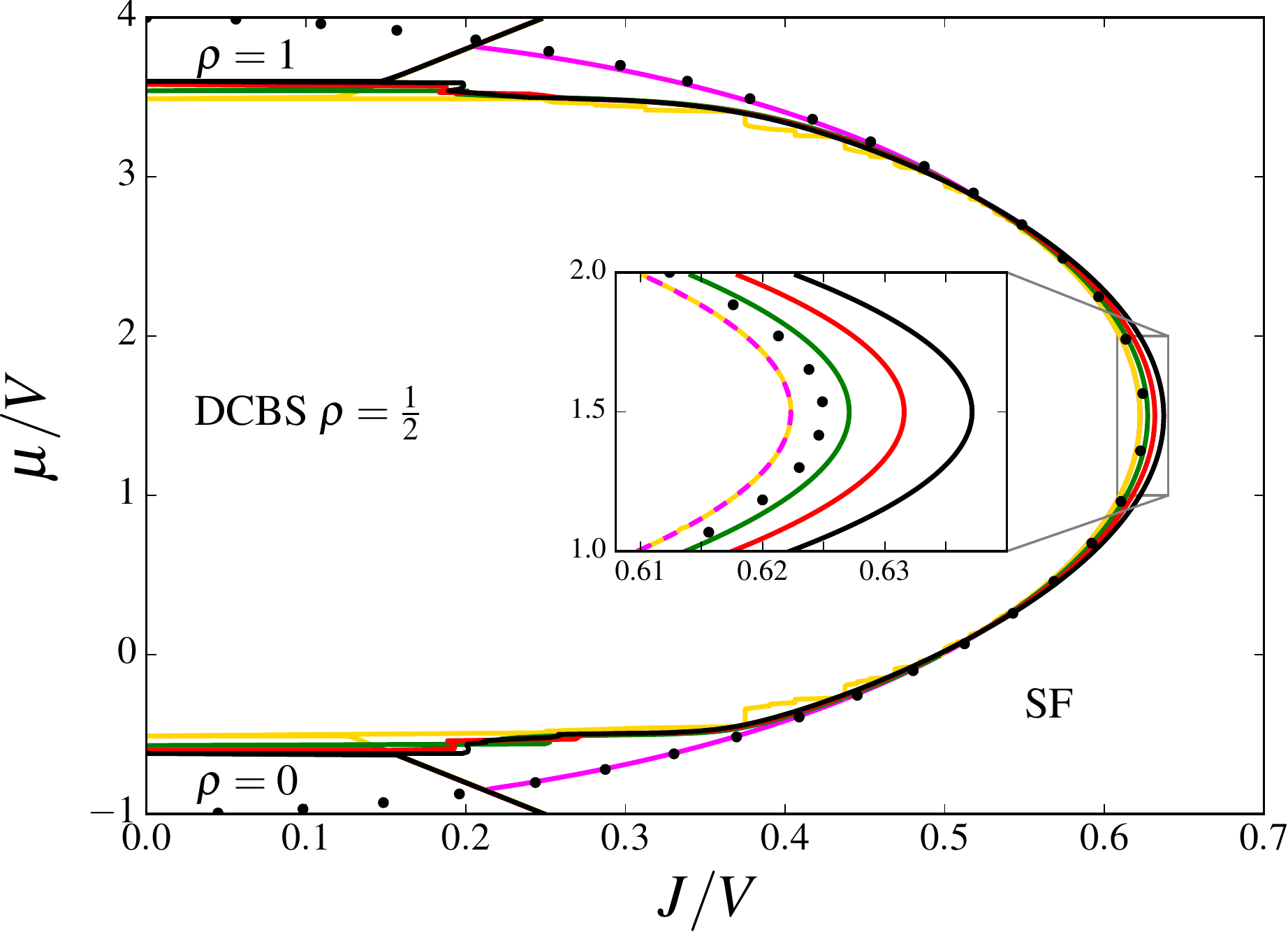}
\caption{Phase diagrams of the bilayer optical lattice in the
	  $J/V-\mu/V$ plane for $J^{\prime}/V = 0$ and $V^{\prime}/V = -1.0$
	  obtained from the clusters of
          different sizes: $1\times1\times1$ (magenta),
          $1\times1\times2$ (gold), $2\times1\times2$ (green), 
          $4\times1\times2$ (red), and $2\times2\times2$ (black). The filled 
          black circles represent the analytically obtained phase boundary 
          obtained from the Eq.(\ref{pb_dcbs_1by2}).}
    \label{jz0_mulclus}
\end{figure}

We first consider the case of  $J^{\prime}/V = 0$, to analyse quantitative 
changes of the DCBS $\rho = 1/2$ lobe with different cluster sizes. The phase 
diagram is shown in Fig.~\ref{jz0_mulclus}. The phase boundary near the tip 
of the lobe shows marginal enlargement with the increase in cluster size. 
For comparison we mark the analytically determined phase boundary by black 
filled circles. This, except for minor deviations around the tip of the
DCBS lobe, is in good agreement with the single-site numerical results. The 
agreement is an expected feature as the site-decoupled mean-field theory 
is equivalent to single-site mean-field theory in determining the 
incompressible-compressible phase boundaries. The deviation around the tip can 
be attributed to the large value of $J/V$ and the numerical threshold in the 
value of the superfluid order parameter. Another interesting aspect of the 
figure is the overlap of the phase boundaries obtained from single-site and 
$1\times1\times2$ clusters. This is because, for $J^{\prime}/V = 0$ the 
inter-layer coupling is only through interlayer interaction. So, in the 
intralayer hopping dominated regime, large $J/V$, the two are expected to 
give similar results. In this domain the results, however, improve with the 
increase in the intralayer cluster size, for example, with 
$2\times1\times2$, $4\times1\times2$ and $2^3$ clusters. On the other hand, 
the phase boundaries obtained from single-site and $1\times1\times2$ clusters 
do not match in the low $J/V$ regime. In this domain the quantum phases are 
determined by the interaction energy, and  is better described by the 
$1\times1\times2$ cluster.

\begin{figure}[ht]
\includegraphics[height = 6.0cm]{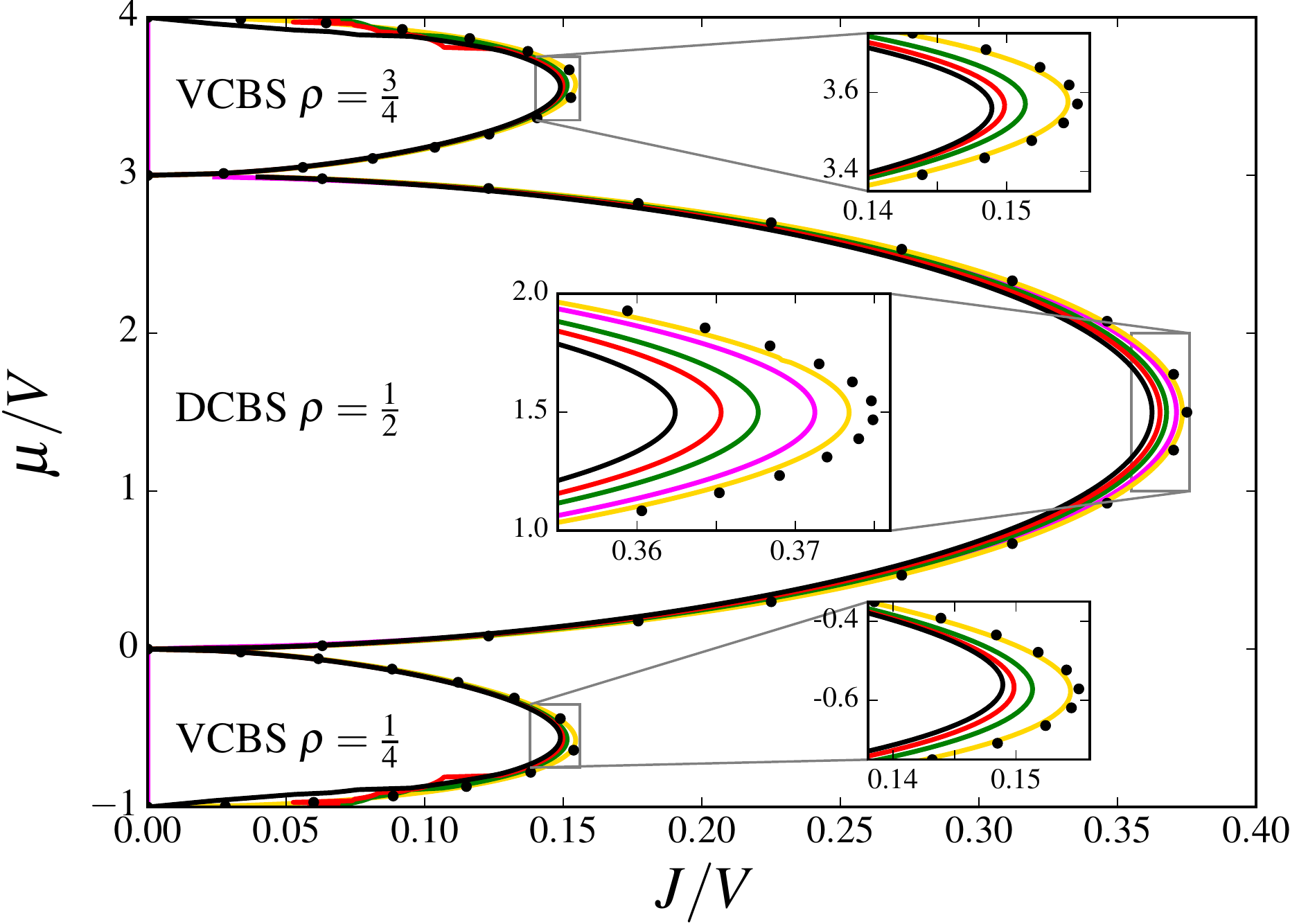}
   \caption{The incompressible to compressible phase boundaries
	    in the $J/V-\mu/V$ plane for $J^{\prime}/V = 1.0$ and 
	    $V^{\prime}/V = -1.0$ obtained from
            clusters of different sizes: $1\times1\times1$ (magenta),
            $1\times1\times2$ (gold), green for $2\times1\times2$ (green),
            $4\times1\times2$ (red), and $2\times2\times2$ (black). The filled 
            black circles indicate the analytical phase boundaries obtained 
	    from Eq.(\ref{pb_dcbs_1by2}), (\ref{pb_vcbs_1by4}) and 
            (\ref{pb_vcbs_3by4}).}
    \label{jz1_mulclus}
\end{figure}

Next, we consider the case of $J^{\prime}/V = 1$, and study the impact of
quantum fluctuations on the VCBS $\rho = 1/4$ and $3/4$, and DCBS $\rho = 1/2$
phases. The VCBS states has checkerboard distribution of the maximally entangled
triplet state. So, it is important to study the effects of the quantum
correlations, better described by the CGMF method with larger cluster sizes, on 
the VCBS phase. For this, we obtain phase diagram by varying the cluster size. 
We observe that the VCBS lobes shrink with the increase in cluster size, shown 
as insets in Fig.~\ref{jz1_mulclus}. Note that the single-site theory
cannot describe the VCBS states due to absence of the minimal intersite 
correlations required to represent the state. Therefore, the 
phase boundaries of the VCBS states are obtained using $1\times1\times2$, 
$2\times1\times2$, $4\times1\times2$, and $2^3$ clusters. In 
addition, we illustrate the analytical phase boundaries obtained by solving 
Eqs.~(\ref{pb_vcbs_1by4}) and~(\ref{pb_vcbs_3by4}), which are in agreement 
with the phase boundaries obtained using $1\times1\times2$ cluster. This is 
because, the unperturbed Hamiltonian in Eq.~(\ref{unperturb_vcbs_hamil}) treats 
the interlayer hopping term exactly, and the intralayer hopping terms are 
considered as perturbation with the SF order parameter as perturbation 
parameter. So, this is similar to the mean-field Hamiltonian 
considered in the CGMF method for $1\times1\times2$ cluster. It is important to
note that, unlike the $J^{\prime}/V = 0$ case, now the DCBS lobe shrinks with 
the increase in cluster size. This is evident from the phase boundaries
shown as inset in Figs.~\ref{jz0_mulclus} and~\ref{jz1_mulclus} for the DCBS 
lobe.

In addition, to verify the robustness of the quantum phases against quantum
correlations and fluctuations, we have performed several computations with 
$4\times4\times2$ cluster as well. But, we do not present phase diagrams 
using this cluster due to exponential growth of the local Hilbert space of a 
cluster with its size. This leads to long computational time to obtain the 
phase boundaries.


\subsection{$V^{\prime} \ne V$ case} \label{vp_ne_v}
 The results discussed so far consider identical intralayer repulsive and 
interlayer attractive NN interactions. Now we consider the regime with 
asymmetric NN interaction strengths. In particular, we consider the cases of
$V^{\prime} = -0.25V$ and $-2V$. The former is motivated by the experimental 
study of Baier \emph{et al.}~\cite{baier_2016}. In the experiment, the 
wavelengths of the lasers generating the optical lattice have the ratios
$\lambda_{x}:\lambda_{y}:\lambda_{z} = 1:1:2$. This implies $a_z = 2a$,
and corresponds to $V^{\prime} = -0.25V$. The case of $V^{\prime} = -2V$
corresponds to $a_z = a$ ( $\lambda_{x}:\lambda_{y}:\lambda_{z} = 1:1:1$) and 
the optical lattice has cubic unit cell.  

\begin{figure}[ht]
\includegraphics[height = 6.2cm]{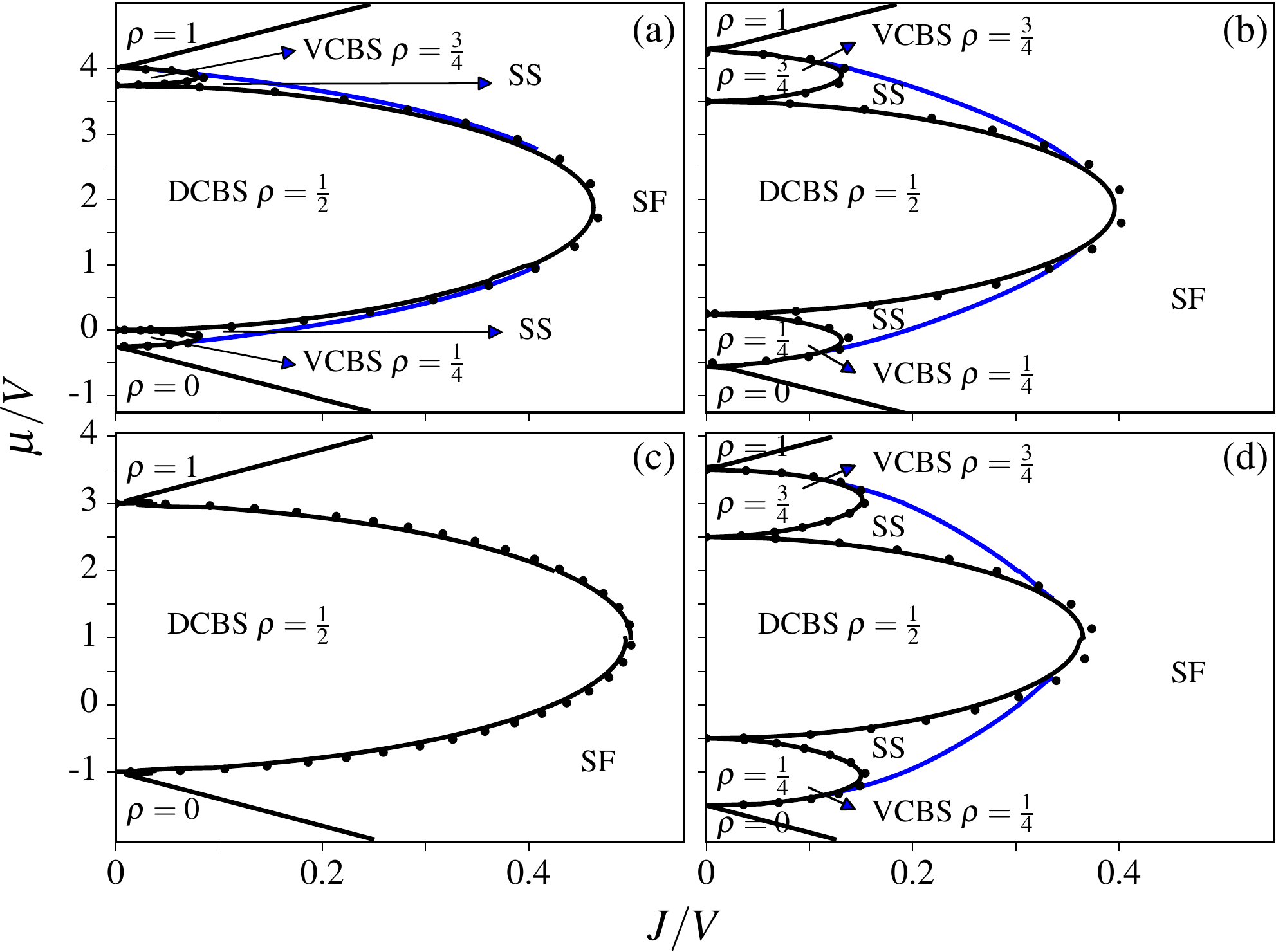}
        \caption{Phase diagram in the
            $J/V-\mu/V$ plane for $V^{\prime}/V = -0.25$ (upper panel) and
            $V^{\prime}/V = -2.0$ (lower panel). The subplots (a) and (b)
            correspond to $J^{\prime}/V = 0.25$ and $J^{\prime}/V = 0.5$,
            respectively. And the subplots (c) and (d) correspond to
            $J^{\prime}/V = 1.0$ and $J^{\prime}/V = 1.5$, respectively. The 
	    black solid lines represent the phase boundaries 
	    between incompressible 
            and compressible phases, and the blue lines indicate the phase
            boundaries between SS and SF phase. The black circles represent the
            analytically obtained phase boundary points obtained from the 
            Eq.(\ref{pb_dcbs_1by2}), (\ref{pb_vcbs_1by4}) and 
            (\ref{pb_vcbs_3by4}).}
    \label{ph_diag_vp_ne_v}
\end{figure}


\subsubsection{$V^{\prime} = -0.25V$}

The phase diagram for $J^{\prime}/V = 0.25$ and $0.50$ are shown in 
Fig.~\ref{ph_diag_vp_ne_v}(a) and (b), respectively. The phase diagrams are 
symmetric about the particle-hole symmetry point $\mu/V = 1.875$. And, we 
obtain the same value from the analytic expression in 
Eq.~(\ref{par_hole_sym}). In the phase diagram, the VCBS $\rho = 1/4$ and 
$\rho = 3/4$ lobes are small. This is due to the weak interlayer hopping 
strength and these lobes emerge when $J^{\prime}/V> 0.13$. Like in 
$V^{\prime} = -V$, we observe domains of SS phase which originate from the
edge of the VCBS lobes, and terminate on the DCBS lobe. On increasing 
$J^{\prime}/V$ to $0.5$, the VCBS lobes are enhanced and so do the
domains of the SS phase. In contrast the DCBS lobe, liker earlier cases, 
shrinks in size. For both the values of $J^{\prime}/V$, the phase boundaries
obtained analytically are in good agreement with the numerical results.


\subsubsection{$V^{\prime} = -2V$}

The phase diagrams for this case are shown in Figs.~\ref{ph_diag_vp_ne_v}(c) 
and (d), and these correspond to the interlayer hopping strengths 
$J^{\prime}/V = 1$ and $1.5$, respectively. An important feature of the phase 
diagram in Figs.~\ref{ph_diag_vp_ne_v}(c) is the absence of the VCBS phase. 
The phase diagram is thus qualitatively similar to the phase diagram in 
Fig.~\ref{comb_bil1}(b). This indicates that a larger interlayer hopping 
$J^{\prime}/V$ is essential for the VCBS phase to appear in the system.
This ensures that one of the sub-lattice has triplet state $|t_0\rangle$ as 
ground state, which is a characteristic of the VCBS phase. We observe the 
VCBS state enters as a possible ground state when $J^{\prime}/V>1$. So, 
based on the phase diagrams for $V^{\prime}/V = -1$, $-0.25$ and $-2$, 
the system may exhibit the VCBS phase when $J^{\prime} > |V^{\prime}|/2$.


\subsection{Finite temperature phase diagram} \label{fint_temp}
\begin{figure}[ht]
\includegraphics[height = 5.0cm]{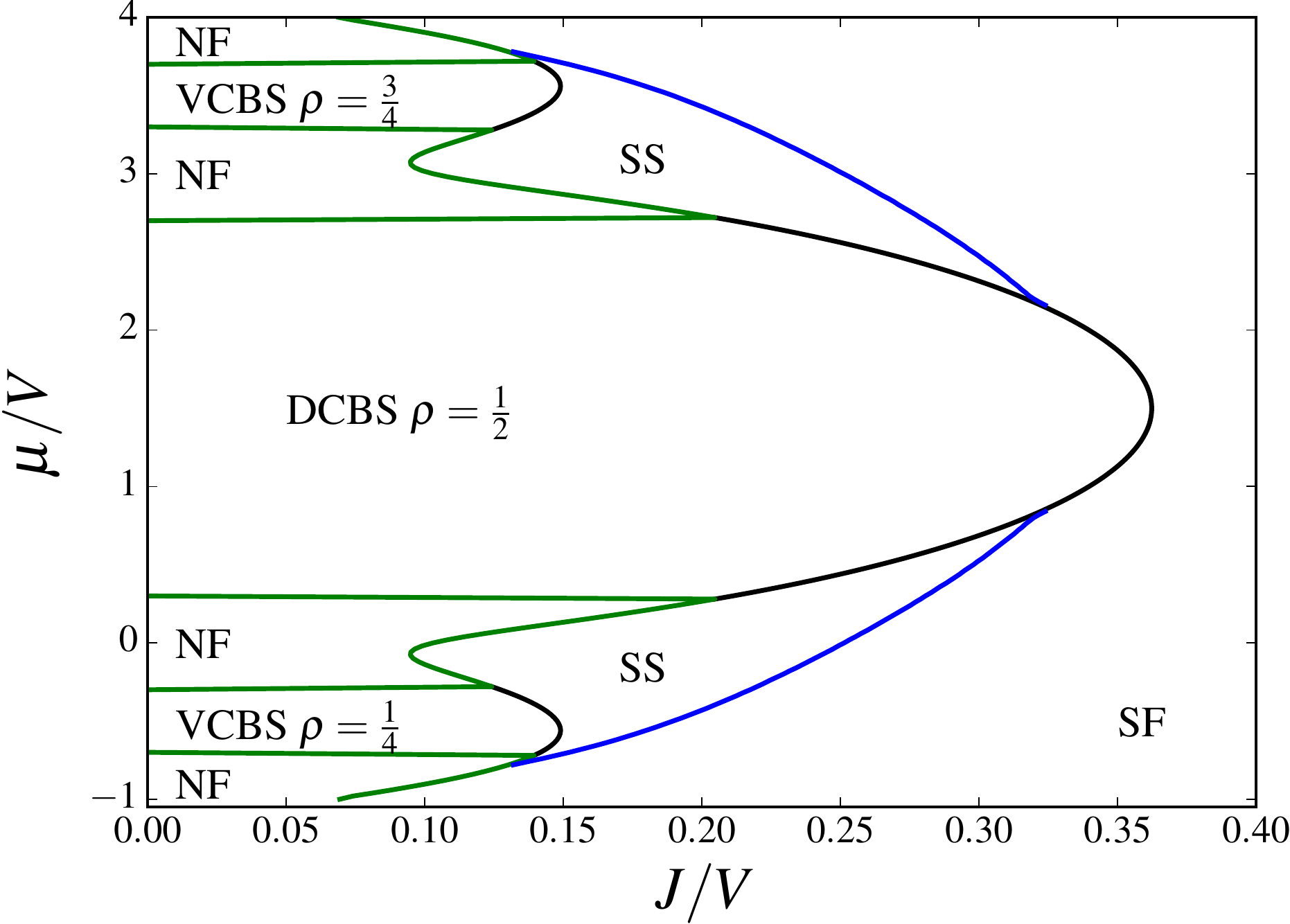}
  \caption{Finite temperature phase diagram for the bilayer 
	   optical lattice in the $J/V-\mu/V$ plane for $J^{\prime}/V = 1.0$, 
	   $V^{\prime}/V = -1.0$ and $k_BT=0.05V$. The green lines 
	   represent the phase boundary between 
           the NF and the compressible phase, while black lines indicate the 
           phase boundaries between incompressible, unmelted phases and 
           compressible phases.}
\label{pd_fint}
\end{figure}

 The results discussed so far are obtained at zero temperature. These provide
qualitative descriptions of the quantum phases present in the system. This 
follows from the characterization of quantum phases and quantum phase 
transitions as zero temperature phenomena. Experimental realizations, however, 
are at finite temperatures. So, we incorporate the effects of 
temperature on the quantum phases, and examine the domains in the phase 
diagrams. The quantum phases are known to ``melt" by the thermal fluctuations
associated with finite temperatures. Thus, at finite temperature the system 
exhibits a normal fluid (NF) phase. It is characterized by zero SF order 
parameter and real occupancy at each 
lattice site~\cite{mahmud_2011, fang_2011, parny_2012, pal_2019, suthar_2020, 
bai_2020}. This is to be contrasted with the incompressible quantum phases, 
which have integer occupancy at each lattice site and zero SF order 
parameter.

 The finite temperature calculations require thermal averaging of the 
observables. This is done by computing the partition function of a cluster as
\begin{equation}
 Z = \sum_{l} e^{-\beta E_{l}},
\end{equation}
where $\beta = 1/k_{\rm B}T$, $k_{\rm B}$ is the Boltzmann constant, $T$ is the
temperature of the system, and $E_{l}$ is the $l$th eigen value of the cluster
Hamiltonian $\hat{H}_{C}$ in Eq.~(\ref{ml_clus_hamil}). Then, the thermal 
average of a local operator $\hat{O}_{p,q}$ at the site $(p,q)$ within the 
cluster is
\begin{equation}
 \langle\langle\hat{O_{p,q}}\rangle\rangle 
        = \frac{1}{Z}{\rm Tr}\left(e^{-\beta\hat{H}_{C}}\hat{O}_{p,q}\right),
\end{equation}
where $\langle\langle...\rangle\rangle$ denotes thermal average, and here the 
trace is calculated with respect to the eigenstates of the cluster 
Hamiltonian. The details of the finite temperature computations with the
CGMF method are given in our previous work~\cite{pal_2019}.

 We consider $J^{\prime}/V = 1$ and $V^{\prime}/V = -1$ as a representative 
case to examine the effects of thermal fluctuations and phase diagram is
shown in Fig.\ref{pd_fint} for $k_BT=0.05V$. We observe the melting of the 
DCBS and VCBS phases at the top and bottom domains of the lobes. These are the 
parameter domains where the lobes close and density fluctuations are 
prominent. So, it is natural for the thermal fluctuation effects to be higher 
in these domains and melting to commence. The phase fluctuations are dominant
around the tip of the lobes and the quantum phases persist. As mentioned 
earlier, the SF order parameter is zero in the NF phase, but the system has 
strong number fluctuation. This leads to incommensurate density distribution. 
As a result, zero SF order parameter is no longer the criterion to identify 
incompressible phases at finite temperatures. Therefore, to
distinguish the NF phase from the incompressible quantum phases, we consider
compressibility, $\kappa$, as the order parameter. The NF phase possesses 
finite $\kappa$, but, it is negligibly small for the incompressible phases. 
One point to be noted is, at finite temperatures the number fluctuations, 
however small, are always present. This leads to finite $\kappa$ throughout 
the parameter domain of the phase diagram. So, we empirically define $\kappa$ 
as the number variance~\cite{mahmud_2011}. Then, we have to make an appropriate
choice of threshold value for $\kappa$ to determine the NF-VCBS or NF-DCBS 
phase boundaries. This is also a characteristic of a continuous phase 
transition. Earlier works \cite{mahmud_2011, fang_2011} have reported the 
condition on $\kappa$ to distinguish the NF phase from the incompressible 
quantum phases. In the present work, we consider the threshold value of 
$\kappa$ as $0.04/V$. That is, $\kappa > 0.04/V$ indicates melting of the 
quantum phases, and thus, the presence of NF phase.
\begin{figure}[ht]
\includegraphics[height = 5.0cm]{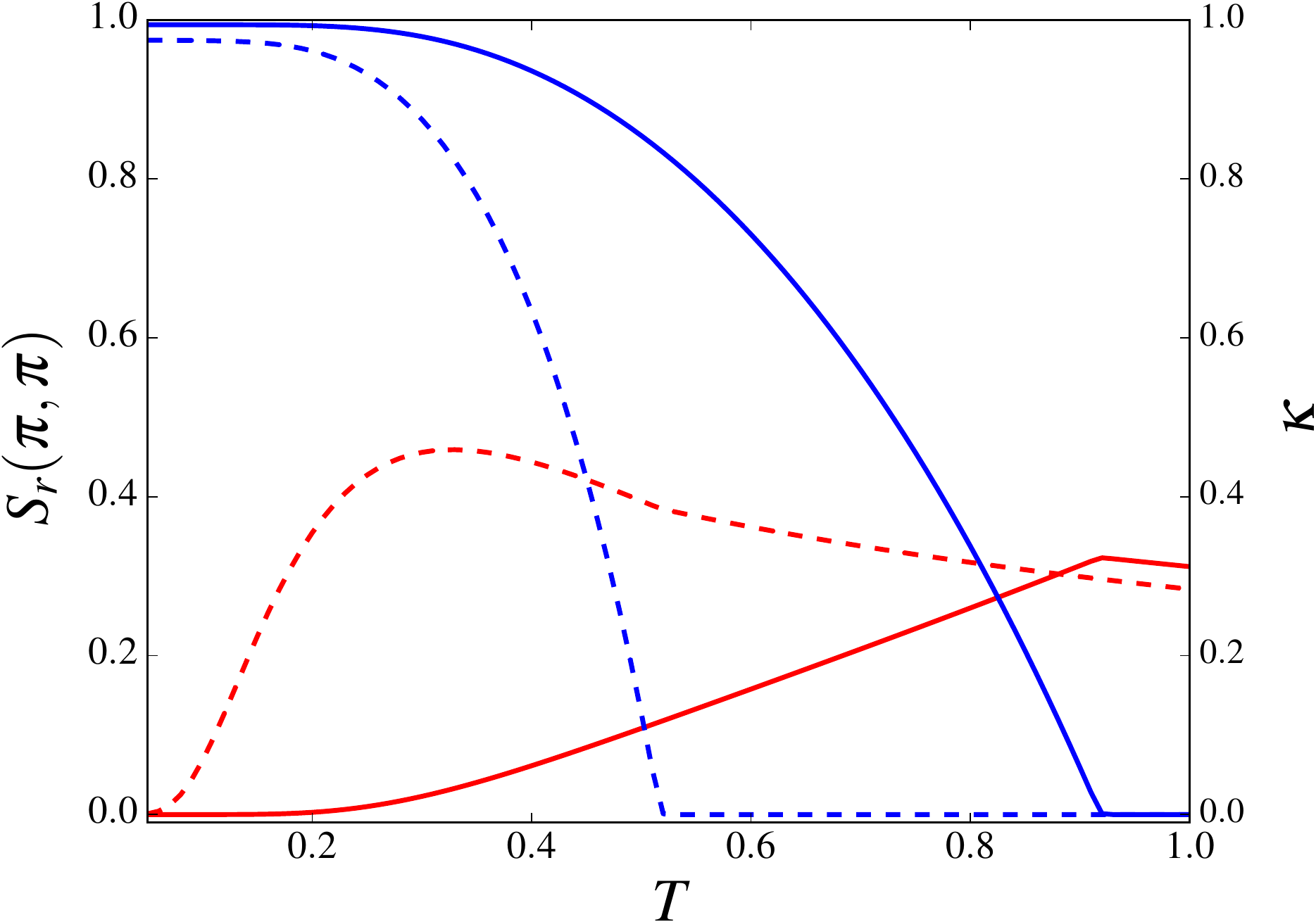}
    \caption{Plot of the $\kappa$ and $S(\pi,\pi)$ as a 
            function of temperature. The blue solid (dashed) line indicates 
            the $S(\pi,\pi)$ for DCBS $\rho = 1/2$ (VCBS $\rho = 1/4$) 
            state at low temperatures. And the red solid (dashed) line 
            indicates the $\kappa$ for DCBS $\rho = 1/2$ (VCBS $\rho = 1/4$) 
	    state. Here the system parameters $J^{\prime}/V = 1.0$ and 
	    $V^{\prime}/V = -1.0$ are considered. The value of $J/V$ is 
	    fixed at $0.08$, and the values of the chemical potential 
	    $\mu/V$ are $-0.5$
	    and $1.5$ for VCBS ($\rho = 1/4$) and DCBS phases, respectively.}
\label{sfpp_kappa}
\end{figure}

 One important feature of the NF phase is, it inherits the density order
of the original incompressible phase.  For example, melting of the checkerboard 
incompressible quantum phases forms a checkerboard NF (CBNF) phase. The 
checkerboard order vanishes at higher temperature, and we obtain uniform 
density NF phase. To illustrate the transition, we plot the $S_{r}(\pi,\pi)$ 
and the $\kappa$ as a function of temperature in Fig.\ref{sfpp_kappa}. The 
CBNF domain is identified by the non-zero $S_{r}(\pi,\pi)$ and $\kappa$ and
occurs at lower temperatures. With the increase in temperature 
$S_{r}(\pi,\pi)$ decrease and becomes zero at $T = 0.92$ and $0.52$ for 
the DCBS and VCBS $\rho = 1/4$ phases, respectively. In this domain, the NF 
phase has uniform density distribution. However, the particle-hole symmetry of 
the phase diagram is robust against thermal fluctuations and persists at finite
temperatures. That is, the melting of the quantum phases is symmetric 
about $\mu/V = 1.5$. It is important to notice that for a fixed value of 
$J/V$, the VCBS states melt at lower temperatures than the DCBS state. This 
can be read off from Fig.~\ref{sfpp_kappa}. The melting of the VCBS phase at 
a lower temperature implies that the VCBS states are entangled and have larger 
quantum correlation embedded. This renders the phase fragile against the 
thermal fluctuations. 
\begin{figure}[ht]
\includegraphics[height = 5.0cm]{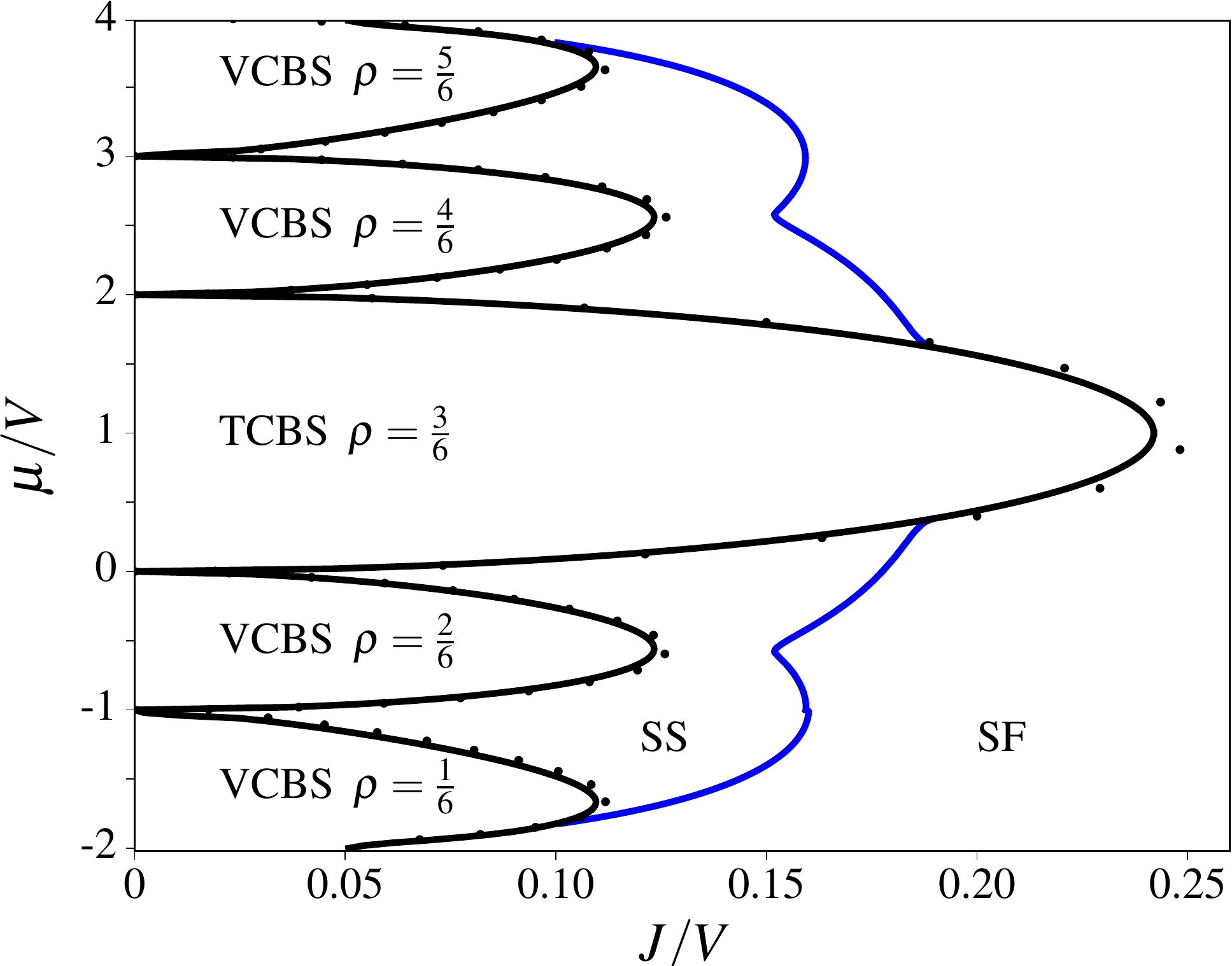}
	\caption{Phase diagram for trilayer optical lattice 
	    in the $J/V-\mu/V$ plane for $J^{\prime}/V = 1$ and
	    $V^{\prime}/V = -1$.  The black solid lines
	    represent the phase boundaries between incompressible and 
	    compressible phases, and the blue lines indicate the phase 
	    boundaries between SS and SF phase. The black circles represent
	    the analytically obtained phase boundary points from the 
	    Eq.(\ref{pb_tcbs_1by2}) - (\ref{pb_vcbs_5by6}).}
\label{pd_tri}
\end{figure}


\subsection{Phase diagrams of trilayer system} \label{trilayer}

We now consider introducing an additional layer to the bilayer system and 
make it a trilayer system. The Hamiltonian of the system is given by the
Eq.(\ref{mult_hamil}) with $M = 3$. We, then, investigate the quantum 
phases emerging due to the competition between the interlayer hopping and 
interlayer interaction. For illustration, we consider $V^{\prime} = -V$, 
and $J^{\prime} = V$, and the phase diagram in the $J/V - \mu/V$ plane,
is shown in Fig.\ref{pd_tri}. 

At low $J/V$, the ground state is either the VCBS state, or a trimer 
checkerboard solid (TCBS) state. The TCBS lobe corresponds to the 
$\rho = 1/2$, and is the central large lobe in the phase diagram. This state
is an analog of the DCBS $\rho = 1/2$ state in the bilayer system, and 
is incompressible. The two lobes below the TCBS lobe are the VCBS $\rho = 1/6$
and VCBS $\rho = 2/6$ lobes, in the increasing order of $\mu/V$. And, the 
two lobes above the TCBS lobe are of the VCBS $\rho = 4/6$ and VCBS 
$\rho = 5/6$ states. Thus, the system displays a rich structure of the 
VCBS states for the trilayer system. As in the case of the bilayer system, 
the VCBS states are also incompressible. They result from the increased 
degrees of freedom associated with the inter-layer hopping. The particle-hole 
symmetry of the trilayer system which can be deduced from the phase diagram
is $\mu/V = 1$. This particle-hole symmetric $\mu/V$ value is also 
analytically obtained from Eq.(\ref{gen_par_hole_sym}), the details are given 
in the Appendix (\ref{append_a}). 

As $J/V$ is increased, the system acquires higher intra-layer hopping energy. 
This, then, leads to co-existence of diagonal and off-diagonal long range 
order in the system and makes the system compressible. The system is then 
in the supersolid phase. The supersolid phases envelopes the VCBS states, 
earlier similar results observed for the  bilayered systems. For an 
$n$-layered system, we can generalize the VCBS states from $\rho = 1/2n$ to 
$\rho = 2n-1/2n$, excluding the $\rho = n/2n = 1/2$ state. We thus observe 
that our results can be generalized for the multilayered systems.


\section{Conclusions}\label{conclusions}
In conclusion, we have examined the quantum phases of the polarized dipolar
atoms in multilayer optical lattices at zero and finite temperatures. The
rich phase diagrams of the model display parameter domains for dimer (trimer)
checkerboard solid and valence bond checkerboard solid phases for the bi-(tri-)
layer lattice system. The interlayer attractive interaction is responsible for
formation of the dimers (trimers), and the intralayer repulsion induces the 
in-plane checkerboard ordering. This stabilizes dimer (trimer) checkerboard
solid phase for average occupancy $\rho = 1/2$. With the increase in
interlayer hopping the dimers (trimers) break to form resonating valence bond
like states. Then, the bilayer lattice can exhibit VCBS phases with average
occupancies $\rho = 1/4$ and $3/4$, respectively. Similarly, the trilayer
lattice supports VCBS phases with $\rho = 1/6$, $2/6$, $4/6$ and $5/6$. These
states are stabilized when $J^{\prime}>|V^{\prime}|/2$, and the corresponding
lobes are enlarged with increasing interlayer hopping strength $J^{\prime}$. On 
the contrary, the parameter domain of the dimer (trimer) checkerboard solid 
shrinks with increasing $J^{\prime}$. In addition to the solid phases, the 
system also exhibits supersolid phases. In the weak and moderate interlayer 
hopping regime, this phase appear in the vicinity of both the DCBS
and VCBS lobes. But, the domains envelope only the VCBS lobes for strong
interlayer hopping, indicating the valance bond nature of the supersolid phase.
With the inclusion of the thermal fluctuations, the quantum phases are observed
to melt to a structured normal fluid, where the melting of the VCBS phase occurs
at a lower temperature than the DCBS phase.

\begin{acknowledgments}
The results presented in the paper are based on the computations using 
Vikram-100, the 100TFLOP HPC Cluster at Physical Research Laboratory, 
Ahmedabad, India. S.B acknowledges the support by Quantum Science and 
Technology in Trento (Q@TN), Provincia Autonoma di Trento, and the ERC Starting
Grant StrEnQTh (Project-ID  804305). SM acknowledges support from the DST, Govt. of India.
\end{acknowledgments}

\appendix
\section{Particle-hole symmetry}\label{append_a}
 The particle-hole symmetry of the Hamiltonian in Eq.(\ref{ml_hamil}) can be
seen by doing the particle hole transformation of the creation and annihilation
operators. The following transformation is used
\begin{eqnarray}
   \hat{a}_{i,r}^{\dagger} &=& \hat{b}_{i,r} \nonumber \\
   \hat{a}_{i,r}           &=& \hat{b}_{i,r}^{\dagger} \nonumber
\end{eqnarray}
where the operators $\hat{a}_{i,r}$ and $\hat{a}_{i,r}^{\dagger}$ are the 
hole annihilation and creation operators. The particle and the hole operators
satisfy the canonical anti-commutation relation
\begin{align*}
	\{\hat{b}_{i,r} \, , \hat{b}_{i,r}^{\dagger} \} = 1 \\ 
	\{\hat{a}_{i,r} \, , \hat{a}_{i,r}^{\dagger} \} = 1 
\end{align*}
With these expressions, the Hamiltonian in Eq.(\ref{ml_hamil}) can be 
rewritten as 
\begin{equation}
 \begin{split}
	\hat{\tilde{H}}_{\rm bi}&=-J\sum_{\langle ij\rangle, r}
	       \Big(\hat{a}_{i,r}\hat{a}_{j, r}^{\dagger} 
               + {\rm H.c.}\Big) - J'\sum_i\left(\hat{a}_{i, 1}
	       \hat{a}_{i, 2}^{\dagger} + {\rm H.c.}\right ) 
           \\
	&+ V\sum_{\langle ij \rangle, r} \hat{a}_{i,r}\hat{a}_{i,r}^{\dagger}
	    \hat{a}_{j,r}\hat{a}_{j,r}^{\dagger}
	   + V' \sum_i\hat{a}_{i, 1}\hat{a}_{i, 1}^{\dagger}\hat{a}_{i, 2}
	     \hat{a}_{i, 2}^{\dagger} \\
	   &- \mu\sum_{i, r}\hat{a}_{i, r}\hat{a}_{i, r}^{\dagger}
 \end{split}
 \label{ml_hamil_hole}           
\end{equation}
This can be further simplified as 
\begin{equation}
 \begin{split}
	\hat{\tilde{H}}_{\rm bi}&=-J\sum_{\langle ij\rangle, r}
	       \Big(\hat{a}_{j, r}^{\dagger}\hat{a}_{i,r}
               + {\rm H.c.}\Big) - J'\sum_i\left(
	       \hat{a}_{i, 2}^{\dagger}\hat{a}_{i, 1} + {\rm H.c.}\right ) \\
	&+ V\sum_{\langle ij \rangle, r} (1 - \hat{\tilde{n}}_{i,r})
	          (1 - \hat{\tilde{n}}_{j,r})\\
	   &+ V' \sum_i(1 - \hat{\tilde{n}}_{i, 1}) (1 - \hat{\tilde{n}}_{i, 2})
	   - \mu\sum_{i, r}(1 - \hat{\tilde{n}}_{i, r}) 
 \end{split}
 \label{ml_hamil_hole1}           
\end{equation}
where the hole number operator is $\hat{\tilde{n}}_{i,r} = 
\hat{a}_{i, r}^{\dagger}\hat{a}_{i,r}$.

\begin{equation}
 \begin{split}
	\hat{\tilde{H}}_{\rm bi}&=-J\sum_{\langle ij\rangle, r}
	         \Big(\hat{a}_{j, r}^{\dagger}\hat{a}_{i,r}
                 + {\rm H.c.}\Big) - J'\sum_i\left(
	         \hat{a}_{i, 2}^{\dagger}\hat{a}_{i, 1} + {\rm H.c.}\right ) \\
	&+ V\sum_{\langle ij \rangle, r} \Big(\hat{\tilde{n}}_{i,r}
	         \hat{\tilde{n}}_{j,r} - \hat{\tilde{n}}_{j,r} -
		 \hat{\tilde{n}}_{i,r} + 1  \Big) \\
	&+ V' \sum_i\Big(\hat{\tilde{n}}_{i, 1}\hat{\tilde{n}}_{i, 2} - 
	         \hat{\tilde{n}}_{i, 1} - \hat{\tilde{n}}_{i, 2} + 1\Big) 
	       - \mu\sum_{i, r}(1 - \hat{\tilde{n}}_{i, r}) 
 \end{split}
 \label{ml_hamil_hole2}           
\end{equation}
\begin{eqnarray}
\hat{\tilde{H}}_{\rm bi} &=& 
             -J \sum_{\langle ij\rangle, r}  \left( \hat{a}^{\dagger}_{j, r}
                \hat{a}_{i, r} + {\rm H.c.} \right)
                -J' \sum_{i} \left( \hat{a}^{\dagger}_{i, 2}
                \hat{a}_{i, 1} + {\rm H.c.} \right)
                \nonumber \\
                &&+V \sum_{\langle ij\rangle, r} \hat{\tilde{n}}_{i, r}
                \hat{\tilde{n}}_{j, r} 
                -zV \sum_{i, r} \hat{\tilde{n}}_{i, r} \nonumber \\
		&&+V' \sum_{i} \hat{\tilde{n}}_{i, 1} \hat{\tilde{n}}_{i, 2} 
                -V' \sum_{i, r} \hat{\tilde{n}}_{i, r}
                +\mu \sum_{i, r} \hat{\tilde{n}}_{i, r} \nonumber \\ 
                &=&
                -J \sum_{\langle ij\rangle, r}  \left( \hat{a}^{\dagger}_{j, r}
                \hat{a}_{i, r} + {\rm H.c.} \right)
                -J' \sum_{i} \left( \hat{a}^{\dagger}_{i, 2}
                \hat{a}_{i, 1} + {\rm H.c.} \right)
                \nonumber \\
                &&+V \sum_{\langle ij\rangle, r} \hat{\tilde{n}}_{i, r}
                \hat{\tilde{n}}_{j, r} 
                +V' \sum_{i} \hat{\tilde{n}}_{i, 1} \hat{\tilde{n}}_{i, 2} 
                -\tilde{\mu} \sum_{i, r} \hat{\tilde{n}}_{i, r} 
                \nonumber
\end{eqnarray}
where $\tilde{\mu} = (-\mu +zV +V')$ \\

Particle hole symmetry point is
\begin{eqnarray}
        \quad \mu &=& \tilde{\mu} \nonumber \\
        \implies \mu &=& -\mu +zV +V' \nonumber \\
        \implies \mu &=& \left( \frac{zV + V'}{2} \right)
 \label{par_hole_sym} 
\end{eqnarray}

Similarly, for a multilayer system, the particle-hole symmetry point is  
\begin{eqnarray}
  \mu &=& \left( \frac{zV + 2V'}{2} \right).
 \label{gen_par_hole_sym} 
\end{eqnarray}
This result is obtained after employing the periodic boundary condition along 
the $z$ direction.


\section{Trilayer phase boundary} \label{append_b}
 We present the derivation of the analytical phase boundaries of the 
quantum phases of the trilayer optical lattice. We first derive the phase 
boundary separating the TCBS $\rho = 1/2$ domain from the compressible phase.

\subsubsection{TCBS phase boundary}
As discussed for the phase boundary of the DCBS phase, we first write the 
site-decoupled unperturbed Hamiltonian for trilayer optical lattice as 

\begin{eqnarray} 
  \hat{h}^{(0)}_{p,q}&=&\sum_{r=1}^{3}\hat{n}_{p,q,r}\left(V'n_{p,q,r+1}
	        +V\overline{n}_{p,q,r} -\mu\right) 
 \label{unperturb_hamil_tri}	     
\end{eqnarray}
It has to be understood that the $r+1$ in the above equation is considered 
with modulo $3$, so that we have interaction between the $1^{\rm st}$ and 
$3^{\rm rd}$ layer.
The unperturbed energies corresponding to the eigenstates 
$|n_1^r, n_2^r, n_3^r \rangle$ of the unperturbed Hamiltonian 
$\left(r = \{A,B\}\right)$ are 
\begin{eqnarray} 
  E^{A}_{n^{A}_{1}n^{A}_{2}n^A_3} &=& V'\left(n^{A}_{1}n^{A}_{2} + 
	                       n^{A}_{2}n^{A}_{3} + n^{A}_{3}n^{A}_{1}\right)
			       \nonumber \\
	&+& \sum_{r=1}^{3}\left(4V n^A_r n^B_r - \mu n^A_r\right),
 \label{subA_energy_tri}	
\end{eqnarray}
for $(p,q)\in A$, and
\begin{eqnarray} 
  E^{B}_{n^{B}_{1}n^{B}_{2}n^B_3} &=& V'\left(n^B_1 n^B_2 + 
			       n^{B}_{2}n^{B}_{3} + n^{B}_{3}n^{B}_{1}\right)
			       \nonumber \\
	&+& \sum_{r=1}^{3}\left( 4Vn^B_r n^A_r - \mu n^B_r\right),
 \label{subB_energy_tri}	
\end{eqnarray}
for $(p,q)\in B$.
The perturbation Hamiltonian is 
\begin{eqnarray}
  \hat{T}^{A} &=& - 4J\sum_{r=1}^{3}\phi_r^B \left (\hat{b}^A_r
	            +{\hat{b}^{A^\dagger}_r}\right ) - 2J'\sum_{r=1}^{3}
		    \phi_{r+1}^{A}
	            \left(\hat{b}^A_r+\hat{b}^{A^{\dagger}}_r\right), 
	            \nonumber \\
 \hat{T}^{B} &=& -4J\sum_{r=1}^{3}\phi_{r}^{A}\left(\hat{b}^{B}_{r}
	   +\hat{b}^{B^{\dagger}}_{r}\right) 
           -2J^{\prime}\sum_{r=1}^{3}\phi_{r+1}^{B}
	    \left(\hat{b}^{B}_{r}+\hat{b}^{B^{\dagger}}_{r}\right). 
        \nonumber 	
\end{eqnarray}
The TCBS state, given by Eq.(\ref{tcbs_1by2}), has occupancies as 
$(n^{A}_{1}, n^{A}_{2}, n^{A}_{3}) = (1,1,1)$ and 
$(n^{B}_{1}, n^{B}_{2}, n^{B}_{3}) = (0,0,0)$. Then, the perturbed 
ground state for sublattice A, similar to the one in Eq.(\ref{perturb_gs}), is 

\begin{equation}
 \begin{split}
	 |\chi^{A}\rangle &= |111\rangle^{A} \\
	  &+ \sum_{m^{A}_{1},m^{A}_{2},m^{A}_{3}} 
	  \frac{\langle m^{A}_{1}m^{A}_{2}m^{A}_{3}|\hat{T}^{A}
	  |111\rangle^{A}}
	  {(E^{A}_{111}
	   -E^{A}_{m_{1}m_{2}m_{3}})} 
	  |m^{A}_{1}m^{A}_{2}m^{A}_{3} \rangle,
 \end{split}
 \label{perturb_gs_tri}	
\end{equation}
where $(m^{A}_{1},m^{A}_{2},m^{A}_{3})\neq(1,1,1)$.
Substituting the perturbation Hamiltonian $\hat{T}^A$ in the previous equation,
we obtain
\begin{equation}
 \begin{split}
	 |\chi^{A}\rangle &= |111\rangle^{A} 
	+ \left(-4J\varphi_B - 2J'\varphi_A\right) \\
	& \left[\frac{|011\rangle^{A}}{(E^{A}_{111}-E^{A}_{011})} +     
	 \frac{|101\rangle^{A}}{(E^{A}_{111}-E^{A}_{101})} +     
	 \frac{|110\rangle^{A}}{(E^{A}_{111}-E^{A}_{110})}\right].  
 \end{split} 
\label{perturb_gs_tri_siteA}	
\end{equation}
And similarly, the perturbed state for sublattice B is 
\begin{equation}
 \begin{split}
  |\chi^{B}\rangle &= |000\rangle^{B} 
	+ \left(-4J\varphi_A - 2J'\varphi_B\right)\\  
	& \left[\frac{|100\rangle^{B}}{(E^{B}_{000}-E^{B}_{100})} +     
	 \frac{|010\rangle^{B}}{(E^{B}_{000}-E^{B}_{010})} +     
	 \frac{|001\rangle^{B}}{(E^{B}_{000}-E^{B}_{001})}\right].  
 \end{split} 
 \label{perturb_gs_tri_siteB}	
\end{equation}
Like in the bilayer case, we have assumed that $\phi_r^{A} = \varphi^{A}$ and 
$\phi_r^{B} = \varphi^{B}$ for all values of $r$.
We can substitute the energy difference denominators by calculating the 
energies, using Eq.(\ref{subA_energy_tri}) and Eq.(\ref{subB_energy_tri}).
Then, we calculate the order parameter for sublattice A and B to get 
\begin{equation}
	\varphi^{A} = -\frac{4J\varphi^{B}}{2V^{\prime}-\mu+2J^{\prime}}, 
 \label{subA_sfop_tri}
 \end{equation}
and
 \begin{equation}
	 \varphi^{B} = -\frac{4J\varphi^{A}}{\mu-4V+2J^{\prime}}.
 \label{subB_sfop_tri}
 \end{equation}
Solving these two equations simultaneously, and taking the limit 
$\{\varphi^{A}, \varphi^{B}\}\rightarrow0^{+}$, we get the phase boundary
separating the TCBS phase from the compressible phase, given in 
Eq.(\ref{pb_tcbs_1by2}).

\subsubsection{VCBS phase boundary}
We first discuss the derivation of the phase boundary between the 
VCBS $\rho = 1/6$ phase and the compressible phase. The unperturbed local 
Hamiltonian for sublattice A is 
\begin{eqnarray}
  \mathcal{H}^{(0)}_{A} &=& -J'\sum_{r=1}^{3}(\hat{b}^{\dagger}_{p,q,r}
	                        \hat{b}_{p,q,r+1}  + {\rm H.c.})
                     + V'\sum_{r=1}^{3}\hat{n}_{p,q,r}\hat{n}_{p,q,r+1}  
		         \nonumber\\
                   &&+ \sum_{r=1}^{3}\left(V\hat{n}_{p,q,r}\overline{n}_{p,q,r}
                     -\mu\hat{n}_{p,q,r}\right),        
 \label{unperturb_vcbs_hamil_tri}   
\end{eqnarray}
The eigenstates of this unperturbed Hamiltonian are
\begin{eqnarray}
\Big\{|000\rangle,& |111\rangle&, |w_0 \rangle = \frac{1}{\sqrt{3}}(|001\rangle + |010\rangle + |100\rangle),  \nonumber \\ 
|\alpha^1\rangle &=& \frac{1}{\sqrt{2}}(|001\rangle - |100\rangle), \;\;
|\beta^1\rangle = \frac{1}{\sqrt{2}}(|010\rangle - |100\rangle), \nonumber \\ 
|w_1 \rangle &=& \frac{1}{\sqrt{3}}(|011\rangle + |110\rangle + |101\rangle),
	        \nonumber \\
|\alpha^2\rangle &=& \frac{1}{\sqrt{2}}(|011\rangle - |110\rangle), \;\;
|\beta^2\rangle = \frac{1}{\sqrt{2}}(|101\rangle - |110\rangle) \Big\}\nonumber
\end{eqnarray}

The perturbation Hamiltonian consists of only the intralayer hopping terms and
has a similar form as that of Eq.(\ref{pert_hamil_vcbs}). For the 
VCBS $\rho = 1/6$ phase, the $|w_0 \rangle$ state is present at sublattice A.
The first order correction to the wavefunction at sublattice A is then given by
\begin{align*}
  |\chi^{A}\rangle = |w_0\rangle^{A} + \sum_{\Gamma\neq w_0} 
	  \frac{\langle \Gamma|\hat{T}^{A}|w_0\rangle^{A}}{(E^{A}_{w_0}
	   -E^{A}_{\Gamma})} |\Gamma \rangle,
 \label{perturb_gs_tri_w0}	
\end{align*}
where the state $|\Gamma \rangle$ in the summation is chosen from the set 
of the eigenstates stated previously. Then, with the substitution of the 
perturbing Hamiltonian, and simplifying steps, we obtain the expression for the 
perturbed state as 
\begin{equation}
  |\chi^{A}\rangle = |w_0\rangle^{A} 
	- 4J\varphi^B\left(\frac{\sqrt{3}}{-\mu-2J'}|000\rangle + 
	  \frac{2}{\mu-V'}|w_1\rangle\right).
 \label{perturbed_gs_tri_w0}	
\end{equation}
The $\varphi_A = \langle \chi^A | \hat{b}^A | \chi^A \rangle$ is then given as
\begin{equation}
 \varphi^A = -4J\phi^B\left(\frac{1}{-\mu-2J'} + \frac{4}{3(\mu-4V')}\right).
\end{equation}

Since the state on the sublattice B is $|000\rangle$ for VCBS $\rho = 1/6$ 
phase, we can use the expression given in Eq.(\ref{subB_sfop_tri}). Then
solving for $\varphi^A$ and $\varphi^B$ simultaneously and taking the limit
$\{\varphi^{A}, \varphi^{B}\}\rightarrow0^{+}$, we get Eq.(\ref{pb_vcbs_1by6}).
A similar analysis can be performed for obtaining the phase boundary of the 
VCBS $\rho = 2/6$ and the compressible phase. The particle-hole symmetry can 
be exploited to obtain the phase boundaries of the VCBS $\rho = 4/6$ and 
$\rho=5/6$ states. We can substitute $\mu \rightarrow -\mu + zV + 2V^{\prime}$
in the phase boundaries of VCBS $\rho=1/6$ and VCBS $\rho=2/6$ states, to get
the phase boundaries of VCBS $\rho = 5/6$ and VCBS $\rho=4/6$ states.

 \bibliography{multlayer}{}

\begin{thebibliography}{81}%
\makeatletter
\providecommand \@ifxundefined [1]{%
 \@ifx{#1\undefined}
}%
\providecommand \@ifnum [1]{%
 \ifnum #1\expandafter \@firstoftwo
 \else \expandafter \@secondoftwo
 \fi
}%
\providecommand \@ifx [1]{%
 \ifx #1\expandafter \@firstoftwo
 \else \expandafter \@secondoftwo
 \fi
}%
\providecommand \natexlab [1]{#1}%
\providecommand \enquote  [1]{``#1''}%
\providecommand \bibnamefont  [1]{#1}%
\providecommand \bibfnamefont [1]{#1}%
\providecommand \citenamefont [1]{#1}%
\providecommand \href@noop [0]{\@secondoftwo}%
\providecommand \href [0]{\begingroup \@sanitize@url \@href}%
\providecommand \@href[1]{\@@startlink{#1}\@@href}%
\providecommand \@@href[1]{\endgroup#1\@@endlink}%
\providecommand \@sanitize@url [0]{\catcode `\\12\catcode `\$12\catcode
  `\&12\catcode `\#12\catcode `\^12\catcode `\_12\catcode `\%12\relax}%
\providecommand \@@startlink[1]{}%
\providecommand \@@endlink[0]{}%
\providecommand \url  [0]{\begingroup\@sanitize@url \@url }%
\providecommand \@url [1]{\endgroup\@href {#1}{\urlprefix }}%
\providecommand \urlprefix  [0]{URL }%
\providecommand \Eprint [0]{\href }%
\providecommand \doibase [0]{http://dx.doi.org/}%
\providecommand \selectlanguage [0]{\@gobble}%
\providecommand \bibinfo  [0]{\@secondoftwo}%
\providecommand \bibfield  [0]{\@secondoftwo}%
\providecommand \translation [1]{[#1]}%
\providecommand \BibitemOpen [0]{}%
\providecommand \bibitemStop [0]{}%
\providecommand \bibitemNoStop [0]{.\EOS\space}%
\providecommand \EOS [0]{\spacefactor3000\relax}%
\providecommand \BibitemShut  [1]{\csname bibitem#1\endcsname}%
\let\auto@bib@innerbib\@empty
\bibitem [{\citenamefont {Greiner}\ \emph
  {et~al.}(2002{\natexlab{a}})\citenamefont {Greiner}, \citenamefont {Mandel},
  \citenamefont {Esslinger}, \citenamefont {H\"ansch},\ and\ \citenamefont
  {Bloch}}]{greiner_2002_1}%
  \BibitemOpen
  \bibfield  {author} {\bibinfo {author} {\bibfnamefont {M.}~\bibnamefont
  {Greiner}}, \bibinfo {author} {\bibfnamefont {O.}~\bibnamefont {Mandel}},
  \bibinfo {author} {\bibfnamefont {T.}~\bibnamefont {Esslinger}}, \bibinfo
  {author} {\bibfnamefont {T.~W.}\ \bibnamefont {H\"ansch}}, \ and\ \bibinfo
  {author} {\bibfnamefont {I.}~\bibnamefont {Bloch}},\ }\href {\doibase
  10.1038/415039a} {\bibfield  {journal} {\bibinfo  {journal} {Nature
  (London)}\ }\textbf {\bibinfo {volume} {415}},\ \bibinfo {pages} {39}
  (\bibinfo {year} {2002}{\natexlab{a}})}\BibitemShut {NoStop}%
\bibitem [{\citenamefont {Greiner}\ \emph
  {et~al.}(2002{\natexlab{b}})\citenamefont {Greiner}, \citenamefont {Mandel},
  \citenamefont {H\"ansch},\ and\ \citenamefont {Bloch}}]{greiner_2002_2}%
  \BibitemOpen
  \bibfield  {author} {\bibinfo {author} {\bibfnamefont {M.}~\bibnamefont
  {Greiner}}, \bibinfo {author} {\bibfnamefont {O.}~\bibnamefont {Mandel}},
  \bibinfo {author} {\bibfnamefont {T.~W.}\ \bibnamefont {H\"ansch}}, \ and\
  \bibinfo {author} {\bibfnamefont {I.}~\bibnamefont {Bloch}},\ }\href
  {\doibase 10.1038/nature00968} {\bibfield  {journal} {\bibinfo  {journal}
  {Nature}\ }\textbf {\bibinfo {volume} {419}},\ \bibinfo {pages} {51}
  (\bibinfo {year} {2002}{\natexlab{b}})}\BibitemShut {NoStop}%
\bibitem [{\citenamefont {St\"oferle}\ \emph {et~al.}(2004)\citenamefont
  {St\"oferle}, \citenamefont {Moritz}, \citenamefont {Schori}, \citenamefont
  {K\"ohl},\ and\ \citenamefont {Esslinger}}]{stoferle_2004}%
  \BibitemOpen
  \bibfield  {author} {\bibinfo {author} {\bibfnamefont {T.}~\bibnamefont
  {St\"oferle}}, \bibinfo {author} {\bibfnamefont {H.}~\bibnamefont {Moritz}},
  \bibinfo {author} {\bibfnamefont {C.}~\bibnamefont {Schori}}, \bibinfo
  {author} {\bibfnamefont {M.}~\bibnamefont {K\"ohl}}, \ and\ \bibinfo {author}
  {\bibfnamefont {T.}~\bibnamefont {Esslinger}},\ }\href {\doibase
  10.1103/PhysRevLett.92.130403} {\bibfield  {journal} {\bibinfo  {journal}
  {Phys. Rev. Lett.}\ }\textbf {\bibinfo {volume} {92}},\ \bibinfo {pages}
  {130403} (\bibinfo {year} {2004})}\BibitemShut {NoStop}%
\bibitem [{\citenamefont {F\"olling}\ \emph {et~al.}(2006)\citenamefont
  {F\"olling}, \citenamefont {Widera}, \citenamefont {M\"uller}, \citenamefont
  {Gerbier},\ and\ \citenamefont {Bloch}}]{folling_2006}%
  \BibitemOpen
  \bibfield  {author} {\bibinfo {author} {\bibfnamefont {S.}~\bibnamefont
  {F\"olling}}, \bibinfo {author} {\bibfnamefont {A.}~\bibnamefont {Widera}},
  \bibinfo {author} {\bibfnamefont {T.}~\bibnamefont {M\"uller}}, \bibinfo
  {author} {\bibfnamefont {F.}~\bibnamefont {Gerbier}}, \ and\ \bibinfo
  {author} {\bibfnamefont {I.}~\bibnamefont {Bloch}},\ }\href {\doibase
  10.1103/PhysRevLett.97.060403} {\bibfield  {journal} {\bibinfo  {journal}
  {Phys. Rev. Lett.}\ }\textbf {\bibinfo {volume} {97}},\ \bibinfo {pages}
  {060403} (\bibinfo {year} {2006})}\BibitemShut {NoStop}%
\bibitem [{\citenamefont {Spielman}\ \emph {et~al.}(2007)\citenamefont
  {Spielman}, \citenamefont {Phillips},\ and\ \citenamefont
  {Porto}}]{spielman_2007}%
  \BibitemOpen
  \bibfield  {author} {\bibinfo {author} {\bibfnamefont {I.~B.}\ \bibnamefont
  {Spielman}}, \bibinfo {author} {\bibfnamefont {W.~D.}\ \bibnamefont
  {Phillips}}, \ and\ \bibinfo {author} {\bibfnamefont {J.~V.}\ \bibnamefont
  {Porto}},\ }\href {\doibase 10.1103/PhysRevLett.98.080404} {\bibfield
  {journal} {\bibinfo  {journal} {Phys. Rev. Lett.}\ }\textbf {\bibinfo
  {volume} {98}},\ \bibinfo {pages} {080404} (\bibinfo {year}
  {2007})}\BibitemShut {NoStop}%
\bibitem [{\citenamefont {Baier}\ \emph {et~al.}(2016)\citenamefont {Baier},
  \citenamefont {Mark}, \citenamefont {Petter}, \citenamefont {Aikawa},
  \citenamefont {Chomaz}, \citenamefont {Cai}, \citenamefont {Baranov},
  \citenamefont {Zoller},\ and\ \citenamefont {Ferlaino}}]{baier_2016}%
  \BibitemOpen
  \bibfield  {author} {\bibinfo {author} {\bibfnamefont {S.}~\bibnamefont
  {Baier}}, \bibinfo {author} {\bibfnamefont {M.~J.}\ \bibnamefont {Mark}},
  \bibinfo {author} {\bibfnamefont {D.}~\bibnamefont {Petter}}, \bibinfo
  {author} {\bibfnamefont {K.}~\bibnamefont {Aikawa}}, \bibinfo {author}
  {\bibfnamefont {L.}~\bibnamefont {Chomaz}}, \bibinfo {author} {\bibfnamefont
  {Z.}~\bibnamefont {Cai}}, \bibinfo {author} {\bibfnamefont {M.}~\bibnamefont
  {Baranov}}, \bibinfo {author} {\bibfnamefont {P.}~\bibnamefont {Zoller}}, \
  and\ \bibinfo {author} {\bibfnamefont {F.}~\bibnamefont {Ferlaino}},\ }\href
  {\doibase 10.1126/science.aac9812} {\bibfield  {journal} {\bibinfo  {journal}
  {Science}\ }\textbf {\bibinfo {volume} {352}},\ \bibinfo {pages} {201}
  (\bibinfo {year} {2016})}\BibitemShut {NoStop}%
\bibitem [{\citenamefont {Jaksch}\ and\ \citenamefont
  {Zoller}(2005)}]{jaksch_2005}%
  \BibitemOpen
  \bibfield  {author} {\bibinfo {author} {\bibfnamefont {D.}~\bibnamefont
  {Jaksch}}\ and\ \bibinfo {author} {\bibfnamefont {P.}~\bibnamefont
  {Zoller}},\ }\href {\doibase https://doi.org/10.1016/j.aop.2004.09.010}
  {\bibfield  {journal} {\bibinfo  {journal} {Annals of Physics}\ }\textbf
  {\bibinfo {volume} {315}},\ \bibinfo {pages} {52} (\bibinfo {year} {2005})},\
  \bibinfo {note} {special Issue}\BibitemShut {NoStop}%
\bibitem [{\citenamefont {Bloch}\ \emph {et~al.}(2008)\citenamefont {Bloch},
  \citenamefont {Dalibard},\ and\ \citenamefont {Zwerger}}]{bloch_2008_1}%
  \BibitemOpen
  \bibfield  {author} {\bibinfo {author} {\bibfnamefont {I.}~\bibnamefont
  {Bloch}}, \bibinfo {author} {\bibfnamefont {J.}~\bibnamefont {Dalibard}}, \
  and\ \bibinfo {author} {\bibfnamefont {W.}~\bibnamefont {Zwerger}},\ }\href
  {\doibase 10.1103/RevModPhys.80.885} {\bibfield  {journal} {\bibinfo
  {journal} {Rev. Mod. Phys.}\ }\textbf {\bibinfo {volume} {80}},\ \bibinfo
  {pages} {885} (\bibinfo {year} {2008})}\BibitemShut {NoStop}%
\bibitem [{\citenamefont {Bloch}(2008)}]{bloch_2008_2}%
  \BibitemOpen
  \bibfield  {author} {\bibinfo {author} {\bibfnamefont {I.}~\bibnamefont
  {Bloch}},\ }\href {\doibase 10.1038/nature07126} {\bibfield  {journal}
  {\bibinfo  {journal} {Nature}\ }\textbf {\bibinfo {volume} {453}},\ \bibinfo
  {pages} {1016} (\bibinfo {year} {2008})}\BibitemShut {NoStop}%
\bibitem [{\citenamefont {Lewenstein}\ \emph {et~al.}(2007)\citenamefont
  {Lewenstein}, \citenamefont {Sanpera}, \citenamefont {Ahufinger},
  \citenamefont {Damski}, \citenamefont {Sen(De)},\ and\ \citenamefont
  {Sen}}]{lewenstein_2007}%
  \BibitemOpen
  \bibfield  {author} {\bibinfo {author} {\bibfnamefont {M.}~\bibnamefont
  {Lewenstein}}, \bibinfo {author} {\bibfnamefont {A.}~\bibnamefont {Sanpera}},
  \bibinfo {author} {\bibfnamefont {V.}~\bibnamefont {Ahufinger}}, \bibinfo
  {author} {\bibfnamefont {B.}~\bibnamefont {Damski}}, \bibinfo {author}
  {\bibfnamefont {A.}~\bibnamefont {Sen(De)}}, \ and\ \bibinfo {author}
  {\bibfnamefont {U.}~\bibnamefont {Sen}},\ }\href {\doibase
  10.1080/00018730701223200} {\bibfield  {journal} {\bibinfo  {journal} {Adv.
  Phys.}\ }\textbf {\bibinfo {volume} {56}},\ \bibinfo {pages} {243} (\bibinfo
  {year} {2007})}\BibitemShut {NoStop}%
\bibitem [{\citenamefont {Gross}\ and\ \citenamefont
  {Bloch}(2017)}]{gross_2017}%
  \BibitemOpen
  \bibfield  {author} {\bibinfo {author} {\bibfnamefont {C.}~\bibnamefont
  {Gross}}\ and\ \bibinfo {author} {\bibfnamefont {I.}~\bibnamefont {Bloch}},\
  }\href {\doibase 10.1126/science.aal3837} {\bibfield  {journal} {\bibinfo
  {journal} {Science}\ }\textbf {\bibinfo {volume} {357}},\ \bibinfo {pages}
  {995} (\bibinfo {year} {2017})},\ \Eprint
  {http://arxiv.org/abs/https://science.sciencemag.org/content/357/6355/995.full.pdf}
  {https://science.sciencemag.org/content/357/6355/995.full.pdf} \BibitemShut
  {NoStop}%
\bibitem [{\citenamefont {Jaksch}\ \emph {et~al.}(1998)\citenamefont {Jaksch},
  \citenamefont {Bruder}, \citenamefont {Cirac}, \citenamefont {Gardiner},\
  and\ \citenamefont {Zoller}}]{jaksch_1998}%
  \BibitemOpen
  \bibfield  {author} {\bibinfo {author} {\bibfnamefont {D.}~\bibnamefont
  {Jaksch}}, \bibinfo {author} {\bibfnamefont {C.}~\bibnamefont {Bruder}},
  \bibinfo {author} {\bibfnamefont {J.~I.}\ \bibnamefont {Cirac}}, \bibinfo
  {author} {\bibfnamefont {C.~W.}\ \bibnamefont {Gardiner}}, \ and\ \bibinfo
  {author} {\bibfnamefont {P.}~\bibnamefont {Zoller}},\ }\href {\doibase
  10.1103/PhysRevLett.81.3108} {\bibfield  {journal} {\bibinfo  {journal}
  {Phys. Rev. Lett.}\ }\textbf {\bibinfo {volume} {81}},\ \bibinfo {pages}
  {3108} (\bibinfo {year} {1998})}\BibitemShut {NoStop}%
\bibitem [{\citenamefont {G\'{o}ral}\ \emph {et~al.}(2002)\citenamefont
  {G\'{o}ral}, \citenamefont {Santos},\ and\ \citenamefont
  {Lewenstein}}]{goral_2002}%
  \BibitemOpen
  \bibfield  {author} {\bibinfo {author} {\bibfnamefont {K.}~\bibnamefont
  {G\'{o}ral}}, \bibinfo {author} {\bibfnamefont {L.}~\bibnamefont {Santos}}, \
  and\ \bibinfo {author} {\bibfnamefont {M.}~\bibnamefont {Lewenstein}},\
  }\href {\doibase 10.1103/PhysRevLett.88.170406} {\bibfield  {journal}
  {\bibinfo  {journal} {Phys. Rev. Lett.}\ }\textbf {\bibinfo {volume} {88}},\
  \bibinfo {pages} {170406} (\bibinfo {year} {2002})}\BibitemShut {NoStop}%
\bibitem [{\citenamefont {Danshita}\ and\ \citenamefont {S\'{a}~de
  Melo}(2009)}]{danshita_2009}%
  \BibitemOpen
  \bibfield  {author} {\bibinfo {author} {\bibfnamefont {I.}~\bibnamefont
  {Danshita}}\ and\ \bibinfo {author} {\bibfnamefont {C.~A.~R.}\ \bibnamefont
  {S\'{a}~de Melo}},\ }\href {\doibase 10.1103/PhysRevLett.103.225301}
  {\bibfield  {journal} {\bibinfo  {journal} {Phys. Rev. Lett.}\ }\textbf
  {\bibinfo {volume} {103}},\ \bibinfo {pages} {225301} (\bibinfo {year}
  {2009})}\BibitemShut {NoStop}%
\bibitem [{\citenamefont {Yi}\ \emph {et~al.}(2007)\citenamefont {Yi},
  \citenamefont {Li},\ and\ \citenamefont {Sun}}]{yi_2007}%
  \BibitemOpen
  \bibfield  {author} {\bibinfo {author} {\bibfnamefont {S.}~\bibnamefont
  {Yi}}, \bibinfo {author} {\bibfnamefont {T.}~\bibnamefont {Li}}, \ and\
  \bibinfo {author} {\bibfnamefont {C.~P.}\ \bibnamefont {Sun}},\ }\href
  {\doibase 10.1103/PhysRevLett.98.260405} {\bibfield  {journal} {\bibinfo
  {journal} {Phys. Rev. Lett.}\ }\textbf {\bibinfo {volume} {98}},\ \bibinfo
  {pages} {260405} (\bibinfo {year} {2007})}\BibitemShut {NoStop}%
\bibitem [{\citenamefont {Zhang}\ \emph {et~al.}(2015)\citenamefont {Zhang},
  \citenamefont {Safavi-Naini}, \citenamefont {Rey},\ and\ \citenamefont
  {Capogrosso-Sansone}}]{zhang_2015}%
  \BibitemOpen
  \bibfield  {author} {\bibinfo {author} {\bibfnamefont {C.}~\bibnamefont
  {Zhang}}, \bibinfo {author} {\bibfnamefont {A.}~\bibnamefont {Safavi-Naini}},
  \bibinfo {author} {\bibfnamefont {A.~M.}\ \bibnamefont {Rey}}, \ and\
  \bibinfo {author} {\bibfnamefont {B.}~\bibnamefont {Capogrosso-Sansone}},\
  }\href {http://stacks.iop.org/1367-2630/17/i=12/a=123014} {\bibfield
  {journal} {\bibinfo  {journal} {New Journal of Physics}\ }\textbf {\bibinfo
  {volume} {17}},\ \bibinfo {pages} {123014} (\bibinfo {year}
  {2015})}\BibitemShut {NoStop}%
\bibitem [{\citenamefont {Bandyopadhyay}\ \emph {et~al.}(2019)\citenamefont
  {Bandyopadhyay}, \citenamefont {Bai}, \citenamefont {Pal}, \citenamefont
  {Suthar}, \citenamefont {Nath},\ and\ \citenamefont
  {Angom}}]{bandyopadhyay_2019}%
  \BibitemOpen
  \bibfield  {author} {\bibinfo {author} {\bibfnamefont {S.}~\bibnamefont
  {Bandyopadhyay}}, \bibinfo {author} {\bibfnamefont {R.}~\bibnamefont {Bai}},
  \bibinfo {author} {\bibfnamefont {S.}~\bibnamefont {Pal}}, \bibinfo {author}
  {\bibfnamefont {K.}~\bibnamefont {Suthar}}, \bibinfo {author} {\bibfnamefont
  {R.}~\bibnamefont {Nath}}, \ and\ \bibinfo {author} {\bibfnamefont
  {D.}~\bibnamefont {Angom}},\ }\href {\doibase 10.1103/PhysRevA.100.053623}
  {\bibfield  {journal} {\bibinfo  {journal} {Phys. Rev. A}\ }\textbf {\bibinfo
  {volume} {100}},\ \bibinfo {pages} {053623} (\bibinfo {year}
  {2019})}\BibitemShut {NoStop}%
\bibitem [{\citenamefont {Pal}\ \emph {et~al.}(2019)\citenamefont {Pal},
  \citenamefont {Bai}, \citenamefont {Bandyopadhyay}, \citenamefont {Suthar},\
  and\ \citenamefont {Angom}}]{pal_2019}%
  \BibitemOpen
  \bibfield  {author} {\bibinfo {author} {\bibfnamefont {S.}~\bibnamefont
  {Pal}}, \bibinfo {author} {\bibfnamefont {R.}~\bibnamefont {Bai}}, \bibinfo
  {author} {\bibfnamefont {S.}~\bibnamefont {Bandyopadhyay}}, \bibinfo {author}
  {\bibfnamefont {K.}~\bibnamefont {Suthar}}, \ and\ \bibinfo {author}
  {\bibfnamefont {D.}~\bibnamefont {Angom}},\ }\href {\doibase
  10.1103/PhysRevA.99.053610} {\bibfield  {journal} {\bibinfo  {journal} {Phys.
  Rev. A}\ }\textbf {\bibinfo {volume} {99}},\ \bibinfo {pages} {053610}
  (\bibinfo {year} {2019})}\BibitemShut {NoStop}%
\bibitem [{\citenamefont {Suthar}\ \emph
  {et~al.}(2020{\natexlab{a}})\citenamefont {Suthar}, \citenamefont {Sable},
  \citenamefont {Bai}, \citenamefont {Bandyopadhyay}, \citenamefont {Pal},\
  and\ \citenamefont {Angom}}]{suthar_2020}%
  \BibitemOpen
  \bibfield  {author} {\bibinfo {author} {\bibfnamefont {K.}~\bibnamefont
  {Suthar}}, \bibinfo {author} {\bibfnamefont {H.}~\bibnamefont {Sable}},
  \bibinfo {author} {\bibfnamefont {R.}~\bibnamefont {Bai}}, \bibinfo {author}
  {\bibfnamefont {S.}~\bibnamefont {Bandyopadhyay}}, \bibinfo {author}
  {\bibfnamefont {S.}~\bibnamefont {Pal}}, \ and\ \bibinfo {author}
  {\bibfnamefont {D.}~\bibnamefont {Angom}},\ }\href {\doibase
  10.1103/PhysRevA.102.013320} {\bibfield  {journal} {\bibinfo  {journal}
  {Phys. Rev. A}\ }\textbf {\bibinfo {volume} {102}},\ \bibinfo {pages}
  {013320} (\bibinfo {year} {2020}{\natexlab{a}})}\BibitemShut {NoStop}%
\bibitem [{\citenamefont {Suthar}\ \emph
  {et~al.}(2020{\natexlab{b}})\citenamefont {Suthar}, \citenamefont {Kraus},
  \citenamefont {Sable}, \citenamefont {Angom}, \citenamefont {Morigi},\ and\
  \citenamefont {Zakrzewski}}]{suthar_2020_2}%
  \BibitemOpen
  \bibfield  {author} {\bibinfo {author} {\bibfnamefont {K.}~\bibnamefont
  {Suthar}}, \bibinfo {author} {\bibfnamefont {R.}~\bibnamefont {Kraus}},
  \bibinfo {author} {\bibfnamefont {H.}~\bibnamefont {Sable}}, \bibinfo
  {author} {\bibfnamefont {D.}~\bibnamefont {Angom}}, \bibinfo {author}
  {\bibfnamefont {G.}~\bibnamefont {Morigi}}, \ and\ \bibinfo {author}
  {\bibfnamefont {J.}~\bibnamefont {Zakrzewski}},\ }\href {\doibase
  10.1103/PhysRevB.102.214503} {\bibfield  {journal} {\bibinfo  {journal}
  {Phys. Rev. B}\ }\textbf {\bibinfo {volume} {102}},\ \bibinfo {pages}
  {214503} (\bibinfo {year} {2020}{\natexlab{b}})}\BibitemShut {NoStop}%
\bibitem [{\citenamefont {Suthar}\ \emph {et~al.}(2015)\citenamefont {Suthar},
  \citenamefont {Roy},\ and\ \citenamefont {Angom}}]{suthar_2015}%
  \BibitemOpen
  \bibfield  {author} {\bibinfo {author} {\bibfnamefont {K.}~\bibnamefont
  {Suthar}}, \bibinfo {author} {\bibfnamefont {A.}~\bibnamefont {Roy}}, \ and\
  \bibinfo {author} {\bibfnamefont {D.}~\bibnamefont {Angom}},\ }\href
  {\doibase 10.1103/PhysRevA.91.043615} {\bibfield  {journal} {\bibinfo
  {journal} {Phys. Rev. A}\ }\textbf {\bibinfo {volume} {91}},\ \bibinfo
  {pages} {043615} (\bibinfo {year} {2015})}\BibitemShut {NoStop}%
\bibitem [{\citenamefont {Suthar}\ and\ \citenamefont
  {Angom}(2016)}]{suthar_2016}%
  \BibitemOpen
  \bibfield  {author} {\bibinfo {author} {\bibfnamefont {K.}~\bibnamefont
  {Suthar}}\ and\ \bibinfo {author} {\bibfnamefont {D.}~\bibnamefont {Angom}},\
  }\href {\doibase 10.1103/PhysRevA.93.063608} {\bibfield  {journal} {\bibinfo
  {journal} {Phys. Rev. A}\ }\textbf {\bibinfo {volume} {93}},\ \bibinfo
  {pages} {063608} (\bibinfo {year} {2016})}\BibitemShut {NoStop}%
\bibitem [{\citenamefont {Suthar}\ and\ \citenamefont
  {Angom}(2017)}]{suthar_2017}%
  \BibitemOpen
  \bibfield  {author} {\bibinfo {author} {\bibfnamefont {K.}~\bibnamefont
  {Suthar}}\ and\ \bibinfo {author} {\bibfnamefont {D.}~\bibnamefont {Angom}},\
  }\href {\doibase 10.1103/PhysRevA.95.043602} {\bibfield  {journal} {\bibinfo
  {journal} {Phys. Rev. A}\ }\textbf {\bibinfo {volume} {95}},\ \bibinfo
  {pages} {043602} (\bibinfo {year} {2017})}\BibitemShut {NoStop}%
\bibitem [{\citenamefont {Krutitsky}\ \emph {et~al.}(2010)\citenamefont
  {Krutitsky}, \citenamefont {Larson},\ and\ \citenamefont
  {Lewenstein}}]{krutitsky_2010}%
  \BibitemOpen
  \bibfield  {author} {\bibinfo {author} {\bibfnamefont {K.~V.}\ \bibnamefont
  {Krutitsky}}, \bibinfo {author} {\bibfnamefont {J.}~\bibnamefont {Larson}}, \
  and\ \bibinfo {author} {\bibfnamefont {M.}~\bibnamefont {Lewenstein}},\
  }\href {\doibase 10.1103/PhysRevA.82.033618} {\bibfield  {journal} {\bibinfo
  {journal} {Phys. Rev. A}\ }\textbf {\bibinfo {volume} {82}},\ \bibinfo
  {pages} {033618} (\bibinfo {year} {2010})}\BibitemShut {NoStop}%
\bibitem [{\citenamefont {Krutitsky}\ and\ \citenamefont
  {Navez}(2011)}]{krutitsky_2011}%
  \BibitemOpen
  \bibfield  {author} {\bibinfo {author} {\bibfnamefont {K.~V.}\ \bibnamefont
  {Krutitsky}}\ and\ \bibinfo {author} {\bibfnamefont {P.}~\bibnamefont
  {Navez}},\ }\href {\doibase 10.1103/PhysRevA.84.033602} {\bibfield  {journal}
  {\bibinfo  {journal} {Phys. Rev. A}\ }\textbf {\bibinfo {volume} {84}},\
  \bibinfo {pages} {033602} (\bibinfo {year} {2011})}\BibitemShut {NoStop}%
\bibitem [{\citenamefont {Saito}\ \emph {et~al.}(2012)\citenamefont {Saito},
  \citenamefont {Danshita}, \citenamefont {Ozaki},\ and\ \citenamefont
  {Nikuni}}]{saito_2012}%
  \BibitemOpen
  \bibfield  {author} {\bibinfo {author} {\bibfnamefont {T.}~\bibnamefont
  {Saito}}, \bibinfo {author} {\bibfnamefont {I.}~\bibnamefont {Danshita}},
  \bibinfo {author} {\bibfnamefont {T.}~\bibnamefont {Ozaki}}, \ and\ \bibinfo
  {author} {\bibfnamefont {T.}~\bibnamefont {Nikuni}},\ }\href {\doibase
  10.1103/PhysRevA.86.023623} {\bibfield  {journal} {\bibinfo  {journal} {Phys.
  Rev. A}\ }\textbf {\bibinfo {volume} {86}},\ \bibinfo {pages} {023623}
  (\bibinfo {year} {2012})}\BibitemShut {NoStop}%
\bibitem [{\citenamefont {Chen}\ \emph {et~al.}(2011)\citenamefont {Chen},
  \citenamefont {White}, \citenamefont {Borries},\ and\ \citenamefont
  {DeMarco}}]{chen_2011}%
  \BibitemOpen
  \bibfield  {author} {\bibinfo {author} {\bibfnamefont {D.}~\bibnamefont
  {Chen}}, \bibinfo {author} {\bibfnamefont {M.}~\bibnamefont {White}},
  \bibinfo {author} {\bibfnamefont {C.}~\bibnamefont {Borries}}, \ and\
  \bibinfo {author} {\bibfnamefont {B.}~\bibnamefont {DeMarco}},\ }\href
  {\doibase 10.1103/PhysRevLett.106.235304} {\bibfield  {journal} {\bibinfo
  {journal} {Phys. Rev. Lett.}\ }\textbf {\bibinfo {volume} {106}},\ \bibinfo
  {pages} {235304} (\bibinfo {year} {2011})}\BibitemShut {NoStop}%
\bibitem [{\citenamefont {Braun}\ \emph {et~al.}(2015)\citenamefont {Braun},
  \citenamefont {Friesdorf}, \citenamefont {Hodgman}, \citenamefont
  {Schreiber}, \citenamefont {Ronzheimer}, \citenamefont {Riera}, \citenamefont
  {del Rey}, \citenamefont {Bloch}, \citenamefont {Eisert},\ and\ \citenamefont
  {Schneider}}]{braun_2015}%
  \BibitemOpen
  \bibfield  {author} {\bibinfo {author} {\bibfnamefont {S.}~\bibnamefont
  {Braun}}, \bibinfo {author} {\bibfnamefont {M.}~\bibnamefont {Friesdorf}},
  \bibinfo {author} {\bibfnamefont {S.~S.}\ \bibnamefont {Hodgman}}, \bibinfo
  {author} {\bibfnamefont {M.}~\bibnamefont {Schreiber}}, \bibinfo {author}
  {\bibfnamefont {J.~P.}\ \bibnamefont {Ronzheimer}}, \bibinfo {author}
  {\bibfnamefont {A.}~\bibnamefont {Riera}}, \bibinfo {author} {\bibfnamefont
  {M.}~\bibnamefont {del Rey}}, \bibinfo {author} {\bibfnamefont
  {I.}~\bibnamefont {Bloch}}, \bibinfo {author} {\bibfnamefont
  {J.}~\bibnamefont {Eisert}}, \ and\ \bibinfo {author} {\bibfnamefont
  {U.}~\bibnamefont {Schneider}},\ }\href {\doibase 10.1073/pnas.1408861112}
  {\bibfield  {journal} {\bibinfo  {journal} {Proceedings of the National
  Academy of Sciences}\ }\textbf {\bibinfo {volume} {112}},\ \bibinfo {pages}
  {3641} (\bibinfo {year} {2015})}\BibitemShut {NoStop}%
\bibitem [{\citenamefont {Shimizu}\ \emph
  {et~al.}(2018{\natexlab{a}})\citenamefont {Shimizu}, \citenamefont {Kuno},
  \citenamefont {Hirano},\ and\ \citenamefont {Ichinose}}]{shimizu_misf_2018}%
  \BibitemOpen
  \bibfield  {author} {\bibinfo {author} {\bibfnamefont {K.}~\bibnamefont
  {Shimizu}}, \bibinfo {author} {\bibfnamefont {Y.}~\bibnamefont {Kuno}},
  \bibinfo {author} {\bibfnamefont {T.}~\bibnamefont {Hirano}}, \ and\ \bibinfo
  {author} {\bibfnamefont {I.}~\bibnamefont {Ichinose}},\ }\href {\doibase
  10.1103/PhysRevA.97.033626} {\bibfield  {journal} {\bibinfo  {journal} {Phys.
  Rev. A}\ }\textbf {\bibinfo {volume} {97}},\ \bibinfo {pages} {033626}
  (\bibinfo {year} {2018}{\natexlab{a}})}\BibitemShut {NoStop}%
\bibitem [{\citenamefont {Shimizu}\ \emph
  {et~al.}(2018{\natexlab{b}})\citenamefont {Shimizu}, \citenamefont {Hirano},
  \citenamefont {Park}, \citenamefont {Kuno},\ and\ \citenamefont
  {Ichinose}}]{shimizu_dwss_18}%
  \BibitemOpen
  \bibfield  {author} {\bibinfo {author} {\bibfnamefont {K.}~\bibnamefont
  {Shimizu}}, \bibinfo {author} {\bibfnamefont {T.}~\bibnamefont {Hirano}},
  \bibinfo {author} {\bibfnamefont {J.}~\bibnamefont {Park}}, \bibinfo {author}
  {\bibfnamefont {Y.}~\bibnamefont {Kuno}}, \ and\ \bibinfo {author}
  {\bibfnamefont {I.}~\bibnamefont {Ichinose}},\ }\href {\doibase
  10.1103/PhysRevA.98.063603} {\bibfield  {journal} {\bibinfo  {journal} {Phys.
  Rev. A}\ }\textbf {\bibinfo {volume} {98}},\ \bibinfo {pages} {063603}
  (\bibinfo {year} {2018}{\natexlab{b}})}\BibitemShut {NoStop}%
\bibitem [{\citenamefont {Shimizu}\ \emph
  {et~al.}(2018{\natexlab{c}})\citenamefont {Shimizu}, \citenamefont {Hirano},
  \citenamefont {Park}, \citenamefont {Kuno},\ and\ \citenamefont
  {Ichinose}}]{shimizu_dwsf_18}%
  \BibitemOpen
  \bibfield  {author} {\bibinfo {author} {\bibfnamefont {K.}~\bibnamefont
  {Shimizu}}, \bibinfo {author} {\bibfnamefont {T.}~\bibnamefont {Hirano}},
  \bibinfo {author} {\bibfnamefont {J.}~\bibnamefont {Park}}, \bibinfo {author}
  {\bibfnamefont {Y.}~\bibnamefont {Kuno}}, \ and\ \bibinfo {author}
  {\bibfnamefont {I.}~\bibnamefont {Ichinose}},\ }\href {\doibase
  10.1088/1367-2630/aad5f9} {\bibfield  {journal} {\bibinfo  {journal} {New
  Journal of Physics}\ }\textbf {\bibinfo {volume} {20}},\ \bibinfo {pages}
  {083006} (\bibinfo {year} {2018}{\natexlab{c}})}\BibitemShut {NoStop}%
\bibitem [{\citenamefont {Zhou}\ \emph {et~al.}(2020)\citenamefont {Zhou},
  \citenamefont {Li}, \citenamefont {Nath},\ and\ \citenamefont
  {Li}}]{zhou_2020}%
  \BibitemOpen
  \bibfield  {author} {\bibinfo {author} {\bibfnamefont {Y.}~\bibnamefont
  {Zhou}}, \bibinfo {author} {\bibfnamefont {Y.}~\bibnamefont {Li}}, \bibinfo
  {author} {\bibfnamefont {R.}~\bibnamefont {Nath}}, \ and\ \bibinfo {author}
  {\bibfnamefont {W.}~\bibnamefont {Li}},\ }\href {\doibase
  10.1103/PhysRevA.101.013427} {\bibfield  {journal} {\bibinfo  {journal}
  {Phys. Rev. A}\ }\textbf {\bibinfo {volume} {101}},\ \bibinfo {pages}
  {013427} (\bibinfo {year} {2020})}\BibitemShut {NoStop}%
\bibitem [{\citenamefont {Sable}\ \emph {et~al.}(2021)\citenamefont {Sable},
  \citenamefont {Gaur}, \citenamefont {Bandyopadhyay}, \citenamefont {Nath},\
  and\ \citenamefont {Angom}}]{sable_21}%
  \BibitemOpen
  \bibfield  {author} {\bibinfo {author} {\bibfnamefont {H.}~\bibnamefont
  {Sable}}, \bibinfo {author} {\bibfnamefont {D.}~\bibnamefont {Gaur}},
  \bibinfo {author} {\bibfnamefont {S.}~\bibnamefont {Bandyopadhyay}}, \bibinfo
  {author} {\bibfnamefont {R.}~\bibnamefont {Nath}}, \ and\ \bibinfo {author}
  {\bibfnamefont {D.}~\bibnamefont {Angom}},\ }\href
  {https://arxiv.org/abs/2106.01725} {\bibfield  {journal} {\bibinfo  {journal}
  {arXiv:2106.01725}\ } (\bibinfo {year} {2021})}\BibitemShut {NoStop}%
\bibitem [{\citenamefont {Kaufman}\ \emph {et~al.}(2016)\citenamefont
  {Kaufman}, \citenamefont {Tai}, \citenamefont {Lukin}, \citenamefont
  {Rispoli}, \citenamefont {Schittko}, \citenamefont {Preiss},\ and\
  \citenamefont {Greiner}}]{kaufman_2016}%
  \BibitemOpen
  \bibfield  {author} {\bibinfo {author} {\bibfnamefont {A.~M.}\ \bibnamefont
  {Kaufman}}, \bibinfo {author} {\bibfnamefont {M.~E.}\ \bibnamefont {Tai}},
  \bibinfo {author} {\bibfnamefont {A.}~\bibnamefont {Lukin}}, \bibinfo
  {author} {\bibfnamefont {M.}~\bibnamefont {Rispoli}}, \bibinfo {author}
  {\bibfnamefont {R.}~\bibnamefont {Schittko}}, \bibinfo {author}
  {\bibfnamefont {P.~M.}\ \bibnamefont {Preiss}}, \ and\ \bibinfo {author}
  {\bibfnamefont {M.}~\bibnamefont {Greiner}},\ }\href {\doibase
  10.1126/science.aaf6725} {\bibfield  {journal} {\bibinfo  {journal}
  {Science}\ }\textbf {\bibinfo {volume} {353}},\ \bibinfo {pages} {794}
  (\bibinfo {year} {2016})}\BibitemShut {NoStop}%
\bibitem [{\citenamefont {Bohrdt}\ \emph {et~al.}(2017)\citenamefont {Bohrdt},
  \citenamefont {Mendl}, \citenamefont {Endres},\ and\ \citenamefont
  {Knap}}]{bohrdt_2017}%
  \BibitemOpen
  \bibfield  {author} {\bibinfo {author} {\bibfnamefont {A.}~\bibnamefont
  {Bohrdt}}, \bibinfo {author} {\bibfnamefont {C.~B.}\ \bibnamefont {Mendl}},
  \bibinfo {author} {\bibfnamefont {M.}~\bibnamefont {Endres}}, \ and\ \bibinfo
  {author} {\bibfnamefont {M.}~\bibnamefont {Knap}},\ }\href {\doibase
  10.1088/1367-2630/aa719b} {\bibfield  {journal} {\bibinfo  {journal} {New
  Journal of Physics}\ }\textbf {\bibinfo {volume} {19}},\ \bibinfo {pages}
  {063001} (\bibinfo {year} {2017})}\BibitemShut {NoStop}%
\bibitem [{\citenamefont {Sierant}\ \emph {et~al.}(2017)\citenamefont
  {Sierant}, \citenamefont {Delande},\ and\ \citenamefont
  {Zakrzewski}}]{sierant_2017}%
  \BibitemOpen
  \bibfield  {author} {\bibinfo {author} {\bibfnamefont {P.}~\bibnamefont
  {Sierant}}, \bibinfo {author} {\bibfnamefont {D.}~\bibnamefont {Delande}}, \
  and\ \bibinfo {author} {\bibfnamefont {J.}~\bibnamefont {Zakrzewski}},\
  }\href {\doibase 10.1103/PhysRevA.95.021601} {\bibfield  {journal} {\bibinfo
  {journal} {Phys. Rev. A}\ }\textbf {\bibinfo {volume} {95}},\ \bibinfo
  {pages} {021601} (\bibinfo {year} {2017})}\BibitemShut {NoStop}%
\bibitem [{\citenamefont {Sierant}\ and\ \citenamefont
  {Zakrzewski}(2018)}]{sierant_2018}%
  \BibitemOpen
  \bibfield  {author} {\bibinfo {author} {\bibfnamefont {P.}~\bibnamefont
  {Sierant}}\ and\ \bibinfo {author} {\bibfnamefont {J.}~\bibnamefont
  {Zakrzewski}},\ }\href {\doibase 10.1088/1367-2630/aabb17} {\bibfield
  {journal} {\bibinfo  {journal} {New Journal of Physics}\ }\textbf {\bibinfo
  {volume} {20}},\ \bibinfo {pages} {043032} (\bibinfo {year}
  {2018})}\BibitemShut {NoStop}%
\bibitem [{\citenamefont {Tomita}\ \emph {et~al.}(2017)\citenamefont {Tomita},
  \citenamefont {Nakajima}, \citenamefont {Danshita}, \citenamefont {Takasu},\
  and\ \citenamefont {Takahashi}}]{tomita_2017}%
  \BibitemOpen
  \bibfield  {author} {\bibinfo {author} {\bibfnamefont {T.}~\bibnamefont
  {Tomita}}, \bibinfo {author} {\bibfnamefont {S.}~\bibnamefont {Nakajima}},
  \bibinfo {author} {\bibfnamefont {I.}~\bibnamefont {Danshita}}, \bibinfo
  {author} {\bibfnamefont {Y.}~\bibnamefont {Takasu}}, \ and\ \bibinfo {author}
  {\bibfnamefont {Y.}~\bibnamefont {Takahashi}},\ }\href {\doibase
  10.1126/sciadv.1701513} {\bibfield  {journal} {\bibinfo  {journal} {Science
  Advances}\ }\textbf {\bibinfo {volume} {3}} (\bibinfo {year} {2017}),\
  10.1126/sciadv.1701513}\BibitemShut {NoStop}%
\bibitem [{\citenamefont {Roy}\ and\ \citenamefont {Saha}(2019)}]{roy_2019}%
  \BibitemOpen
  \bibfield  {author} {\bibinfo {author} {\bibfnamefont {A.}~\bibnamefont
  {Roy}}\ and\ \bibinfo {author} {\bibfnamefont {K.}~\bibnamefont {Saha}},\
  }\href {\doibase 10.1088/1367-2630/ab4da0} {\bibfield  {journal} {\bibinfo
  {journal} {New Journal of Physics}\ }\textbf {\bibinfo {volume} {21}},\
  \bibinfo {pages} {103050} (\bibinfo {year} {2019})}\BibitemShut {NoStop}%
\bibitem [{\citenamefont {Fisher}\ \emph {et~al.}(1989)\citenamefont {Fisher},
  \citenamefont {Weichman}, \citenamefont {Grinstein},\ and\ \citenamefont
  {Fisher}}]{fisher_1989}%
  \BibitemOpen
  \bibfield  {author} {\bibinfo {author} {\bibfnamefont {M.~P.~A.}\
  \bibnamefont {Fisher}}, \bibinfo {author} {\bibfnamefont {P.~B.}\
  \bibnamefont {Weichman}}, \bibinfo {author} {\bibfnamefont {G.}~\bibnamefont
  {Grinstein}}, \ and\ \bibinfo {author} {\bibfnamefont {D.~S.}\ \bibnamefont
  {Fisher}},\ }\href {\doibase 10.1103/PhysRevB.40.546} {\bibfield  {journal}
  {\bibinfo  {journal} {Phys. Rev. B}\ }\textbf {\bibinfo {volume} {40}},\
  \bibinfo {pages} {546} (\bibinfo {year} {1989})}\BibitemShut {NoStop}%
\bibitem [{\citenamefont {Hubbard}(1963)}]{hubbard_1963}%
  \BibitemOpen
  \bibfield  {author} {\bibinfo {author} {\bibfnamefont {J.}~\bibnamefont
  {Hubbard}},\ }\href {\doibase 10.1098/rspa.1963.0204} {\bibfield  {journal}
  {\bibinfo  {journal} {Proc. Royal Soc. A}\ }\textbf {\bibinfo {volume}
  {276}},\ \bibinfo {pages} {238} (\bibinfo {year} {1963})}\BibitemShut
  {NoStop}%
\bibitem [{\citenamefont {Scarola}\ and\ \citenamefont
  {Das~Sarma}(2005)}]{scarola_2005}%
  \BibitemOpen
  \bibfield  {author} {\bibinfo {author} {\bibfnamefont {V.~W.}\ \bibnamefont
  {Scarola}}\ and\ \bibinfo {author} {\bibfnamefont {S.}~\bibnamefont
  {Das~Sarma}},\ }\href {\doibase 10.1103/PhysRevLett.95.033003} {\bibfield
  {journal} {\bibinfo  {journal} {Phys. Rev. Lett.}\ }\textbf {\bibinfo
  {volume} {95}},\ \bibinfo {pages} {033003} (\bibinfo {year}
  {2005})}\BibitemShut {NoStop}%
\bibitem [{\citenamefont {Kovrizhin}\ \emph {et~al.}(2005)\citenamefont
  {Kovrizhin}, \citenamefont {Pai},\ and\ \citenamefont
  {Sinha}}]{kovrizhin_2005}%
  \BibitemOpen
  \bibfield  {author} {\bibinfo {author} {\bibfnamefont {D.~L.}\ \bibnamefont
  {Kovrizhin}}, \bibinfo {author} {\bibfnamefont {G.~V.}\ \bibnamefont {Pai}},
  \ and\ \bibinfo {author} {\bibfnamefont {S.}~\bibnamefont {Sinha}},\ }\href
  {http://stacks.iop.org/0295-5075/72/i=2/a=162} {\bibfield  {journal}
  {\bibinfo  {journal} {EPL (Europhysics Letters)}\ }\textbf {\bibinfo {volume}
  {72}},\ \bibinfo {pages} {162} (\bibinfo {year} {2005})}\BibitemShut
  {NoStop}%
\bibitem [{\citenamefont {Sengupta}\ \emph {et~al.}(2005)\citenamefont
  {Sengupta}, \citenamefont {Pryadko}, \citenamefont {Alet}, \citenamefont
  {Troyer},\ and\ \citenamefont {Schmid}}]{sengupta_2005}%
  \BibitemOpen
  \bibfield  {author} {\bibinfo {author} {\bibfnamefont {P.}~\bibnamefont
  {Sengupta}}, \bibinfo {author} {\bibfnamefont {L.~P.}\ \bibnamefont
  {Pryadko}}, \bibinfo {author} {\bibfnamefont {F.}~\bibnamefont {Alet}},
  \bibinfo {author} {\bibfnamefont {M.}~\bibnamefont {Troyer}}, \ and\ \bibinfo
  {author} {\bibfnamefont {G.}~\bibnamefont {Schmid}},\ }\href {\doibase
  10.1103/PhysRevLett.94.207202} {\bibfield  {journal} {\bibinfo  {journal}
  {Phys. Rev. Lett.}\ }\textbf {\bibinfo {volume} {94}},\ \bibinfo {pages}
  {207202} (\bibinfo {year} {2005})}\BibitemShut {NoStop}%
\bibitem [{\citenamefont {Mazzarella}\ \emph {et~al.}(2006)\citenamefont
  {Mazzarella}, \citenamefont {Giampaolo},\ and\ \citenamefont
  {Illuminati}}]{mazzarella_2006}%
  \BibitemOpen
  \bibfield  {author} {\bibinfo {author} {\bibfnamefont {G.}~\bibnamefont
  {Mazzarella}}, \bibinfo {author} {\bibfnamefont {S.~M.}\ \bibnamefont
  {Giampaolo}}, \ and\ \bibinfo {author} {\bibfnamefont {F.}~\bibnamefont
  {Illuminati}},\ }\href {\doibase 10.1103/PhysRevA.73.013625} {\bibfield
  {journal} {\bibinfo  {journal} {Phys. Rev. A}\ }\textbf {\bibinfo {volume}
  {73}},\ \bibinfo {pages} {013625} (\bibinfo {year} {2006})}\BibitemShut
  {NoStop}%
\bibitem [{\citenamefont {Iskin}(2011)}]{iskin_2011}%
  \BibitemOpen
  \bibfield  {author} {\bibinfo {author} {\bibfnamefont {M.}~\bibnamefont
  {Iskin}},\ }\href {\doibase 10.1103/PhysRevA.83.051606} {\bibfield  {journal}
  {\bibinfo  {journal} {Phys. Rev. A}\ }\textbf {\bibinfo {volume} {83}},\
  \bibinfo {pages} {051606} (\bibinfo {year} {2011})}\BibitemShut {NoStop}%
\bibitem [{\citenamefont {Dutta}\ \emph {et~al.}(2015)\citenamefont {Dutta},
  \citenamefont {Gajda}, \citenamefont {Hauke}, \citenamefont {Lewenstein},
  \citenamefont {Lühmann}, \citenamefont {Malomed}, \citenamefont
  {Sowi{\'{n}}ski},\ and\ \citenamefont {Zakrzewski}}]{dutta_2015}%
  \BibitemOpen
  \bibfield  {author} {\bibinfo {author} {\bibfnamefont {O.}~\bibnamefont
  {Dutta}}, \bibinfo {author} {\bibfnamefont {M.}~\bibnamefont {Gajda}},
  \bibinfo {author} {\bibfnamefont {P.}~\bibnamefont {Hauke}}, \bibinfo
  {author} {\bibfnamefont {M.}~\bibnamefont {Lewenstein}}, \bibinfo {author}
  {\bibfnamefont {D.-S.}\ \bibnamefont {Lühmann}}, \bibinfo {author}
  {\bibfnamefont {B.~A.}\ \bibnamefont {Malomed}}, \bibinfo {author}
  {\bibfnamefont {T.}~\bibnamefont {Sowi{\'{n}}ski}}, \ and\ \bibinfo {author}
  {\bibfnamefont {J.}~\bibnamefont {Zakrzewski}},\ }\href {\doibase
  10.1088/0034-4885/78/6/066001} {\bibfield  {journal} {\bibinfo  {journal}
  {Reports on Progress in Physics}\ }\textbf {\bibinfo {volume} {78}},\
  \bibinfo {pages} {066001} (\bibinfo {year} {2015})}\BibitemShut {NoStop}%
\bibitem [{\citenamefont {Boninsegni}\ and\ \citenamefont
  {Prokof'ev}(2012)}]{boninsegni_2012}%
  \BibitemOpen
  \bibfield  {author} {\bibinfo {author} {\bibfnamefont {M.}~\bibnamefont
  {Boninsegni}}\ and\ \bibinfo {author} {\bibfnamefont {N.~V.}\ \bibnamefont
  {Prokof'ev}},\ }\href {\doibase 10.1103/RevModPhys.84.759} {\bibfield
  {journal} {\bibinfo  {journal} {Rev. Mod. Phys.}\ }\textbf {\bibinfo {volume}
  {84}},\ \bibinfo {pages} {759} (\bibinfo {year} {2012})}\BibitemShut
  {NoStop}%
\bibitem [{\citenamefont {Menotti}\ \emph {et~al.}(2007)\citenamefont
  {Menotti}, \citenamefont {Trefzger},\ and\ \citenamefont
  {Lewenstein}}]{menotti_2007}%
  \BibitemOpen
  \bibfield  {author} {\bibinfo {author} {\bibfnamefont {C.}~\bibnamefont
  {Menotti}}, \bibinfo {author} {\bibfnamefont {C.}~\bibnamefont {Trefzger}}, \
  and\ \bibinfo {author} {\bibfnamefont {M.}~\bibnamefont {Lewenstein}},\
  }\href {\doibase 10.1103/PhysRevLett.98.235301} {\bibfield  {journal}
  {\bibinfo  {journal} {Phys. Rev. Lett.}\ }\textbf {\bibinfo {volume} {98}},\
  \bibinfo {pages} {235301} (\bibinfo {year} {2007})}\BibitemShut {NoStop}%
\bibitem [{\citenamefont {Wu}\ and\ \citenamefont {Tu}(2020)}]{wu_2020}%
  \BibitemOpen
  \bibfield  {author} {\bibinfo {author} {\bibfnamefont {H.-K.}\ \bibnamefont
  {Wu}}\ and\ \bibinfo {author} {\bibfnamefont {W.-L.}\ \bibnamefont {Tu}},\
  }\href {\doibase 10.1103/PhysRevA.102.053306} {\bibfield  {journal} {\bibinfo
   {journal} {Phys. Rev. A}\ }\textbf {\bibinfo {volume} {102}},\ \bibinfo
  {pages} {053306} (\bibinfo {year} {2020})}\BibitemShut {NoStop}%
\bibitem [{\citenamefont {Cao}\ \emph {et~al.}(2018{\natexlab{a}})\citenamefont
  {Cao}, \citenamefont {Fatemi}, \citenamefont {Fang}, \citenamefont
  {Watanabe}, \citenamefont {Taniguchi}, \citenamefont {Kaxiras},\ and\
  \citenamefont {Jarillo-Herrero}}]{cao_2018_1}%
  \BibitemOpen
  \bibfield  {author} {\bibinfo {author} {\bibfnamefont {Y.}~\bibnamefont
  {Cao}}, \bibinfo {author} {\bibfnamefont {V.}~\bibnamefont {Fatemi}},
  \bibinfo {author} {\bibfnamefont {S.}~\bibnamefont {Fang}}, \bibinfo {author}
  {\bibfnamefont {K.}~\bibnamefont {Watanabe}}, \bibinfo {author}
  {\bibfnamefont {T.}~\bibnamefont {Taniguchi}}, \bibinfo {author}
  {\bibfnamefont {E.}~\bibnamefont {Kaxiras}}, \ and\ \bibinfo {author}
  {\bibfnamefont {P.}~\bibnamefont {Jarillo-Herrero}},\ }\href
  {https://doi.org/10.1038/nature26160} {\bibfield  {journal} {\bibinfo
  {journal} {Nature}\ }\textbf {\bibinfo {volume} {556}},\ \bibinfo {pages}
  {43} (\bibinfo {year} {2018}{\natexlab{a}})},\ \bibinfo {note}
  {article}\BibitemShut {NoStop}%
\bibitem [{\citenamefont {Cao}\ \emph {et~al.}(2018{\natexlab{b}})\citenamefont
  {Cao}, \citenamefont {Fatemi}, \citenamefont {Demir}, \citenamefont {Fang},
  \citenamefont {Tomarken}, \citenamefont {Luo}, \citenamefont
  {Sanchez-Yamagishi}, \citenamefont {Watanabe}, \citenamefont {Taniguchi},
  \citenamefont {Kaxiras}, \citenamefont {Ashoori},\ and\ \citenamefont
  {Jarillo-Herrero}}]{cao_2018_2}%
  \BibitemOpen
  \bibfield  {author} {\bibinfo {author} {\bibfnamefont {Y.}~\bibnamefont
  {Cao}}, \bibinfo {author} {\bibfnamefont {V.}~\bibnamefont {Fatemi}},
  \bibinfo {author} {\bibfnamefont {A.}~\bibnamefont {Demir}}, \bibinfo
  {author} {\bibfnamefont {S.}~\bibnamefont {Fang}}, \bibinfo {author}
  {\bibfnamefont {S.~L.}\ \bibnamefont {Tomarken}}, \bibinfo {author}
  {\bibfnamefont {J.~Y.}\ \bibnamefont {Luo}}, \bibinfo {author} {\bibfnamefont
  {J.~D.}\ \bibnamefont {Sanchez-Yamagishi}}, \bibinfo {author} {\bibfnamefont
  {K.}~\bibnamefont {Watanabe}}, \bibinfo {author} {\bibfnamefont
  {T.}~\bibnamefont {Taniguchi}}, \bibinfo {author} {\bibfnamefont
  {E.}~\bibnamefont {Kaxiras}}, \bibinfo {author} {\bibfnamefont {R.~C.}\
  \bibnamefont {Ashoori}}, \ and\ \bibinfo {author} {\bibfnamefont
  {P.}~\bibnamefont {Jarillo-Herrero}},\ }\href
  {https://doi.org/10.1038/nature26154} {\bibfield  {journal} {\bibinfo
  {journal} {Nature}\ }\textbf {\bibinfo {volume} {556}},\ \bibinfo {pages}
  {80} (\bibinfo {year} {2018}{\natexlab{b}})}\BibitemShut {NoStop}%
\bibitem [{\citenamefont {Lian}\ \emph {et~al.}(2019)\citenamefont {Lian},
  \citenamefont {Wang},\ and\ \citenamefont {Bernevig}}]{lian_2019}%
  \BibitemOpen
  \bibfield  {author} {\bibinfo {author} {\bibfnamefont {B.}~\bibnamefont
  {Lian}}, \bibinfo {author} {\bibfnamefont {Z.}~\bibnamefont {Wang}}, \ and\
  \bibinfo {author} {\bibfnamefont {B.~A.}\ \bibnamefont {Bernevig}},\ }\href
  {\doibase 10.1103/PhysRevLett.122.257002} {\bibfield  {journal} {\bibinfo
  {journal} {Phys. Rev. Lett.}\ }\textbf {\bibinfo {volume} {122}},\ \bibinfo
  {pages} {257002} (\bibinfo {year} {2019})}\BibitemShut {NoStop}%
\bibitem [{\citenamefont {Wu}\ \emph {et~al.}(2018)\citenamefont {Wu},
  \citenamefont {MacDonald},\ and\ \citenamefont {Martin}}]{wu_2018}%
  \BibitemOpen
  \bibfield  {author} {\bibinfo {author} {\bibfnamefont {F.}~\bibnamefont
  {Wu}}, \bibinfo {author} {\bibfnamefont {A.~H.}\ \bibnamefont {MacDonald}}, \
  and\ \bibinfo {author} {\bibfnamefont {I.}~\bibnamefont {Martin}},\ }\href
  {\doibase 10.1103/PhysRevLett.121.257001} {\bibfield  {journal} {\bibinfo
  {journal} {Phys. Rev. Lett.}\ }\textbf {\bibinfo {volume} {121}},\ \bibinfo
  {pages} {257001} (\bibinfo {year} {2018})}\BibitemShut {NoStop}%
\bibitem [{\citenamefont {Gonz\'alez}\ and\ \citenamefont
  {Stauber}(2019)}]{gonzalez_2019_1}%
  \BibitemOpen
  \bibfield  {author} {\bibinfo {author} {\bibfnamefont {J.}~\bibnamefont
  {Gonz\'alez}}\ and\ \bibinfo {author} {\bibfnamefont {T.}~\bibnamefont
  {Stauber}},\ }\href {\doibase 10.1103/PhysRevLett.122.026801} {\bibfield
  {journal} {\bibinfo  {journal} {Phys. Rev. Lett.}\ }\textbf {\bibinfo
  {volume} {122}},\ \bibinfo {pages} {026801} (\bibinfo {year}
  {2019})}\BibitemShut {NoStop}%
\bibitem [{\citenamefont {Gonz\'alez-Tudela}\ and\ \citenamefont
  {Cirac}(2019)}]{gonzalez_2019_2}%
  \BibitemOpen
  \bibfield  {author} {\bibinfo {author} {\bibfnamefont {A.}~\bibnamefont
  {Gonz\'alez-Tudela}}\ and\ \bibinfo {author} {\bibfnamefont {J.~I.}\
  \bibnamefont {Cirac}},\ }\href {\doibase 10.1103/PhysRevA.100.053604}
  {\bibfield  {journal} {\bibinfo  {journal} {Phys. Rev. A}\ }\textbf {\bibinfo
  {volume} {100}},\ \bibinfo {pages} {053604} (\bibinfo {year}
  {2019})}\BibitemShut {NoStop}%
\bibitem [{\citenamefont {Pizarro}\ \emph {et~al.}(2019)\citenamefont
  {Pizarro}, \citenamefont {Calder{\'{o}}n},\ and\ \citenamefont
  {Bascones}}]{pizarro_2019}%
  \BibitemOpen
  \bibfield  {author} {\bibinfo {author} {\bibfnamefont {J.~M.}\ \bibnamefont
  {Pizarro}}, \bibinfo {author} {\bibfnamefont {M.~J.}\ \bibnamefont
  {Calder{\'{o}}n}}, \ and\ \bibinfo {author} {\bibfnamefont {E.}~\bibnamefont
  {Bascones}},\ }\href {\doibase 10.1088/2399-6528/ab0fa9} {\bibfield
  {journal} {\bibinfo  {journal} {Journal of Physics Communications}\ }\textbf
  {\bibinfo {volume} {3}},\ \bibinfo {pages} {035024} (\bibinfo {year}
  {2019})}\BibitemShut {NoStop}%
\bibitem [{\citenamefont {Mahapatra}\ \emph {et~al.}(2020)\citenamefont
  {Mahapatra}, \citenamefont {Ghawri}, \citenamefont {Garg}, \citenamefont
  {Mandal}, \citenamefont {Watanabe}, \citenamefont {Taniguchi}, \citenamefont
  {Jain}, \citenamefont {Mukerjee},\ and\ \citenamefont
  {Ghosh}}]{mahapatra_2020}%
  \BibitemOpen
  \bibfield  {author} {\bibinfo {author} {\bibfnamefont {P.~S.}\ \bibnamefont
  {Mahapatra}}, \bibinfo {author} {\bibfnamefont {B.}~\bibnamefont {Ghawri}},
  \bibinfo {author} {\bibfnamefont {M.}~\bibnamefont {Garg}}, \bibinfo {author}
  {\bibfnamefont {S.}~\bibnamefont {Mandal}}, \bibinfo {author} {\bibfnamefont
  {K.}~\bibnamefont {Watanabe}}, \bibinfo {author} {\bibfnamefont
  {T.}~\bibnamefont {Taniguchi}}, \bibinfo {author} {\bibfnamefont
  {M.}~\bibnamefont {Jain}}, \bibinfo {author} {\bibfnamefont {S.}~\bibnamefont
  {Mukerjee}}, \ and\ \bibinfo {author} {\bibfnamefont {A.}~\bibnamefont
  {Ghosh}},\ }\href {\doibase 10.1103/PhysRevLett.125.226802} {\bibfield
  {journal} {\bibinfo  {journal} {Phys. Rev. Lett.}\ }\textbf {\bibinfo
  {volume} {125}},\ \bibinfo {pages} {226802} (\bibinfo {year}
  {2020})}\BibitemShut {NoStop}%
\bibitem [{\citenamefont {Luo}\ and\ \citenamefont {Zhang}(2021)}]{luo_2021}%
  \BibitemOpen
  \bibfield  {author} {\bibinfo {author} {\bibfnamefont {X.-W.}\ \bibnamefont
  {Luo}}\ and\ \bibinfo {author} {\bibfnamefont {C.}~\bibnamefont {Zhang}},\
  }\href {https://arxiv.org/abs/2008.01351} {\bibfield  {journal} {\bibinfo
  {journal} {arXiv:2008.01351 (accepted in PRL)}\ } (\bibinfo {year}
  {2021})}\BibitemShut {NoStop}%
\bibitem [{\citenamefont {Trefzger}\ \emph {et~al.}(2009)\citenamefont
  {Trefzger}, \citenamefont {Menotti},\ and\ \citenamefont
  {Lewenstein}}]{trefzger_2009}%
  \BibitemOpen
  \bibfield  {author} {\bibinfo {author} {\bibfnamefont {C.}~\bibnamefont
  {Trefzger}}, \bibinfo {author} {\bibfnamefont {C.}~\bibnamefont {Menotti}}, \
  and\ \bibinfo {author} {\bibfnamefont {M.}~\bibnamefont {Lewenstein}},\
  }\href {\doibase 10.1103/PhysRevLett.103.035304} {\bibfield  {journal}
  {\bibinfo  {journal} {Phys. Rev. Lett.}\ }\textbf {\bibinfo {volume} {103}},\
  \bibinfo {pages} {035304} (\bibinfo {year} {2009})}\BibitemShut {NoStop}%
\bibitem [{\citenamefont {Safavi-Naini}\ \emph {et~al.}(2013)\citenamefont
  {Safavi-Naini}, \citenamefont {S\"oyler}, \citenamefont {Pupillo},
  \citenamefont {Sadeghpour},\ and\ \citenamefont
  {Capogrosso-Sansone}}]{naini_2013}%
  \BibitemOpen
  \bibfield  {author} {\bibinfo {author} {\bibfnamefont {A.}~\bibnamefont
  {Safavi-Naini}}, \bibinfo {author} {\bibfnamefont {{\c{S}}.~G.}\ \bibnamefont
  {S\"oyler}}, \bibinfo {author} {\bibfnamefont {G.}~\bibnamefont {Pupillo}},
  \bibinfo {author} {\bibfnamefont {H.~R.}\ \bibnamefont {Sadeghpour}}, \ and\
  \bibinfo {author} {\bibfnamefont {B.}~\bibnamefont {Capogrosso-Sansone}},\
  }\href {\doibase 10.1088/1367-2630/15/1/013036} {\bibfield  {journal}
  {\bibinfo  {journal} {New Journal of Physics}\ }\textbf {\bibinfo {volume}
  {15}},\ \bibinfo {pages} {013036} (\bibinfo {year} {2013})}\BibitemShut
  {NoStop}%
\bibitem [{\citenamefont {Macia}\ \emph {et~al.}(2014)\citenamefont {Macia},
  \citenamefont {Astrakharchik}, \citenamefont {Mazzanti}, \citenamefont
  {Giorgini},\ and\ \citenamefont {Boronat}}]{macia_2014}%
  \BibitemOpen
  \bibfield  {author} {\bibinfo {author} {\bibfnamefont {A.}~\bibnamefont
  {Macia}}, \bibinfo {author} {\bibfnamefont {G.~E.}\ \bibnamefont
  {Astrakharchik}}, \bibinfo {author} {\bibfnamefont {F.}~\bibnamefont
  {Mazzanti}}, \bibinfo {author} {\bibfnamefont {S.}~\bibnamefont {Giorgini}},
  \ and\ \bibinfo {author} {\bibfnamefont {J.}~\bibnamefont {Boronat}},\ }\href
  {\doibase 10.1103/PhysRevA.90.043623} {\bibfield  {journal} {\bibinfo
  {journal} {Phys. Rev. A}\ }\textbf {\bibinfo {volume} {90}},\ \bibinfo
  {pages} {043623} (\bibinfo {year} {2014})}\BibitemShut {NoStop}%
\bibitem [{\citenamefont {Ng}(2015)}]{ng_2015}%
  \BibitemOpen
  \bibfield  {author} {\bibinfo {author} {\bibfnamefont {K.-K.}\ \bibnamefont
  {Ng}},\ }\href {\doibase 10.1103/PhysRevB.91.054516} {\bibfield  {journal}
  {\bibinfo  {journal} {Phys. Rev. B}\ }\textbf {\bibinfo {volume} {91}},\
  \bibinfo {pages} {054516} (\bibinfo {year} {2015})}\BibitemShut {NoStop}%
\bibitem [{\citenamefont {Buonsante}\ \emph {et~al.}(2004)\citenamefont
  {Buonsante}, \citenamefont {Penna},\ and\ \citenamefont
  {Vezzani}}]{Buonsante_04}%
  \BibitemOpen
  \bibfield  {author} {\bibinfo {author} {\bibfnamefont {P.}~\bibnamefont
  {Buonsante}}, \bibinfo {author} {\bibfnamefont {V.}~\bibnamefont {Penna}}, \
  and\ \bibinfo {author} {\bibfnamefont {A.}~\bibnamefont {Vezzani}},\ }\href
  {\doibase 10.1103/PhysRevA.70.061603} {\bibfield  {journal} {\bibinfo
  {journal} {Phys. Rev. A}\ }\textbf {\bibinfo {volume} {70}},\ \bibinfo
  {pages} {061603} (\bibinfo {year} {2004})}\BibitemShut {NoStop}%
\bibitem [{\citenamefont {Yamamoto}(2009)}]{Yamamoto_09}%
  \BibitemOpen
  \bibfield  {author} {\bibinfo {author} {\bibfnamefont {D.}~\bibnamefont
  {Yamamoto}},\ }\href {\doibase 10.1103/PhysRevB.79.144427} {\bibfield
  {journal} {\bibinfo  {journal} {Phys. Rev. B}\ }\textbf {\bibinfo {volume}
  {79}},\ \bibinfo {pages} {144427} (\bibinfo {year} {2009})}\BibitemShut
  {NoStop}%
\bibitem [{\citenamefont {Pisarski}\ \emph {et~al.}(2011)\citenamefont
  {Pisarski}, \citenamefont {Jones},\ and\ \citenamefont
  {Gooding}}]{Pisarski_11}%
  \BibitemOpen
  \bibfield  {author} {\bibinfo {author} {\bibfnamefont {P.}~\bibnamefont
  {Pisarski}}, \bibinfo {author} {\bibfnamefont {R.~M.}\ \bibnamefont {Jones}},
  \ and\ \bibinfo {author} {\bibfnamefont {R.~J.}\ \bibnamefont {Gooding}},\
  }\href {\doibase 10.1103/PhysRevA.83.053608} {\bibfield  {journal} {\bibinfo
  {journal} {Phys. Rev. A}\ }\textbf {\bibinfo {volume} {83}},\ \bibinfo
  {pages} {053608} (\bibinfo {year} {2011})}\BibitemShut {NoStop}%
\bibitem [{\citenamefont {McIntosh}\ \emph {et~al.}(2012)\citenamefont
  {McIntosh}, \citenamefont {Pisarski}, \citenamefont {Gooding},\ and\
  \citenamefont {Zaremba}}]{McIntosh_12}%
  \BibitemOpen
  \bibfield  {author} {\bibinfo {author} {\bibfnamefont {T.}~\bibnamefont
  {McIntosh}}, \bibinfo {author} {\bibfnamefont {P.}~\bibnamefont {Pisarski}},
  \bibinfo {author} {\bibfnamefont {R.~J.}\ \bibnamefont {Gooding}}, \ and\
  \bibinfo {author} {\bibfnamefont {E.}~\bibnamefont {Zaremba}},\ }\href
  {\doibase 10.1103/PhysRevA.86.013623} {\bibfield  {journal} {\bibinfo
  {journal} {Phys. Rev. A}\ }\textbf {\bibinfo {volume} {86}},\ \bibinfo
  {pages} {013623} (\bibinfo {year} {2012})}\BibitemShut {NoStop}%
\bibitem [{\citenamefont {L\"uhmann}(2013{\natexlab{a}})}]{luhmann_13}%
  \BibitemOpen
  \bibfield  {author} {\bibinfo {author} {\bibfnamefont {D.-S.}\ \bibnamefont
  {L\"uhmann}},\ }\href {\doibase 10.1103/PhysRevA.87.043619} {\bibfield
  {journal} {\bibinfo  {journal} {Phys. Rev. A}\ }\textbf {\bibinfo {volume}
  {87}},\ \bibinfo {pages} {043619} (\bibinfo {year}
  {2013}{\natexlab{a}})}\BibitemShut {NoStop}%
\bibitem [{\citenamefont {van Oosten}\ \emph {et~al.}(2001)\citenamefont {van
  Oosten}, \citenamefont {van~der Straten},\ and\ \citenamefont
  {Stoof}}]{oosten_2001}%
  \BibitemOpen
  \bibfield  {author} {\bibinfo {author} {\bibfnamefont {D.}~\bibnamefont {van
  Oosten}}, \bibinfo {author} {\bibfnamefont {P.}~\bibnamefont {van~der
  Straten}}, \ and\ \bibinfo {author} {\bibfnamefont {H.~T.~C.}\ \bibnamefont
  {Stoof}},\ }\href {\doibase 10.1103/PhysRevA.63.053601} {\bibfield  {journal}
  {\bibinfo  {journal} {Phys. Rev. A}\ }\textbf {\bibinfo {volume} {63}},\
  \bibinfo {pages} {053601} (\bibinfo {year} {2001})}\BibitemShut {NoStop}%
\bibitem [{\citenamefont {Iskin}\ and\ \citenamefont
  {Freericks}(2009)}]{iskin_2009}%
  \BibitemOpen
  \bibfield  {author} {\bibinfo {author} {\bibfnamefont {M.}~\bibnamefont
  {Iskin}}\ and\ \bibinfo {author} {\bibfnamefont {J.~K.}\ \bibnamefont
  {Freericks}},\ }\href {\doibase 10.1103/PhysRevA.79.053634} {\bibfield
  {journal} {\bibinfo  {journal} {Phys. Rev. A}\ }\textbf {\bibinfo {volume}
  {79}},\ \bibinfo {pages} {053634} (\bibinfo {year} {2009})}\BibitemShut
  {NoStop}%
\bibitem [{\citenamefont {Bai}\ \emph {et~al.}(2020)\citenamefont {Bai},
  \citenamefont {Gaur}, \citenamefont {Sable}, \citenamefont {Bandyopadhyay},
  \citenamefont {Suthar},\ and\ \citenamefont {Angom}}]{bai_2020}%
  \BibitemOpen
  \bibfield  {author} {\bibinfo {author} {\bibfnamefont {R.}~\bibnamefont
  {Bai}}, \bibinfo {author} {\bibfnamefont {D.}~\bibnamefont {Gaur}}, \bibinfo
  {author} {\bibfnamefont {H.}~\bibnamefont {Sable}}, \bibinfo {author}
  {\bibfnamefont {S.}~\bibnamefont {Bandyopadhyay}}, \bibinfo {author}
  {\bibfnamefont {K.}~\bibnamefont {Suthar}}, \ and\ \bibinfo {author}
  {\bibfnamefont {D.}~\bibnamefont {Angom}},\ }\href {\doibase
  10.1103/PhysRevA.102.043309} {\bibfield  {journal} {\bibinfo  {journal}
  {Phys. Rev. A}\ }\textbf {\bibinfo {volume} {102}},\ \bibinfo {pages}
  {043309} (\bibinfo {year} {2020})}\BibitemShut {NoStop}%
\bibitem [{\citenamefont {Rokhsar}\ and\ \citenamefont
  {Kotliar}(1991)}]{rokhsar_1991}%
  \BibitemOpen
  \bibfield  {author} {\bibinfo {author} {\bibfnamefont {D.~S.}\ \bibnamefont
  {Rokhsar}}\ and\ \bibinfo {author} {\bibfnamefont {B.~G.}\ \bibnamefont
  {Kotliar}},\ }\href {\doibase 10.1103/PhysRevB.44.10328} {\bibfield
  {journal} {\bibinfo  {journal} {Phys. Rev. B}\ }\textbf {\bibinfo {volume}
  {44}},\ \bibinfo {pages} {10328} (\bibinfo {year} {1991})}\BibitemShut
  {NoStop}%
\bibitem [{\citenamefont {Sheshadri}\ \emph {et~al.}(1993)\citenamefont
  {Sheshadri}, \citenamefont {Krishnamurthy}, \citenamefont {Pandit},\ and\
  \citenamefont {Ramakrishnan}}]{sheshadri_1993}%
  \BibitemOpen
  \bibfield  {author} {\bibinfo {author} {\bibfnamefont {K.}~\bibnamefont
  {Sheshadri}}, \bibinfo {author} {\bibfnamefont {H.~R.}\ \bibnamefont
  {Krishnamurthy}}, \bibinfo {author} {\bibfnamefont {R.}~\bibnamefont
  {Pandit}}, \ and\ \bibinfo {author} {\bibfnamefont {T.~V.}\ \bibnamefont
  {Ramakrishnan}},\ }\href {\doibase
  https://doi.org/10.1209/0295-5075/22/4/004} {\bibfield  {journal} {\bibinfo
  {journal} {EPL}\ }\textbf {\bibinfo {volume} {22}},\ \bibinfo {pages} {257}
  (\bibinfo {year} {1993})}\BibitemShut {NoStop}%
\bibitem [{\citenamefont {Bai}\ \emph {et~al.}(2018)\citenamefont {Bai},
  \citenamefont {Bandyopadhyay}, \citenamefont {Pal}, \citenamefont {Suthar},\
  and\ \citenamefont {Angom}}]{bai_2018}%
  \BibitemOpen
  \bibfield  {author} {\bibinfo {author} {\bibfnamefont {R.}~\bibnamefont
  {Bai}}, \bibinfo {author} {\bibfnamefont {S.}~\bibnamefont {Bandyopadhyay}},
  \bibinfo {author} {\bibfnamefont {S.}~\bibnamefont {Pal}}, \bibinfo {author}
  {\bibfnamefont {K.}~\bibnamefont {Suthar}}, \ and\ \bibinfo {author}
  {\bibfnamefont {D.}~\bibnamefont {Angom}},\ }\href {\doibase
  10.1103/PhysRevA.98.023606} {\bibfield  {journal} {\bibinfo  {journal} {Phys.
  Rev. A}\ }\textbf {\bibinfo {volume} {98}},\ \bibinfo {pages} {023606}
  (\bibinfo {year} {2018})}\BibitemShut {NoStop}%
\bibitem [{\citenamefont {Malakar}\ \emph {et~al.}(2020)\citenamefont
  {Malakar}, \citenamefont {Ray}, \citenamefont {Sinha},\ and\ \citenamefont
  {Angom}}]{malakar_2020}%
  \BibitemOpen
  \bibfield  {author} {\bibinfo {author} {\bibfnamefont {M.}~\bibnamefont
  {Malakar}}, \bibinfo {author} {\bibfnamefont {S.}~\bibnamefont {Ray}},
  \bibinfo {author} {\bibfnamefont {S.}~\bibnamefont {Sinha}}, \ and\ \bibinfo
  {author} {\bibfnamefont {D.}~\bibnamefont {Angom}},\ }\href {\doibase
  10.1103/PhysRevB.102.184515} {\bibfield  {journal} {\bibinfo  {journal}
  {Phys. Rev. B}\ }\textbf {\bibinfo {volume} {102}},\ \bibinfo {pages}
  {184515} (\bibinfo {year} {2020})}\BibitemShut {NoStop}%
\bibitem [{\citenamefont {L\"uhmann}(2013{\natexlab{b}})}]{luhmann_2013}%
  \BibitemOpen
  \bibfield  {author} {\bibinfo {author} {\bibfnamefont {D.-S.}\ \bibnamefont
  {L\"uhmann}},\ }\href {\doibase 10.1103/PhysRevA.87.043619} {\bibfield
  {journal} {\bibinfo  {journal} {Phys. Rev. A}\ }\textbf {\bibinfo {volume}
  {87}},\ \bibinfo {pages} {043619} (\bibinfo {year}
  {2013}{\natexlab{b}})}\BibitemShut {NoStop}%
\bibitem [{\citenamefont {D\"ur}\ \emph {et~al.}(2000)\citenamefont {D\"ur},
  \citenamefont {Vidal},\ and\ \citenamefont {Cirac}}]{dur_2000}%
  \BibitemOpen
  \bibfield  {author} {\bibinfo {author} {\bibfnamefont {W.}~\bibnamefont
  {D\"ur}}, \bibinfo {author} {\bibfnamefont {G.}~\bibnamefont {Vidal}}, \ and\
  \bibinfo {author} {\bibfnamefont {J.~I.}\ \bibnamefont {Cirac}},\ }\href
  {\doibase 10.1103/PhysRevA.62.062314} {\bibfield  {journal} {\bibinfo
  {journal} {Phys. Rev. A}\ }\textbf {\bibinfo {volume} {62}},\ \bibinfo
  {pages} {062314} (\bibinfo {year} {2000})}\BibitemShut {NoStop}%
\bibitem [{\citenamefont {Safavi-Naini}\ \emph {et~al.}(2012)\citenamefont
  {Safavi-Naini}, \citenamefont {von Stecher}, \citenamefont
  {Capogrosso-Sansone},\ and\ \citenamefont {Rittenhouse}}]{naini_2012}%
  \BibitemOpen
  \bibfield  {author} {\bibinfo {author} {\bibfnamefont {A.}~\bibnamefont
  {Safavi-Naini}}, \bibinfo {author} {\bibfnamefont {J.}~\bibnamefont {von
  Stecher}}, \bibinfo {author} {\bibfnamefont {B.}~\bibnamefont
  {Capogrosso-Sansone}}, \ and\ \bibinfo {author} {\bibfnamefont {S.~T.}\
  \bibnamefont {Rittenhouse}},\ }\href {\doibase
  10.1103/PhysRevLett.109.135302} {\bibfield  {journal} {\bibinfo  {journal}
  {Phys. Rev. Lett.}\ }\textbf {\bibinfo {volume} {109}},\ \bibinfo {pages}
  {135302} (\bibinfo {year} {2012})}\BibitemShut {NoStop}%
\bibitem [{\citenamefont {Mahmud}\ \emph {et~al.}(2011)\citenamefont {Mahmud},
  \citenamefont {Duchon}, \citenamefont {Kato}, \citenamefont {Kawashima},
  \citenamefont {Scalettar},\ and\ \citenamefont {Trivedi}}]{mahmud_2011}%
  \BibitemOpen
  \bibfield  {author} {\bibinfo {author} {\bibfnamefont {K.~W.}\ \bibnamefont
  {Mahmud}}, \bibinfo {author} {\bibfnamefont {E.~N.}\ \bibnamefont {Duchon}},
  \bibinfo {author} {\bibfnamefont {Y.}~\bibnamefont {Kato}}, \bibinfo {author}
  {\bibfnamefont {N.}~\bibnamefont {Kawashima}}, \bibinfo {author}
  {\bibfnamefont {R.~T.}\ \bibnamefont {Scalettar}}, \ and\ \bibinfo {author}
  {\bibfnamefont {N.}~\bibnamefont {Trivedi}},\ }\href {\doibase
  10.1103/PhysRevB.84.054302} {\bibfield  {journal} {\bibinfo  {journal} {Phys.
  Rev. B}\ }\textbf {\bibinfo {volume} {84}},\ \bibinfo {pages} {054302}
  (\bibinfo {year} {2011})}\BibitemShut {NoStop}%
\bibitem [{\citenamefont {Fang}\ \emph {et~al.}(2011)\citenamefont {Fang},
  \citenamefont {Chung}, \citenamefont {Ma}, \citenamefont {Chen},\ and\
  \citenamefont {Wang}}]{fang_2011}%
  \BibitemOpen
  \bibfield  {author} {\bibinfo {author} {\bibfnamefont {S.}~\bibnamefont
  {Fang}}, \bibinfo {author} {\bibfnamefont {C.-M.}\ \bibnamefont {Chung}},
  \bibinfo {author} {\bibfnamefont {P.~N.}\ \bibnamefont {Ma}}, \bibinfo
  {author} {\bibfnamefont {P.}~\bibnamefont {Chen}}, \ and\ \bibinfo {author}
  {\bibfnamefont {D.-W.}\ \bibnamefont {Wang}},\ }\href {\doibase
  10.1103/PhysRevA.83.031605} {\bibfield  {journal} {\bibinfo  {journal} {Phys.
  Rev. A}\ }\textbf {\bibinfo {volume} {83}},\ \bibinfo {pages} {031605}
  (\bibinfo {year} {2011})}\BibitemShut {NoStop}%
\bibitem [{\citenamefont {de~Forges~de Parny}\ \emph
  {et~al.}(2012)\citenamefont {de~Forges~de Parny}, \citenamefont {H\'ebert},
  \citenamefont {Rousseau},\ and\ \citenamefont {Batrouni}}]{parny_2012}%
  \BibitemOpen
  \bibfield  {author} {\bibinfo {author} {\bibfnamefont {L.}~\bibnamefont
  {de~Forges~de Parny}}, \bibinfo {author} {\bibfnamefont {F.}~\bibnamefont
  {H\'ebert}}, \bibinfo {author} {\bibfnamefont {V.~G.}\ \bibnamefont
  {Rousseau}}, \ and\ \bibinfo {author} {\bibfnamefont {G.~G.}\ \bibnamefont
  {Batrouni}},\ }\href {\doibase 10.1140/epjb/e2012-30055-9} {\bibfield
  {journal} {\bibinfo  {journal} {Eur. Phys. J. B}\ }\textbf {\bibinfo {volume}
  {85}},\ \bibinfo {pages} {169} (\bibinfo {year} {2012})}\BibitemShut
  {NoStop}%
\end{thebibliography}%
\bibliographystyle{apsrev4-1}

\end{document}